\documentclass[11pt]{article}

\usepackage[usenames,dvipsnames,table]{xcolor}
\usepackage[utf8]{inputenc}

\pdfoutput=1 
\usepackage{jheppub} 

\usepackage{graphicx}
\usepackage{amsmath, amssymb}
\usepackage{youngtab}
\usepackage{changepage}

\newcommand{\be}{\begin{eqnarray}}
\newcommand{\ee}{\end{eqnarray}}
\newcommand{\bea}{\begin{eqnarray}}
\newcommand{\eea}{\end{eqnarray}}

\newcommand{\nn}{\nonumber}
\newcommand{\bn}{\begin{enumerate}}
\newcommand{\en}{\end{enumerate}}

\def\Tr{\mathop{\text{Tr}}\nolimits}

\def\e{\mathrm{e}}

\newcommand{\ud}{\,\mathrm{d}} 
\newcommand{\udl}[1]{\mathrm{d} #1 \,}

\newcommand{\sbfunc}[1]{s_b\left( #1\right)}

\newcommand{\Gpq}[1]{\Gamma_e\left( #1\right)}

\def\ga{\alpha}
\def\gb{\beta}

\def\Gc{\Gamma}
\def\Gd{\Delta}
\def\gd{\delta}

\def\Gp{\Phi}

\title{Dualities from dualities: the sequential deconfinement technique}

\author[a]{Lea E.~Bottini,}
\author[b]{Chiung Hwang,}
\author[c,d]{Sara Pasquetti,}
\author[a]{and Matteo Sacchi}

\affiliation[a]{Mathematical Institute, University of Oxford, Woodstock Road, Oxford, OX2 6GG, United Kingdom}
\affiliation[b]{Department of Applied Mathematics and Theoretical Physics, University of Cambridge, Cambridge CB3 0WA, United Kingdom}
\affiliation[c]{Dipartimento di Fisica, Università di Milano-Bicocca,
Piazza della Scienza 3, I-20126 Milano, Italy}
\affiliation[d]{INFN, sezione di Milano-Bicocca, Piazza della Scienza 3, I-20126 Milano, Italy}

\emailAdd{lea.bottini@maths.ox.ac.uk}
\emailAdd{ch911@cam.ac.uk}
\emailAdd{sara.pasquetti@gmail.com} 
\emailAdd{matteo.sacchi@maths.ox.ac.uk}

\abstract{It is an interesting question whether a given infra-red duality between quantum field theories can be explained in terms of other more elementary dualities. For example recently it has been shown that mirror dualities can be obtained by iterative applications of Seiberg-like dualities. In this paper we continue this line of investigation focusing on theories with tensor matter. In such cases one can apply the idea of deconfinement, which consists of trading the tensor matter for extra gauge nodes by means of a suitable elementary duality. This gives an auxiliary dual frame which can then be manipulated with further dualizations, in an iterative procedure eventually yielding an interesting dual description of the original theory. The sequential deconfinement technique has avatars in different areas of mathematical physics, such as the study of hypergeometric and elliptic hypergeometric integral identities or of $2d$ free field correlators. We discuss various examples in the context $4d$ $\mathcal{N}=1$ supersymmetric theories, which are related to elliptic hypergeometric integrals. These include a new self-duality involving a quiver theory which exhibits a non-trivial global symmetry enhancement to $E_6$.
}

\begin{document} 

\maketitle
\flushbottom

\section{Introduction}

One of the most interesting non-perturbative phenomena that can characterize a quantum field theory is the one of infra-red (IR) duality. This occurs when two distinct microscopic theories flow to the same fixed point at low energies. A famous example of this is the Seiberg duality \cite{Seiberg:1994pq}. This applies to SQCD in $4d$ with the minimal amount of supersymmetry, namely $\mathcal{N}=1$, which is also the set-up we will consider in this paper. The advantage of considering supersymmetric models is that we can compute exactly various Renormalization Group (RG) flow invariants, which we can then match between the two theories to provide strong evidence of the duality. One type of such invariants are supersymmetric partition functions on various compact manifolds, which can be expressed in the much more manageable form of ordinary matrix integrals using localization techniques (see \cite{Pestun:2016zxk} for a review and references therein).

An intriguing question in this context is whether there is some minimal set of fundamental dualities from which all the others can be derived. For example, if we consider quiver gauge theories, we can apply the Seiberg duality or variants thereof locally on some gauge nodes so to find new dual frames. This sort of manipulations can be performed very explicitly at the level of the integrals of supersymmetric partition functions. Very recently in \cite{Hwang:2021ulb}, building on the results of \cite{Bottini:2021vms}, it was shown that following such a strategy the $3d$ $\mathcal{N}=4$ mirror symmetry \cite{Intriligator:1996ex} as well as its $4d$ $\mathcal{N}=1$ ancestor \cite{Hwang:2020wpd} can be derived by a
piecewise dualization algorithm based on  two  duality moves, which can, in turn, be derived by sequentially applying the Aharony duality in $3d$ \cite{Aharony:1997gp} or the Intriligator--Pouliot (IP) duality in $4d$ \cite{Intriligator:1995ne}.

The set of possible manipulations that we can perform on a gauge theory using fundamental dualities can be drastically enlarged if we employ a technique called \emph{deconfinement}. This consists of trading a matter field in a rank-2 representation of the gauge group for an auxiliary gauge node by using an s-confining duality, in the sense of \cite{Csaki:1996zb}\footnote{According to \cite{Csaki:1996zb}, a theory is called s-confining when it admits an IR dual description everywhere in the moduli space in terms of a Wess--Zumino (WZ) model whose fields are mapped to gauge invariant operators in the original gauge theory and with a dynamically generated superpotential among them that preserves the same global symmetries as those of the gauge theory.}. In other words, we can find a dual frame where the rank-2 matter is absent, but we have an additional gauge node, and the two theories are related by an application of the s-confining duality on the new node. One can then perform new dualizations of the auxiliary quiver theory obtained from the deconfinement, which can lead to a non-trivial dual frame of the original theory.

Some examples of deconfinements in $4d$ $\mathcal{N}=1$ theories first appeared in the physics literature in \cite{Berkooz:1995km,Pouliot:1995me,Luty:1996cg,Garcia-Etxebarria:2012ypj,Garcia-Etxebarria:2013tba,Etxebarria:2021lmq}. 
Deconfinement appeared more recently also in lower dimensions. For example,  various $3d$ $\mathcal{N}=2$ dualities \cite{Pasquetti:2019uop,Pasquetti:2019tix,Benvenuti:2020gvy,Benvenuti:2021nwt,Nii:2016jzi} have been derived with 
an iterative (or sequential) application of the deconfinement procedure.
An example of deconfinement also appeared in the context of $2d$ $\mathcal{N}=(0,2)$ theories in \cite{Sacchi:2020pet}.

Interestingly the technique of deconfinement has also appeared elsewhere in the literature, even if in a different disguise. One strong evidence for a duality is the matching of the supersymmetric indices or partition functions of dual  theories. For example, the $4d$ supersymmetric index  \cite{Romelsberger:2005eg,Kinney:2005ej,Dolan:2008qi,Rastelli:2016tbz} can be written in terms of  elliptic functions, and each duality implies a highly non-trivial identity between elliptic hypergeometric integrals which are extensively discussed in the math literature, see for example \cite{Spiridonov:2009za,Spiridonov:2011hf}. In particular, in \cite{2003math......9252R,spiridonov2004theta} it was shown that the integral identities for various of the s-confining dualities of \cite{Csaki:1996zb} can be  derived by iterating integral identities  for some more elementary dualities like the Seiberg duality or the IP duality, with a strategy that is nothing but the deconfinement technique we described before in the field theory.

Another incarnation of the  technique of deconfinement  was observed in  \cite{Pasquetti:2019uop} where 
it was established a correspondence between the $\mathbb{S}^2\times\mathbb{S}^1$ partition function of certain $3d$ $\mathcal{N}=2$ gauge theories and the correlation functions of $2d$ CFTs in the free field realization. 
$2d$ free field correlators are expressed in terms of Dotsenko--Fateev  integrals which can be manipulated by iterating a fundamental set of integral identities. 
In  particular, in \cite{Pasquetti:2019uop} it was shown how the manipulations of the  Liouville 
3-point free field correlator leading to the DOZZ evaluation formula \cite{Dorn:1994xn,Zamolodchikov:1995aa} by iterative use of fundamental integral identities
discussed in  \cite{Fateev:2007qn} have a field theory avatar as the sequential deconfinement procedure for  the $3d$ $\mathcal{N}=2$ s-confining duality relating the $U(N)$ gauge theory with one adjoint and one fundamental flavors to a Wess--Zumino (WZ)  proposed in \cite{Benvenuti:2018bav}.
Building on the correspondence between $3d$ supersymmetric indices  and free field correlators, in \cite{Pasquetti:2019tix} various new $3d$ $\mathcal{N}=2$ dualities were found by uplifting known integral identities for $2d$ free field correlators in Liouville CFT  appearing in  \cite{Fateev:2007qn}. In particular, the integral kernel function that shows up in the multipoint Liouville  correlator has been related to the $3d$ $M[SU(N)]$ theory, which admits a further uplift to the $4d$ $\mathcal{N}=1$ $E[USp(2N)]$  theory, which was shown to provide the building block for E-string compactifications on a torus in \cite{Pasquetti:2019hxf}.

In this paper, we discuss various new $4d$ $\mathcal{N}=1$ IR dualities, most of which are derived using the technique of deconfinement. 
We start in Section \ref{sec:confining} by revisiting the known s-confining duality for the $USp(2N)$ gauge theory with one antisymmetric and six fundamental chiral fields \cite{Csaki:1996zb}\footnote{See also \cite{Hwang:2021xyw} for a geometric derivation of this duality in terms of a sphere compactification of the $6d$ $\mathcal{N}=(1,0)$ E-string SCFT.} and showing how to derive it by deconfining the antisymmetric field via the IP duality.  

We then consider various possible generalizations of this duality. In Section \ref{sec:confquiv}, we consider a linear quiver gauge theory with $USp$ nodes and matter in the fundamental, bifundamental and antisymmetric representation. Using deconfinement, we show that this theory is also dual to a simple Wess--Zumino model. 

In Section \ref{sec:crossleg}, we consider a similar linear quiver theory, but with a different arrangement of the fundamental fields, and show using deconfinement that it enjoys a self-duality that we call \emph{cross-leg duality}. Moreover, we propose that by suitably adding some singlet fields to this theory, it enjoys a global symmetry enhancement from the manifest $SU(3)^3$ global symmetry to $E_6$, and we perform some checks of this claim.

Finally, in Section \ref{sec:rankstab}, we propose a duality between the $USp(2N)$ gauge theory with one antisymmetric and $6+2k$ fundamental chirals, $2k$ of which interact with the antisymmetric with a cubic superpotential, and a quiver gauge theory that can be obtained by gauging one of the $USp(2k)$ symmetries of the $E[USp(2k)]$ theory of \cite{Pasquetti:2019hxf} with six fundamental chiral fields and some singlets. This duality is a $4d$ ancestor of the $3d$ \emph{rank stabilization duality} of \cite{Pasquetti:2019tix}, to which it reduces upon circle compactification followed by suitable real mass deformations. This is the only duality in this paper for which we don't present a derivation using deconfinement.

Even if we won't investigate the relation to  free field correlators
further in this paper, all the dualities that we are going to discuss can be shown to reduce in $3d$ to dualities that naturally arise from such connection with $2d$ free field correlators. The simplest example is the confining duality for $USp(2N)$ with one antisymmetric and six fundamentals of \cite{Csaki:1996zb}. In \cite{Benvenuti:2018bav} it was shown that this duality reduces in $3d$ to the confining duality for a $U(N)$ gauge theory with one adjoint and one fundamental flavors, which as we mentioned above is related to the  3-point correlator of Liouville theory. Similarly, the dualities that we will consider in Sections \ref{sec:confquiv} and \ref{sec:crossleg} can be related to integral identities for the free field correlators for the 3-point function of the Toda CFT \cite{Fateev:2007ab,Fateev:2008bm}.
Interestingly  the manipulations performed in CFT on the free field integrals to derive such identities are completely analogous to the deconfinement procedure we will discuss in sections Sections \ref{sec:confquiv} and \ref{sec:crossleg}.
Finally,  the rank stabilization duality is  related to the correlation function involving 3 primaries and $k$ degenerate operators in Liouville theory \cite{Pasquetti:2019tix,Fateev:2007qn}.

\newpage

The interplay between physics and math perspective fostered by localization has led to a significant progress in our understanding of supersymmetric dualities. The fact that the same tools, or different avatars of the same tool as we have seen for the deconfinement, appear both in physics and math indicates an even deeper connection and reinforces the hope that this interplay can help us organize the vast landscape of dualities.
\\

\noindent \textbf{Note added}: while completing this work, we became aware of \cite{stephane-sergio} which has some overlap with the results of Section \ref{sec:confining}. We thank the authors of \cite{stephane-sergio} for coordinating the submission.

\section{Confining duality for $USp(2N)$ gauge theory with antisymmetric matter}
\label{sec:confining}

In order to explain  the deconfinement strategy that we are going to use to derive new dualities in the next sections, we first apply it to re-derive the well-known duality  by Csaki, Skiba and Schmaltz between a $USp(2N)$ gauge theory and a WZ model \cite{Csaki:1996zb}.

\begin{figure}[t]
\centering
\includegraphics[width = 5 cm]{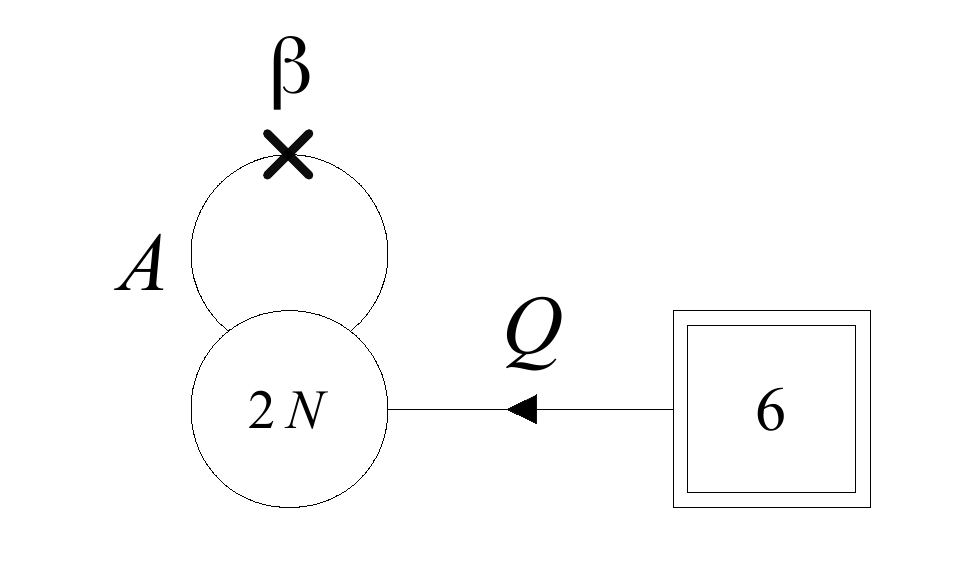}
\caption{Quiver diagram of the original gauge theory. The round node denotes the $USp(2N)$ gauge symmetry, while the square node the $SU(6)$ flavor symmetry. The line connecting them represents the chiral in the bifundamental representation of the two groups, while the arc denotes the chiral in the antisymmetric representation of the gauge group. The dash over the arc denotes the tower of $\beta_i$ singlets.}
\label{fig4}
\end{figure}

More precisely, the electric theory is a  $USp(2N)$ gauge theory with six fundamental chirals $Q_a$, one antisymmetric chiral $A$ and $N$ chiral singlets $\beta_i$ with superpotential\footnote{For $USp(2N)$ groups, we define the trace with the contraction of indices done with the antisymmetric tensor $J^{(N)}= \mathbb{I}_{N \times N} \otimes i \sigma_2$. For example, we have $\text{Tr}_N A = A^{i}_{\text{ } i} = J^{(N)}_{ij} A^{ji}$.}
\begin{equation}
	\mathcal{W}= \sum_{i=1}^N  \beta_i \text{Tr}_N A^i \,.
\end{equation}
The content of the theory is schematically represented in Figure \ref{fig4}.
The charges under the global symmetry group $SU(6)_x \times U(1)_t$\footnote{Throughout the paper, we label symmetries with the corresponding fugacities in the index.} are given in the following table:
\begin{table}[h]
\centering
\begin{tabular}{|c|c|c|c|}
\hline
 & $SU(6)_x$ & $U(1)_t$ & $U(1)_{R_0}$ \\
 	\hline
 	$A$ & \textbf{1} & 1 & 0 \\
 	$Q_a$ & {\bf 6} &  $\frac{1-N}{3}$ & $\frac{1}{3}$ \\
 	$\beta_i$ & \textbf{1} & $-i$ & 2 \\ \hline
\end{tabular}
\caption{Transformation properties of the matter fields under the gauge group.}
\label{tab:chargeUSpconf}
\end{table}

\noindent Here $U(1)_{R_0}$ is a possible choice of UV trial $R$-symmetry that is non-anomalous and consistent with the superpotential. The charges of the fields under $U(1)_t$ are fixed requiring that $U(1)_R$ is not anomalous, where $U(1)_R$ is defined taking into account the possible mixing of the $R$-symmetry with the other abelian symmetry
\begin{equation} \label{6_1}
	R = R_0 + \mathfrak{t} q_t \, ,
\end{equation}
where  $q_t$ is the charge under $U(1)_t$ and $\mathfrak{t}$ the mixing coefficient\footnote{
We recall that the requirement for the $U(1)_R$ symmetry to be anomaly free translates into the condition $
	\sum_f T(\mathcal{R}_f) R_f = 0 \,, $
where the trace is taken over all the fermions of the theory. The Dynkin indices for the $USp(2N)$ representations of our interest are $
	T({\bf 2N}) = 1/2 \; , \;
	T({\bf N(2N+1)}) = N+1 \; , \;
	T({\bf N(2N-1)}) = N-1 $. }. 
The superconformal $R$-symmetry can be determined using $a$-maximization \cite{Intriligator:2003jj}, which sets $\mathfrak{t}=0$. 	

As gauge invariant operators, we can construct the mesons, possibly dressed with powers of the antisymmetric
\begin{align}\label{elecmes}
	M_k = \Tr_N\left[Q A^k Q\right] \quad , \quad k=0,1,2,\cdots,N-1 \, ,
	\end{align}
with the following transformation properties under the global symmetries:
\begin{table}[h]
\centering
\begin{tabular}{|c|c|c|c|}
\hline
 & $SU(6)_x$ & $U(1)_t$ & $U(1)_{R_0}$ \\
 	\hline
	$M_k$ & {\bf 15} & $k+\frac{2}{3} (1-N)$ & $\frac{2}{3}$  \\ \hline
\end{tabular}
\caption{Transformation properties of the gauge invariant operators.}
\end{table}

\noindent Notice that the superconformal $R$-symmetry implies that these operator have dimension one and are free.
The gauge invariant operators $T_k = \text{Tr}_N A^k ,k=1,2,3,\cdots,N $ 
are instead flipped by the $\beta_k$ fields since otherwise  they  would fall below the unitarity bound.

In \cite{Csaki:1996zb} it was argued  that this theory is dual to a WZ model of $15 N$ chiral singlets.  Following  \cite{Benvenuti:2018bav},  we can write the superpotential by collecting the dual fields in a matrix $\mu_{ab,i} = -\mu_{ba,i}$, 
$i=1,...,N$, $ 1 \leq a < b \leq 6 $ as
\begin{equation}
\label{sspp}
	\hat{\mathcal{W}}= \sum_{i,j,k=1}^N \sum_{a,b,c,d,e,f=1}^6 \epsilon_{abcdef} \; \mu_{ab,i} \, \mu_{cd,j} \, \mu_{ef,k} \, \delta_{i+j+k,2N+1} \, .
\end{equation}
The gauge invariant operators $Q_a A^{i-1} Q_b$ of the electric theory are mapped into the gauge singlet fields $\mu_{ab,i}$ of the magnetic WZ model.

The singlets $\beta_k$ are not mapped into any operator of the dual WZ model, since
 we can argue that these fields can't take a VEV as in \cite{Benvenuti:2017bpg,Benvenuti:2017kud}. Indeed, a VEV for $\beta_j$, for example, would correspond to turning on $\text{Tr}_N A^{j}$  in the superpotential and the effect of such deformation can be understood by considering the duality discussed in \cite{Intriligator:1995ff}. This duality
relates an $USp(2N)$ theory with one antisymmetric chiral $A$, $2N_f$ chirals and  $\mathcal{W}=\text{Tr}_N A^{K+1}$  to a dual theory with  $USp(K(N_f -2) - N)$  gauge group. The condition to have a stable supersymmetric vacuum moduli space which includes the origin is that the dual rank must be greater or equal then zero, namely we require $N_f \geq N/K + 2$. We can then see that for $N_f=3$ the theory with the deformation $\text{Tr}_N A^{j}$ has no stable vacuum, since $j\le N$. \\

We can look at the duality at the level of the supersymmetric index. 
The index will depend on fugacities for the $SU(6)_x \times U(1)_t$ global symmetry, which we denote by $x_a$ and $t$ respectively. 
For convenience, we define the fugacities $x_a$ for the $SU(6)_x$ flavor symmetry such that they satisfy the balancing condition\footnote{In order to recover the charge assignment of Table \ref{tab:chargeUSpconf} we need to shift $x_a\to x_a t^{\frac{1-N}{3}}(pq)^{\frac{1}{6}}$, so that \eqref{balUSpconf} becomes the standard $SU(6)_x$ tracelessness condition $\prod_{a=1}^6x_a=1$.}
\begin{equation}\label{balUSpconf}
	t^{2N-2} \prod_{a=1}^6 x_a = pq \,,
\end{equation}
which can be understood as a consequence of the requirement that $U(1)_R$ is not anomalous.
With these conventions, the duality is expressed by the identity\footnote{
This elliptic hypergeometric identity was first conjectured in \cite{Diejen2000AnEM}.
}
\begin{align} \label{csaki}
	\mathcal{I}_A(t,x_a) &= \prod_{i=1}^N \Gamma_e (pq t^{-i}) \oint d  \vec{z}_N \Gamma_e(t)^N \prod_{i<j}^N \Gamma_e (t z_i^{\pm 1} z_j^{\pm 1}) \prod_{i=1}^N \prod_{a=1}^6 \Gamma_e (z_i^{\pm 1} x_a) \nonumber \\
	&= \prod_{i=1}^N \prod_{a<b}^6 \Gamma_e (t^{i-1}x_a x_b) = \mathcal{I}_B(t,x_a) \, ,
\end{align}
where we included the contribution of the vector multiplet in the integration measure 
\begin{align}
		d \vec{z}_N = \frac{[(p;p)(q;q)]^N}{2^N N!} \prod_{i=1}^N \frac{d z_i}{2 \pi i z_i} \frac{1}{\prod_{i<j}^N \Gamma_e (z_i^{\pm 1} z_j^{\pm 1}) \prod_{i=1}^N \Gamma_e (z_i^{\pm 2})} \, .
\end{align}

\begin{figure}[t]
\centering
\includegraphics[width = 4 cm]{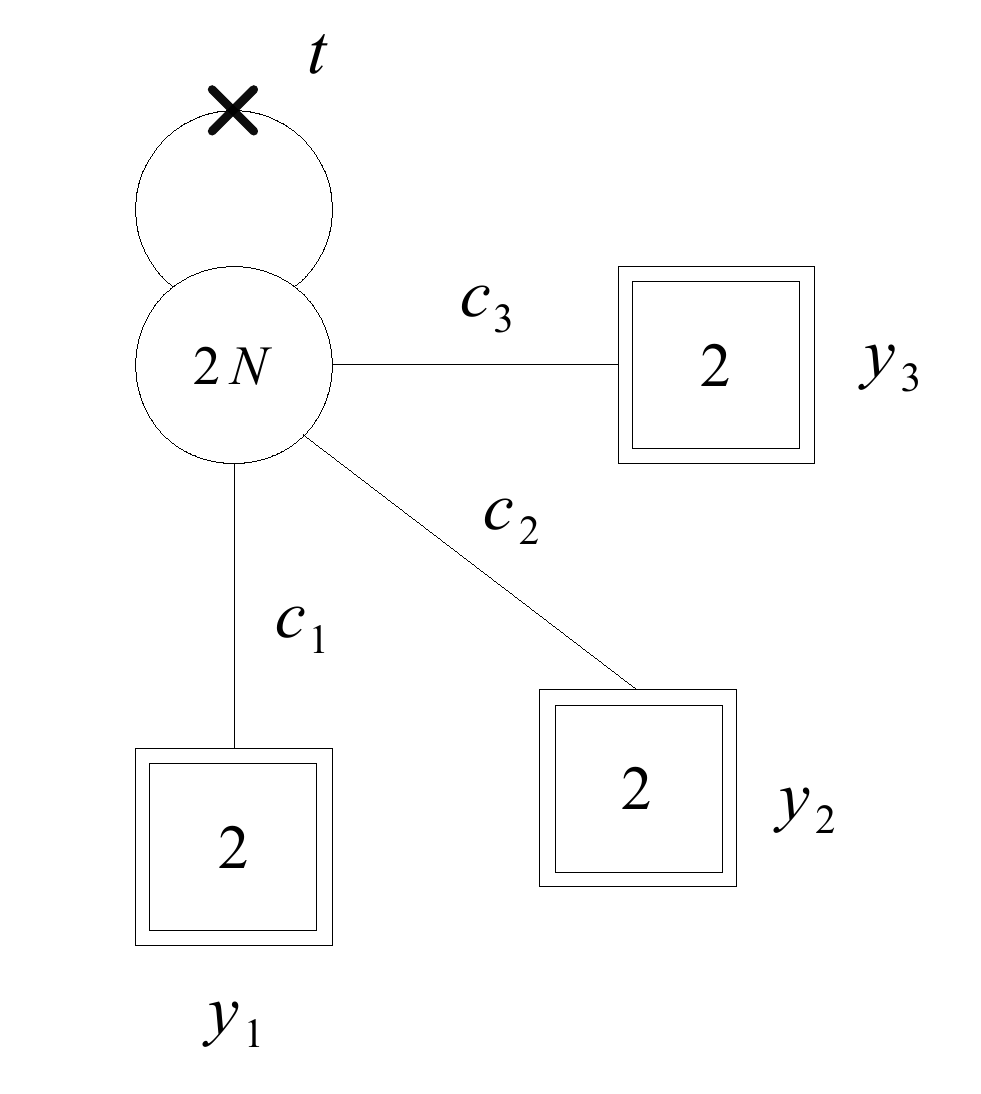}
\caption{Equivalent representation of the theory. This form is more suitable for the derivation of the duality we want to present. }
\label{fig5}
\end{figure}

In the following it will be useful to use an equivalent parameterisation where we split the 6  flavors  as in Figure \ref{fig5}.
We accordingly use the new fugacities $c_i, \; y_i$, which are simply related to the previous ones by 
\begin{equation}
	x_{1,2} = c_1 y_1^{\pm 1} \quad , \quad x_{3,4} = c_2  	y_2^{\pm 1} \quad , \quad x_{5,6} = c_3 y_3^{\pm 1} \,, \quad {\rm with} \qquad
	t^{2N-2} (c_1  c_2 c_3)^2 = pq \,.
\end{equation}

As we have already mentioned in the introduction, the electric theory -- once we compactify to $3d$ and take a suitable combination of Coulomb branch VEV and real mass deformations -- flows to a
$U(N)$ theory with one adjoint, three flavors and with a superpotential 
$\mathcal{M}^+ +\mathcal{M}^-$, where $\mathcal{M}^\pm $ are monopole operators with unit magnetic charge. A further real mass deformation for $U(1)_{c_1}$  makes two of the flavors massive removing the monopole superpotential and we obtain a  $U(N)$ theory with one flavor and one adjoint, which is dual to a WZ model \cite{Benvenuti:2018bav}\footnote{
The limit reducing the  $\mathbb{S}^3\times\mathbb{S}^1$ index   to the $\mathbb{S}^3_b$ partition function 
is implemented by  redefining:
\be
\label{para}
&&z_j=\e^{2\pi irZ_j},\quad j=1,\cdots,N,\nn\\
&&y_i=\e^{2\pi irY_i},  \quad c_i=\e^{2\pi ir\Gd_i}   \quad i=1,\cdots,3,\nn\\
&&\quad t=\e^{2\pi ir(iQ-2m_A)},\quad p=\e^{-2\pi rb},\quad q=\e^{-2\pi rb^{-1}}\,,
\ee
where $r$ is the radius of $\mathbb{S}^1$ and  the new parameters in capital letters are in $\left[-\frac{1}{2r},\frac{1}{2r}\right]$, and using that:
\be
\lim_{r\to0}\Gc_e\left(\e^{2\pi irx};p=\e^{-2\pi rb},q=\e^{-2\pi rb^{-1}}\right)=\e^{-\frac{i\pi}{6r}\left(i\frac{Q}{2}-x\right)}\sbfunc{i\frac{Q}{2}-x}\,,
\ee
where $Q=b+b^{-1}$ and $s_b$ is  the double-sine function, in terms of which the  $\mathbb{S}^3_b$ partition function can be written (we follow the conventions of Section 5 and Appendix A.1  of \cite{Bottini:2021vms}
for the $\mathbb{S}^3_b$ partition function).
The first  deformation to the $U(N)$ theory with monopole superpotential is obtained by redefining 
\be
Z_i\to Z_i+s,\quad Y_i\to Y_i+s\,.
\ee
and sending $s\to \infty$, while the real mass for $U(1)_{c_1}$  corresponds to sending $\Delta_1\to \Delta_1+t$
and $t\to \infty$.
}.
In \cite{Pasquetti:2019uop} it was  argued how this $3d$ duality can be derived using  a  sequential deconfinement procedure which parallels the manipulations leading to the evaluation formula for the 3-point free field correlator in Liouville theory. The idea is to start with an auxiliary quiver theory where we trade the adjoint matter for an extra $U(N-1)$ gauge node with a
a linear monopole superpotential. We then apply iteratively the one-monopole \cite{Benini:2017dud} and Aharony duality \cite{Aharony:1997gp} to obtain a recursion in which the rank of the gauge group is lowered at each step. \

\begin{figure}[!h]
\centering
\includegraphics[width = 15 cm]{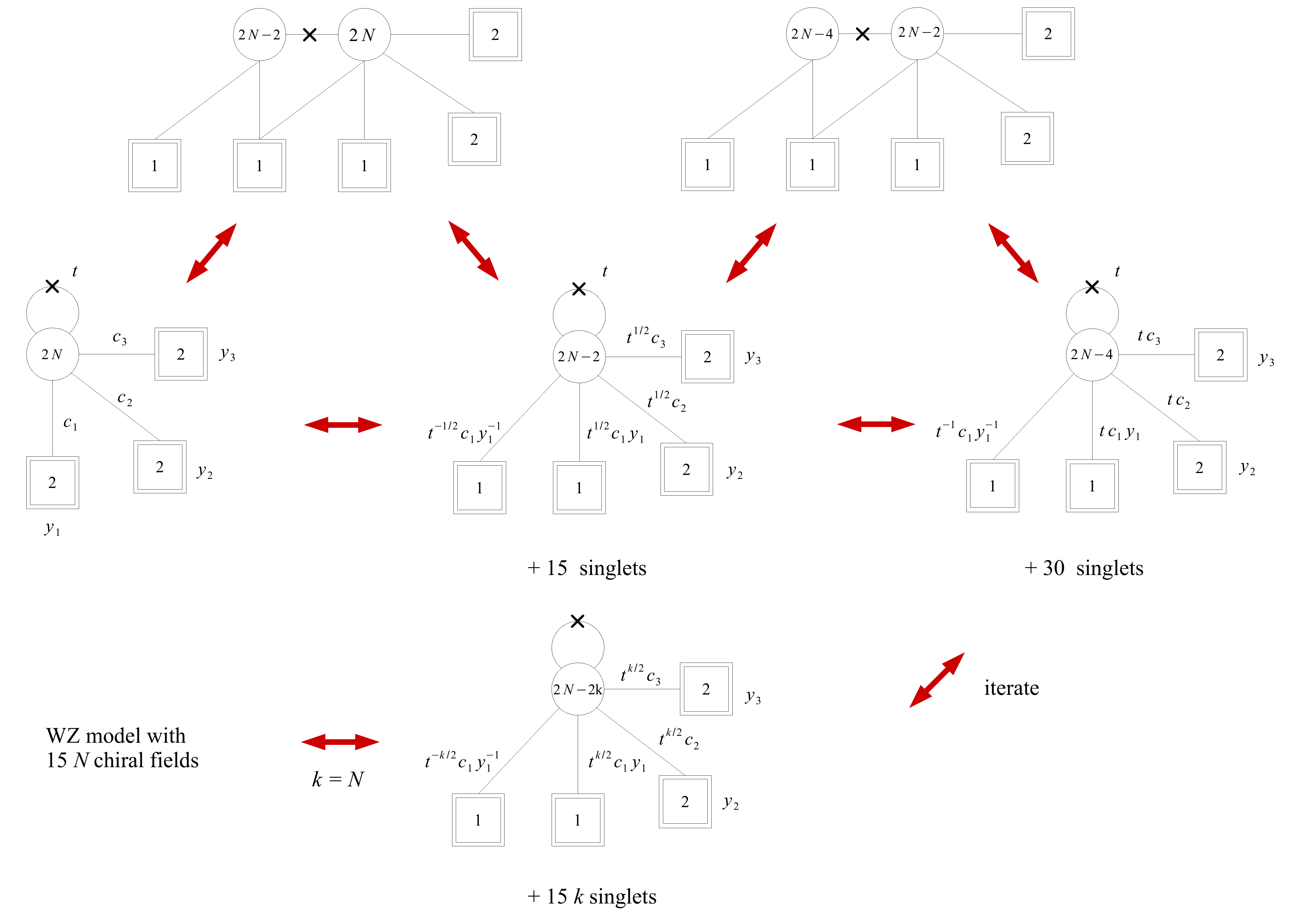}
\caption{Sequential deconfinement procedure. Applying the IP duality to the $USp(2N-2)$ node of the auxiliary quiver 
in the the top left corner we move to the first quiver in the second line, the original electric theory. Applying the IP duality to the $USp(2N)$ node we move to the second theory in the second line,
with rank decreased by one unit and 15 extra singlets. We can now use the  auxiliary quiver in the top right corner. If we apply the IP to the 
$USp(2N-4)$ we go back to the second theory in the second line, while if we dualize the $USp(2N-2)$ node we instead move 
to the third theory in the second line, with rank decreased by  two units and 30 extra singlets. In the last line we show the result after $k$ iterations and the result after $N$ iterations, the WZ dual frame. }
\label{fig6}
\end{figure}

For the $4d$ $USp(2N)$ electric theory we can analogously implement the sequential deconfinement procedure by trading the tensor matter for a new gauge node, and this is how we obtain the auxiliary quiver shown in the top left of Figure \ref{fig6}, where the antisymmetric is traded for an extra $USp(2N-2)$ gauge node. The auxiliary quiver theory has a triangle superpotential involving the $USp(2N-2)\times USp(2N)$ bifundamental $q$. There are also  $N-1$ singlets $\beta_i$ entering the superpotential as $\sum_{i=1}^{N-1}\beta_i \text{Tr}_N \text{Tr}_{N-1}(qq)^i$.
Notice that only $U(1)_{y_1}\subset SU(2)_{y_1}$ is manifest in this frame, but the number of fundamentals attached to each gauge node is always even to avoid Witten's anomaly \cite{Witten:1982fp}. 
This specific arrangement of fields is suggested by the fact that  our $4d$ auxiliary quiver, upon taking the $3d$ limit  combined with the deformations discussed above, reduces precisely to the  auxiliary quiver with linear monopole turned on at the first gauge node  used in the $3d$ deconfinement procedure. 

Starting from the auxiliary quiver on the top left corner of  Figure \ref{fig6} we can move into two directions by applying  the Intriligator--Pouliot  (IP) duality  \cite{Intriligator:1995ne}. We recall that the IP duality relates a $USp(2N)$ gauge theory with $2K$ fundamental chirals and no superpotential to a $USp(2K-2N-4)$ gauge theory with $2K$ fundamental chirals, $K(2K-1)$ singlets $\Gp_{\ga\gb}$ and superpotential $\hat{\mathcal{W}}=\Gp^{\ga\gb}q_\ga q_\gb$.

If we apply the IP duality
to the first node, this confines; the singlets appearing in the magnetic side of the IP duality reconstruct the antisymmetric chiral on the $USp(2N)$ node and the $\beta_N$ singlet, so that we recover exactly the original theory, which is the first quiver in the second line of Figure \ref{fig6}.
In this sense, we call the procedure of going from the original quiver to the auxiliary one \textit{deconfinement}. We can also apply the duality to the second $USp(2N)$ node, which also confines and we reach the second quiver in the second line. Now compared to the original theory the rank of the gauge group is lowered by one and 15 extra singlets have been generated. From here, we can construct another auxiliary quiver, the second one in the first line, this time with gauge group $USp(2N-4) \times USp(2N-2)$ and repeat the same procedure. After applying the duality to the second $USp(2N-2)$ node, we obtain a quiver with the rank lowered by 2 and $30$ extra singlets. We see that at each iteration  we basically just lower by one unit the rank and produce a bunch of extra singlets. At  the $k$-th setp we will have an  $USp(2N-2k)$ theory with one antisymmetric, six fundamental chirals and $15k$ extra singlets. Notice that in Figure \ref{fig6} we don't recombine the two chirals charged under $U(1)_{y_1}$ into a $SU(2)_{y_1}$ because this symmetry is broken by superpotential terms involving the singlets fields at the intermediate steps. Nevertheless, we still obtain a theory with $USp$ gauge group, one antisymmetric and six fundamentals, and in this sense we can say that original theory is \textit{stable} under this sequence of dualizations.
If we iterate the whole procedure $N$ times, the gauge node confines and we end up with a WZ model of $15N$ singlets, which is the dual theory we are looking for. 

All these steps can be performed at the level of superconformal index.
The basic identity for the IP duality
was proven in \cite{2003math......9252R}
\begin{align}
\oint\udl{\vec{z}_{N}}\prod_{i=1}^{N}\prod_{\ga=1}^{2K}\Gpq{v_\ga z_i^{\pm1}}=\prod_{\ga<\gb}^{2K}\Gpq{v_\ga v_\gb}\oint\udl{\vec{z}_{K-N-2}} \prod_{i=1}^{K-N-2}\prod_{\ga=1}^{2K}\Gpq{(pq)^{1/2} v_\ga^{-1} z_i^{\pm1}}\, ,\nn\\
\label{IP}
\end{align}
and it holds provided that the balancing condition
\be
\prod_{\ga=1}^{2K}v_\ga=(pq)^{K-N-1}\,,
\label{balancingIP}
\ee
corresponding to the fact that the anomaly cancellation is satisfied.
The starting point is the index of the auxiliary quiver in the top left corner of Figure \ref{fig6}, which is given by
\begin{align} \label{6_4}
	\mathcal{I}_{\text{aux}}(t,c_i,y_i) &= \prod_{i=1}^{N-1}\Gamma_e (pq t^{-i}) \oint d \vec{w}_{N-1} d \vec{z}_N  \prod_{i=1}^{N-1} \Gamma_e (c_1 y_1^{-1} t^{-1/2} w_i^{\pm 1}) \prod_{i=1}^{N-1} \Gamma_e (pq t^{\frac{1}{2}-N} c_1^{-1}y_1w_i^{\pm 1}) \nonumber \\
	&\times \prod_{i=1}^{N} \Gamma_e (t^{N-1} c_1 y_1^{-1} z_i^{\pm 1}) \prod_{i=1}^{N} \Gamma_e (c_1 y_1 z_i^{\pm 1}) \prod_{i=1}^{N} \Gamma_e (c_2 y_2^{\pm 1} z_i^{\pm 1}) \prod_{i=1}^{N} \Gamma_e (c_3 y_3^{\pm 1} z_i^{\pm 1})   \nn \\
	&\times \prod_{i=1}^{N-1} \prod_{j=1}^N \Gamma_e (t^{\frac{1}{2}} w_i^{\pm 1} z_j^{\pm 1})\,. 
\end{align}
The charges of fields in the auxiliary quiver satisfy the constraints coming from anomaly cancellations and
cubic superpotential and are chosen in a such a way that when we apply the IP duality to the first node,
which corresponds to replacing the terms in \eqref{6_4} depending on the $USp(2N-2)$ fugacities $w_i$ as
\begin{align}
\label{ipfn}
	&\oint d \vec{w}_{N-1} \prod_{i=1}^{N-1} \prod_{j=1}^N \Gamma_e (t^{\frac{1}{2}} w_i^{\pm 1} z_j^{\pm 1}) \prod_{i=1}^{N-1} \Gamma_e (c_1 y_1^{-1} t^{-1/2} w_i^{\pm 1}) \prod_{i=1}^{N-1} \Gamma_e (pq t^{\frac{1}{2}-N} c_1^{-1}y_1w_i^{\pm 1})  \nonumber \\
	&= \Gamma_e(t)^N \prod_{i < j}^N \Gamma_e (t z_i^{\pm 1}z_j^{\pm 1}) \Gamma_e (pq t^{-N}) \prod_{i=1}^N \Gamma_e (c_1 y_1^{-1} z_i^{\pm 1}) \prod_{i=1}^N \Gamma_e (pq t^{1-N} c_1^{-1} y_1 z_i^{\pm 1}) \, ,
\end{align}
we recover the index of the original electric theory in \eqref{csaki}.
Notice that part of the matrix of gauge singlets of the IP  dual \eqref{ipfn} reconstructs the antisymmetric on the $USp(2N)$ gauge node and the $\beta_N$ singlet. The $SU(2)_{y_1}$ flavor is reconstructed since no superpotential breaking it is generated. 

Now we can apply the duality to the second gauge node of the auxiliary quiver, 
which corresponds to replacing the terms in \eqref{6_4} depending on the $USp(2N)$ fugacities $z_j$ as
\begin{align}
	&\oint d \vec{z}_N \prod_{i=1}^{N-1} \prod_{j=1}^N \Gamma_e (t^{\frac{1}{2}} w_i^{\pm 1} z_j^{\pm 1}) \prod_{i=1}^{N} \Gamma_e (t^{N-1} c_1 y_1^{-1} z_i^{\pm 1}) \prod_{i=1}^{N} \Gamma_e (c_1 y_1 z_i^{\pm 1}) \prod_{i=1}^{N}\Gamma_e (c_2 y_2^{\pm 1} z_i^{\pm 1}) \nonumber \\
	&\times  \prod_{i=1}^{N} \Gamma_e (c_3 y_3^{\pm 1} z_i^{\pm 1}) = \Gamma_e(c_2^2) \Gamma_e(c_3^2) \Gamma_e (t^{N-1} c_1^2) \Gamma_e (c_2 c_3 y_2^{\pm 1} y_3^{\pm 1})  \Gamma_e (c_1 c_2 y_1 y_2^{\pm 1})\nonumber \\
	&\times  \Gamma_e (t^{N-1} c_1 c_2 y_1^{-1} y_2^{\pm 1}) \Gamma_e (c_1 c_3 y_1 y_3^{\pm 1}) \Gamma_e (t^{N-1} c_1 c_3 y_1^{-1} y_3^{\pm 1}) \Gamma_e(t)^{N-1} \prod_{i<j}^{N-1} \Gamma_e (t w_i^{\pm 1}w_j^{\pm 1}) \nonumber \\
	&\times \prod_{i=1}^{N-1} \Gamma_e (t^{N-\frac{1}{2}} c_1 y_1^{-1} w_i^{\pm 1}) \prod_{i=1}^{N-1} \Gamma_e (t^\frac{1}{2} c_1 y_1 w_i^{\pm 1}) 
	\prod_{i=1}^{N-1} \Gamma_e (t^\frac{1}{2} c_2 y_2^{\pm 1} w_i^{\pm 1})
	\prod_{i=1}^{N-1} \Gamma_e (t^\frac{1}{2} c_3 y_3^{\pm 1} w_i^{\pm 1}) \,,
\end{align}
Plugging this into \eqref{6_4}, we obtain the index
\begin{align} \label{6_5}
	&S \times \prod_{i=1}^{N-1} \Gamma_e (pq t^{-i})
	\nonumber \oint d \vec{w}_{N-1} \Gamma_e(t)^{N-1} \prod_{i<j}^{N-1} \Gamma_e (t w_i^{\pm 1}w_j^{\pm 1}) \\
	&\times \prod_{i=1}^{N-1} \Gamma_e (c_1 (t^{1/2} y_1)^{\pm 1} w_i^{\pm 1})\prod_{i=1}^{N-1} \Gamma_e (t^{\frac{1}{2}} c_2 y_2^{\pm 1}w_i^{\pm 1}) \prod_{i=1}^{N-1} \Gamma_e (t^{\frac{1}{2}} c_3 y_3^{\pm 1} w_i^{\pm 1})	\,,
\end{align}
where $S$ is the contribution of the gauge singlets produced after the dualization, which is given by
\begin{align}
\label{eq:singlets}
S &= \Gamma_e(c_2^2) \Gamma_e(c_3^2) \Gamma_e (t^{N-1} c_1^2) \Gamma_e (c_2 c_3 y_2^{\pm 1} y_3^{\pm 1}) \nonumber \\
& \times \Gamma_e (c_1 c_2 y_1 y_2^{\pm 1}) \Gamma_e (t^{N-1} c_1 c_2 y_1^{-1} y_2^{\pm 1})  \Gamma_e (c_1 c_3 y_1 y_3^{\pm 1}) \Gamma_e (t^{N-1} c_1 c_3 y_1^{-1} y_3^{\pm 1}) \,.
\end{align}
This is the index of the theory in the middle of the second line of  Figure \ref{fig6},
an  $USp(2N-2)$ theory with one antisymmetric, 6 chirals and
$15$ extra singlets.  We call this step one. 
Notice that $SU(2)_{y_1}$ is still broken in this frame and the $SU(6)_x$ symmetry 
emerges only in the IR.

At this point, we can repeat the whole procedure and consider the auxiliary quiver  with gauge nodes $USp(2N-4) \times USp(2N-2)$ in the top right corner of Figure \ref{fig6}. 
 Its index is given by
\begin{align}
	\mathcal{I}_{\text{aux}^\prime} &= S \times \prod_{i=1}^{N-2} \Gamma_e (pq t^{-i}) \oint d \vec{w}_{N-1} d \vec{k}_{N-2} \prod_{i=1}^{N-2} \prod_{j=1}^{N-1} \Gamma_e (t^\frac{1}{2} k_i^{\pm 1} w_j^{\pm 1}) \prod_{i=1}^{N-2} \Gamma_e (t^{-1} c_1 y_1^{-1}  k_i^{\pm 1}) \nn\\
	&\times\prod_{i=1}^{N-2} \Gamma_e (pq t^{2-N} c_1^{-1} y_1 k_i^{\pm 1})  \prod_{i=1}^{N-1} \Gamma_e (t^{N-\frac{5}{2}} c_1 y_1^{-1} w_i^{\pm 1}) \prod_{i=1}^{N-1} \Gamma_e (t^\frac{1}{2} c_1 y_1 w_i^{\pm 1}) \nonumber \\
	&\times\prod_{i=1}^{N-1} \Gamma_e (t^\frac{1}{2} c_2 y_2^{\pm 1} w_i^{\pm 1}) \prod_{i=1}^{N-1} \Gamma_e (t^\frac{1}{2} c_3 y_3^{\pm 1} w_i^{\pm 1}) \,,
\end{align}
where $S$ is the contribution of the 15 singlets given in \eqref{eq:singlets}.
Applying the IP duality to the first $USp(2N-4)$ node, we recover the expression \eqref{6_5}.
We can also apply the duality to the second gauge node. If we do so, we obtain the index of the third quiver in the second line of   Figure \ref{fig6}
\begin{align}
	&\Gamma_e (t c_2^2) \Gamma_e (t c_3^2) \Gamma_e (t^{N-2} c_1^2)  \Gamma_e (t c_2 c_3 y_2^{\pm 1} y_3^{\pm 1}) \Gamma_e (t c_1 c_2 y_1 y_2^{\pm 1}) \Gamma_e (t^{N-2} c_1 c_2 y_1^{-1} y_2^{\pm 1}) \nonumber \\
	&\times  \Gamma_e (t c_1 c_3 y_1 y_3^{\pm 1}) \Gamma_e (t^{N-2} c_1 c_3 y_1^{-1} y_3^{\pm 1}) 
	\times S  \times	\prod_{i=1}^{N-2} \Gamma_e (pq t^{-i})\oint d \vec{k}_{N-2} \Gamma_e(t)^{N-2}\nn\\
&\times \prod_{i<j}^{N-2} \Gamma_e(t\, k_i^{\pm 1} k_j^{\pm 1})  \prod_{i=1}^{N-2} \Gamma_e (c_1 (t y_1)^{\pm 1} k_i^{\pm 1}) \prod_{i=1}^{N-2} \Gamma_e (t c_2 y_2^{\pm 1} k_i^{\pm 1}) \prod_{i=1}^{N-2} \Gamma_e (t c_3 y_3^{\pm 1} k_i^{\pm 1}) \,,
\end{align}
where we have the contribution of additional 15 singlets on top of the contribution $S$ obtained in the previous step.
Compared to the original electric theory the rank is now lowered by  two units  and a total of $30$ extra singlets have been generated.

After $k$ iterations we reach a frame with $USp(2(N-k))$ gauge group and  $15k$ singlets. The contribution of the singlets at this step is given by
\begin{align}
	&\prod_{i=1}^k \Gamma_e (t^{i-1} c_2^2) \prod_{i=1}^k \Gamma_e (t^{i-1} c_3^2) \prod_{i=1}^k \Gamma_e (t^{N-i} c_1^2) \prod_{i=1}^k \Gamma_e (t^{i-1} c_2 c_3 y_2^{\pm 1} y_3^{\pm 1}) \prod_{i=1}^k \Gamma_e (t^{i-1} c_1 c_2 y_1 y_2^{\pm 1}) \nn \\
	&\prod_{i=1}^k \Gamma_e (t^{i-1} c_1 c_3 y_1 y_3^{\pm 1})\prod_{i=1}^k \Gamma_e (t^{N-i} c_1 c_2 y_1^{-1} y_2^{\pm 1}) \prod_{i=1}^k \Gamma_e (t^{N-i} c_1 c_3 y_1^{-1} y_3^{\pm 1}) \,.
\end{align}

Then we see that if we iterate the procedure $k=N$ times, the gauge node confines and
on the r.h.s. we are just left with the contribution of $15N$ chiral fields in the antisymmetric of $SU(6)_x$, which is now fully manifest. Going back to the $x_a$ variables, the index is given by
\begin{equation}
	\prod_{i=1}^N \prod_{a<b}^6 \Gamma_e (t^{i-1} x_a x_b) \,,
\end{equation}
which is what we intended to prove. 

The index manipulations described  above are exactly the steps leading to  the proof of the elliptic hypergeometric identity \eqref{csaki} provided in  \cite{2003math......9252R}. As shown in  \cite{spiridonov2004theta} similar manipulations can be used to prove integral identities corresponding to other s-confining dualities and
reinterpreted in field theory as  sequential deconfinements procedures \cite{stephane-sergio}.

\section{Confining duality for a linear quiver theory}
\label{sec:confquiv}

In the following sections we aim to give various generalizations of the confining duality between the $USp(2N)$ gauge theory with six fundamental chirals and one antisymmetric chiral and the WZ model. The strategy that we will use to derive these new dualities is again based on the deconfinement procedure and on the iterative application of the IP duality. 

 \begin{figure}[t]
\centering
\includegraphics[width = 15.1 cm]{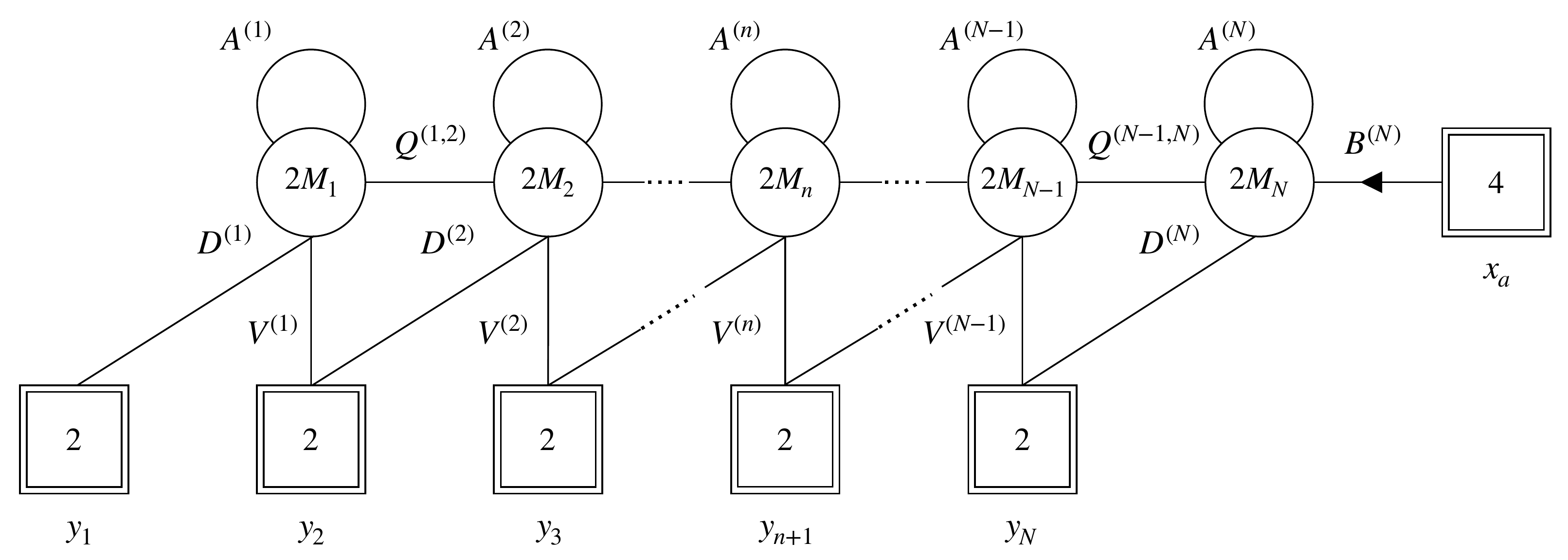}
\caption{\label{fig:longquiver}Electric theory in the case of N gauge nodes. Singlet fields are not included in the figure. }
\end{figure}

The first immediate generalization is to consider many gauge nodes, instead of the single one of the original duality. As our electric theory we will then consider the quiver theory given in the Figure \ref{fig:longquiver},
which clearly reduces to the previous case for $N=1$\footnote{Notice that in this section $N$ is the number of gauge nodes in the quiver, while in the previous section it was the rank of the only gauge node. Hence, setting $N=1$ in the quiver theory we are considering now will give us the theory considered in the previous section with a single $USp(2M_1)$ gauge node.}. 
The gauge group of this theory is $\prod_{i=1}^N USp(2M_i)$, where we assume that the ranks are ordered as $M_1\leq M_2\leq \cdots \leq M_N$. The reason for this choice will be clear when we will discuss the derivation by deconfinement of the duality that we are going to present for this theory, since this works straightforwardly if this conditions is assumed, and we shall briefly comment on what happens otherwise later. 
The matter content is given by the following fields in the fundamental, bifundamental and antisymmetric representations:

\begin{itemize}
	\item a chiral field $Q^{(n,n+1)}$ in the bifundamental representation of $USp(2M_n) \times USp(2M_{n+1})$, $n=1,...,N-1$;
	\item a chiral field $B^{(N)}$ in the fundamental representation of the last $USp(2M_N)$ gauge symmetry and in the fundamental of the $SU(4)_x$ flavor symmetry;
	\item a chiral field $D^{(n)}$ in the fundamental representation of $USp(2M_n)$ and in the fundamental of $n$-th $SU(2)_{y_n}$ flavor symmetry of the saw, $n=1,...,N$;
	\item a chiral field $V^{(n)}$ in the fundamental representation of $USp(2M_n)$ and in the fundamental of $(n+1)$-th $SU(2)_{y_{n+1}}$ flavor symmetry of the saw, $n=1,...,N-1$;
	\item a chiral field $A^{(n)}$ in the antisymmetric representation of $USp(2M_n)$, $n=1,...,N$;
	\item gauge singlets $\beta_i^{(n)}$, coupled to the trace of powers of $A^{(n)}$, $i=1,...,M_n-M_{n-1}$, $n=1,...,N$;
	\item gauge singlets $\gamma_i^{(n)}$, coupled to the gauge invariant mesons built from the diagonal chirals of the saw $D^{(n)}$, also dressed with powers of the antisymmetric of the corresponding gauge node, $i=1,...,M_n-M_{n-1}$, $n=1,...,N$;
\end{itemize}
To write the superpotential in a compact form, we define
\begin{equation}
	\mathbb{Q}_{abij}^{(n,n+1)} = Q_{ai}^{(n,n+1)} Q_{bj}^{(n,n+1)} \,.
\end{equation}
The superpotential consists of a coupling between the bifundamentals and the antisymmetrics, a cubic coupling between the chirals appearing in each one of the triangles and, finally, flip terms for the diagonal mesons and for the powers of the traces of the antisymmetrics\footnote{\label{foot:vanish}The reason why we flip only these specific traces of powers of the antisymmetrics will become clear once we will discuss the duality. In short, the dual theory is a WZ model so we expect all gauge invariant operators in the quiver theory to get mapped to some simple chiral fields and composites of them. Nevertheless, we will see that there is no field in the dual WZ that corresponds to these higher powers of the antisymmetrics, which leads us to conclude that they should vanish in the chiral ring because of quantum effects. The analytic matching of the indices between the quiver theory and the WZ model that we will provide is strong evidence for this. Also note that here we assume the $\beta_i$ fields vanish in the chiral ring as well, for the same reason as in the one node case, which will also be clear from the fact that there are no operators on the WZ side that $\beta_i$ are mapped to.}
\begin{align}\label{qsdsup}
	&\mathcal{W}_{\text{gauge}} = \sum_{n=1}^{N-1} \text{Tr}_n \big[ A^{(n)} \big( \text{Tr}_{n+1} \mathbb{Q}^{(n,n+1)} - \text{Tr}_{n-1} \mathbb{Q}^{(n-1,n)} \big) \Big] - \text{Tr}_N \big[ A^{(N)} \big( \text{Tr}_{N-1} \mathbb{Q}^{(N-1,N)} \big) \Big] +\nonumber \\
	&+ \sum_{n=1}^{N-1} \text{Tr}_{y_{n+1}} \text{Tr}_{n} \text{Tr}_{n+1} \big( V^{(n)} Q^{(n,n+1)} D^{(n+1)}\big) + \sum_{n=1}^N \sum_{i=1}^{M_n-M_{n-1}} \gamma_i^{(n)} \text{Tr}_{y_n} \text{Tr}_n (D^{(n)} D^{(n)} A^{(n) \, i-1}) + \nonumber \\
	&+ \sum_{n=1}^N \sum_{i=1}^{M_n - M_{n-1}} \beta_i^{(n)} \text{Tr}_n \, A^{(n) \, i} \,.
\end{align}
In the expression above $\text{Tr}_n$ denotes the trace over the gauge indices of the $n$-th gauge node and $\text{Tr}_{y_{n}}$ denotes the trace over the $n$-th $SU(2)$ flavor symmetry.  

\begin{table}[t]
\centering
\begin{tabular}{|c|c|c|c|}
\hline
 & $R_0$ & $U(1)_c$ & $U(1)_t$ \\
\hline
$Q^{(n,n+1)}$ & 0 & 0 & $\frac{1}{2}$ \\
$A^{(n)}$ & 2 & 0 & $-1$ \\
$D^{(n)}$ & $M_N+M_{n-1}-M_{N-1}-M_n$ & $1$ & $\frac{M_{N-1}+M_n-M_N -M_{n-1}-N+n}{2}$ \\
$V^{(n)}$ & $2+M_{N-1}+M_{n+1}-M_N - M_n$ & $-1$ & $\frac{M_N+M_n-M_{N-1}-M_{n+1}+N-n-2}{2}$ \\
$B^{(N)}$ & $\frac{3+M_{N-1}-2M_N}{2}$ & $-\frac{1}{2}$ & $\frac{2M_N - M_{N-1}-2}{4}$ \\
$\gb_i^{(n)}$ & $2-2i$ & 0 & $i$\\
$\gamma_i^{(n)}$ & $4-2i+2M_{N-1}+$ & $-2$ & $i-1+N-n+M_N+$\\
 & $+2M_n-2M_N-2M_{n-1}$ &  & $+M_{n-1}-M_{N-1}-M_n$\\
\hline
\end{tabular}
\caption{\label{tab:gauge th charges} Charges under the abelian symmetries of the matter fields of the quiver gauge theory.}
\end{table}

From this Lagrangian description, the manifest global symmetry of the theory is 
\begin{equation}
	\prod_{i=1}^{N} SU(2)_{y_i} \times SU(4)_{x} \times U(1)_t \times U(1)_c \;.
\end{equation}
The charges under the $U(1)$ symmetries are as usual fixed by the superpotential and by the fact that $U(1)_R$ must be non-anomalous at each gauge node, where again $U(1)_R$ is defined taking into account a possible mixing of a trial UV $R$-symmetry $U(1)_{R_0}$ and the other abelian symmetries. The trial $R$-charges and the charges under the abelian symmetries that we assign are shown in Table \ref{tab:gauge th charges}\footnote{Notice that the same charge assignment, in particular the one for the fields $\beta_i^{(n)}$, can be obtained by turning on in the superpotential other combinations of interaction terms of the form $\beta_i^{(n)} \text{Tr}_n \, A^{(m) \, i}$ than those with $m=n$ that appear in \eqref{qsdsup}. Our analysis, which is based on the supersymmetric index, will be blind to this different choice, since the index is only sensible to the charge assignment of the fields and not the specific superpotential terms that enforce them. Nevertheless, we expect that a full analysis of the superpotential and its mapping at each step of the dervation of the duality that we are going to discuss would require that the terms in the last line of \eqref{qsdsup} are the one and only that are turned on.}. 

\begin{figure}
\centering
\includegraphics[width = 10.2 cm]{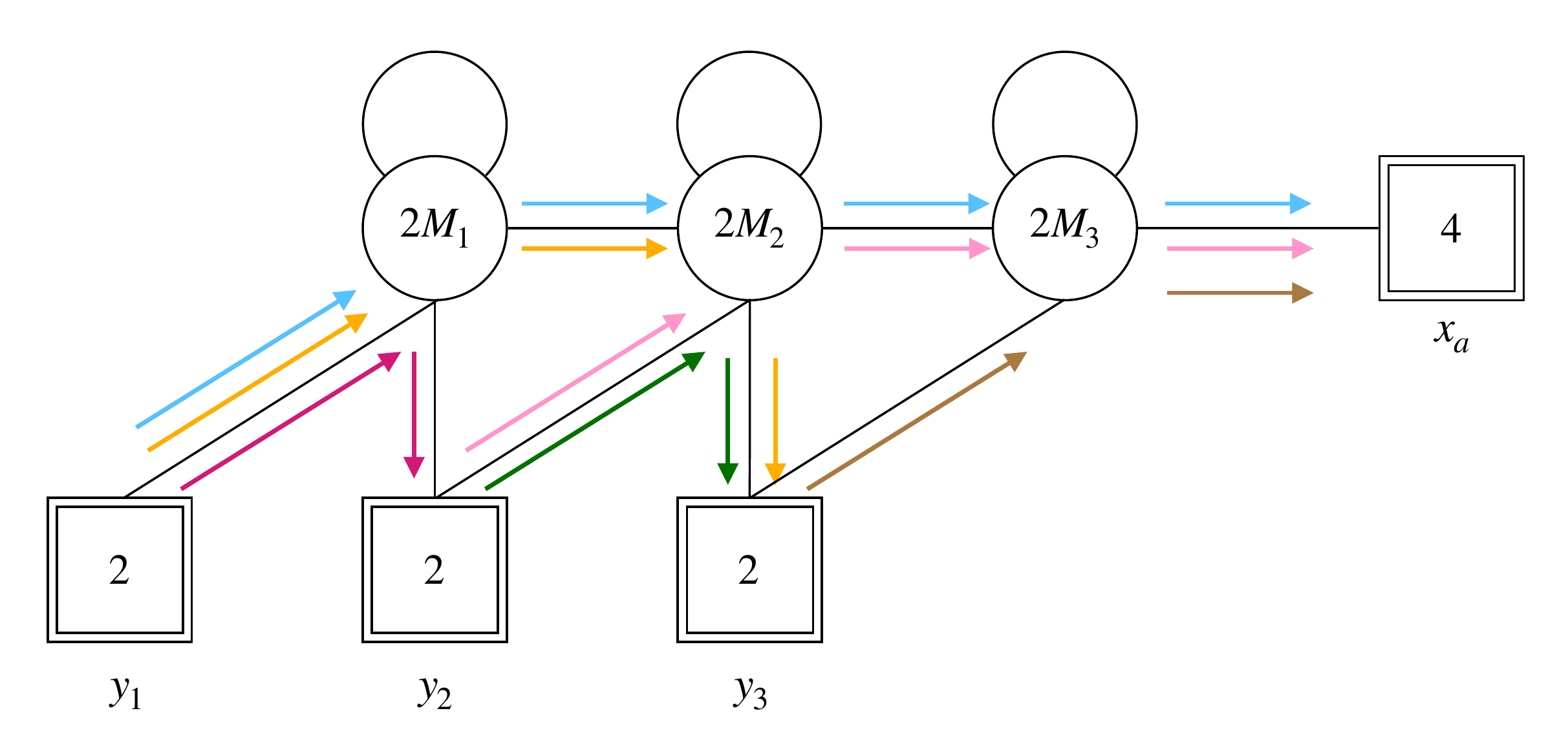}
\caption{Construction of some of the gauge invariant operators for $N=3$.
The  mesons  in violet, orange and green correspond to $E^{(12)}_0$, $E^{(13)}_0$ and $E^{(23)}_0$.
Operators   $V^{(1)}_0$, $V^{(2)}_0$, $V^{(3)}_0$ are respectively in blue, pink and brown.
}
\label{opc}
\end{figure}

There are three interesting kinds of  gauge invariant operators which we will be able to map to the singlets of the WZ dual model. We have $\frac{N(N-1)}{2}$ mesons $E_i^{(nm)}$ in the bifundamental representation  of  $SU(2)_{y_n} \times SU(2)_{y_m}$ with $n=1,...,N-1$, $m=n+1,...,N$. They are constructed starting with a diagonal chiral, going along the tail with the bifundamentals and ending on a vertical chiral. 
Then we have $N$ gauge invariant operators constructed starting from one diagonal flavor and going along
the tail including horizontal bifundamentals and the   last chiral $B^{(N)}$ in the fundamental of $SU(4)_{x}$. These are in the bifundamental representation of $SU(2)_{y_n} \times SU(4)_{x}$ and we denote them by $V^{(n)}_i$
with $n=1,\cdots ,N$.
For $N=3$ these operators are depicted in Figure \ref{opc}.
 Finally, there is the meson $\tilde{B}_i$ constructed from $B^{(N)}$ in the antisymmetric representation of $SU(4)_{x}$.
These operators carry also a label $i$ which indicates the level of dressing with  powers of the antisymmetrics. Because of  the cubic superpotential between the antisymmetric and the horizontal bifundamental chirals in an operator like $E_i^{(nm)}$  we can take any combination  $A_n^{k_n}A_{{n+1}}^{k_{n+1}}\cdots A_m^{k_m}$ of the antisymmetrics of the nodes $USp(2M_n), USp(2M_{n+1}),\cdots, USp(2M_m)$  with the condition $k_n+k_{n+1}+\cdots+ k_m=i$. For the moment we will remain agnostic on the range of the index $i$ that gives non-trivial independent chiral ring operators. This will be determined once we will discuss the operator map with the WZ dual, since only for some values of $i$ we will get operators that correspond to chiral fields in the WZ model (this is similar to what discussed in Footnote \ref{foot:vanish}).
 
\begin{figure}[t]
\centering
\includegraphics[width = 4.5 cm]{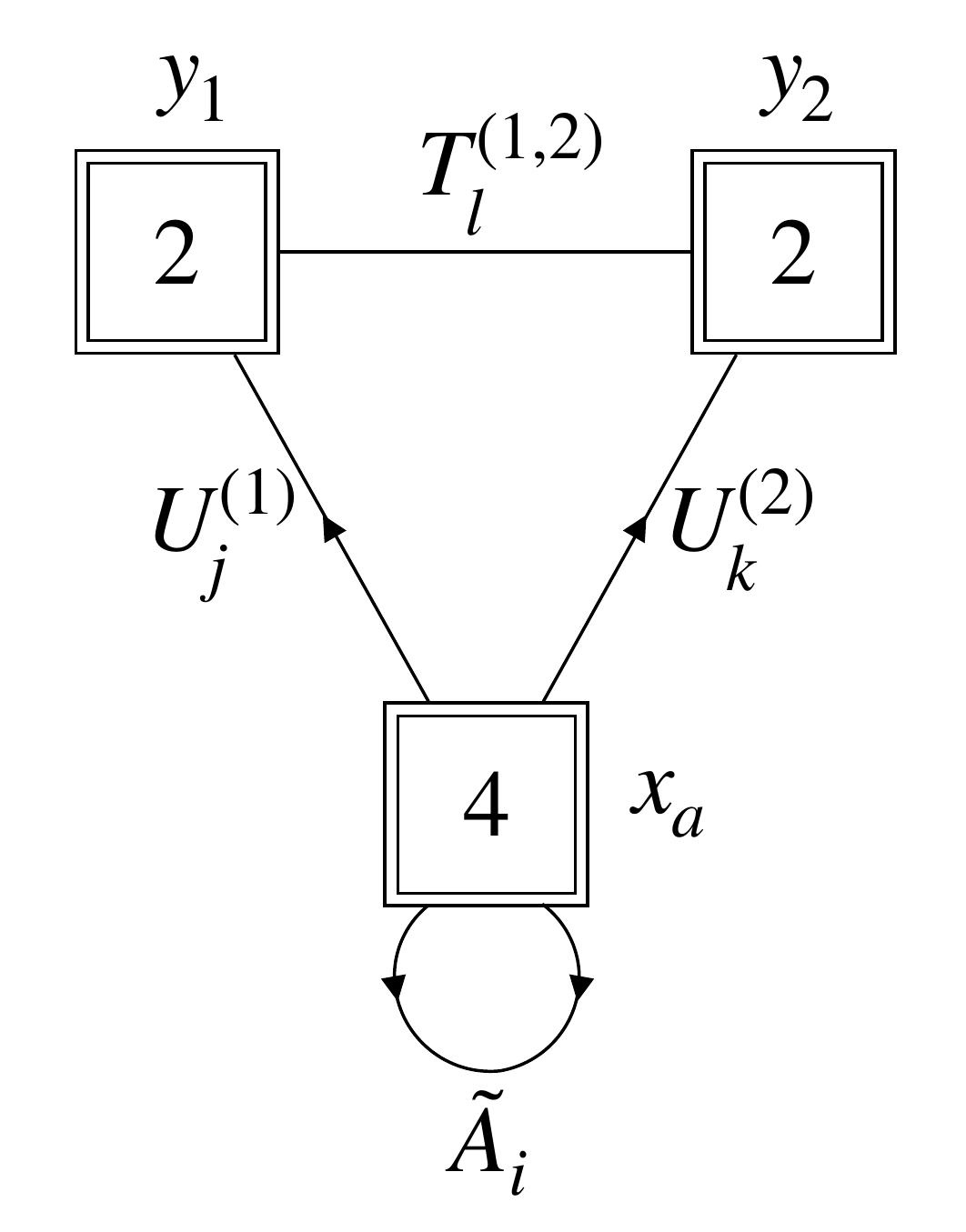}
\caption{Quiver of the dual WZ theory for $N=2$.}
\label{wzfig}
\end{figure}

We claim that this theory is dual to a WZ model of of $4 \sum_{i=1}^{N-1} M_i + 14 M_N$ chiral fields
transforming as follows\footnote{For $N=1$ we have a duality between a $USp(2M_1)$ theory and a  WZ model with $14 M_1$  fields
which is differs from  the Csaki-Skiba-Schmaltz   duality discussed in the previous section by the presence
of an extra set of singlets $\gamma_i^{(1)}$, $i=1,\cdots, M_1 $ on the electric side.}:
\begin{itemize}
	\item a tower of chiral fields $T_{i}^{(nm)}$, $i=1,...,M_n - M_{n-1}$ in the bifundamental representation of $SU(2)_{y_n} \times SU(2)_{y_m}$, $n=1,...,N-1$, $m=n+1,...,N$;
	\item a tower of chiral fields $\tilde{A}_{i}$, $i=1,...,M_N$, in the antisymmetric representation of $SU(4)_{x_a}$;
	\item a tower of chiral fields $U_{i}^{(n)}$, $i=1,...,M_n - M_{n-1}$, in the bifundamental representation of $SU(2)_{y_n} \times SU(4)_{x_a}$, $n=1,...,N$.
\end{itemize}
These fields interact with the superpotential
	\begin{align}
	&\mathcal{W}_{\text{WZ}} = \sum_{n=1}^N \sum_{i=1}^{M_N} \sum_{j,k=1}^{M_n-M_{n-1}}  \epsilon_{abcd} \,\tilde{A}_i^{ab}  \, \text{Tr}_{y_n} U_j^{(n)\, c} U_k^{(n)\, d} \, \delta_{i+j+k,1+M_N+M_n-M_{n-1}} +\nonumber \\
	&+ \sum_{n=1}^{N-1} \sum_{m=n+1}^N \sum_{i=1}^{M_N} \sum_{j,l=1}^{M_n-M_{n-1}} \sum_{k=1}^{M_m-M_{m-1}} \epsilon_{abcd} \,\tilde{A}_i^{ab} \, \text{Tr}_{y_n} \text{Tr}_{y_m} U_j^{(n)\, c} U_k^{(m)\, d} T_l^{(nm)} \,\delta_{i+j+k+l,1+M_N+M_n-M_{n-1}} \,,
\end{align}
preserving the same global symmetry of the electric theory $
	\prod_{i=1}^{N} SU(2)_{y_i} \times SU(4)_{x_a} \times U(1)_t \times U(1)_c$. For $N=2$ the WZ model is shown in Figure \ref{wzfig}.
The charge assignments under the abelian symmetries for the chiral fields in the WZ theory are shown in Table \ref{tab:wz charges}, along with the charges of the gauge invariant operators of the gauge theory side that we described above. 

\begin{table}[b]
\centering
\begin{tabular}{|c|c|c|c|c|}
\hline
Gauge th.  &$U(1)_{R_0}$ & $U(1)_c$ & $U(1)_t$ & WZ th.  \\
\hline
$E_i^{(nm)}$ &  $2i+M_{n-1}-M_n - M_{m-1}+M_m$ & 0 & $\frac{M_n-M_{n-1}+M_{m-1}-M_m - 2i}{2}$ & $T_i^{(nm)}$ \\
$V^{(n)}_i$ & $\frac{4i-1+2M_{n-1}-2M_n-M_{N-1}}{2}$ & $\frac{1}{2}$ & $\frac{2-4i+M_{N-1}+2M_n-2M_{n-1}}{4}$ & $U_i^{(n)}$ \\
$\tilde{B}_i$ & $1+2i+M_{N-1}-2M_N$ & $-1$ & $\frac{-2i+2M_N-M_{N-1}}{2}$ & $\tilde{A}_i$ \\
\hline
\end{tabular}
\caption{\label{tab:wz charges} Charges under the abelian symmetries of the gauge invariant operators of the quiver theory and of the corresponding singlets of the WZ theory.}
\end{table}

From Table \ref{tab:wz charges} we deduce that the mapping of operators across the duality is:
\begin{align}
	E_i^{(nm)} &\longleftrightarrow T_i^{(nm)}\qquad i=1,...,M_n - M_{n-1} \nn\\
	V_i^{(n)} &\longleftrightarrow U_i^{(n)} \,\, \,\qquad i=1,...,M_n - M_{n-1} \nn\\
	\tilde{B}_i &\longleftrightarrow \tilde{A}_i \, \, \qquad \quad i=1,...,M_N \,.
\end{align}
This in particular tells us, as we mentioned previously, that some of the gauge invariant operators in the quiver theory should vanish quantum mechanically in the chiral ring, since there is no corresponding chiral field on the WZ side.

The strategy that we use to prove this duality is again based on sequential deconfinement. For simplicity, in the following discussion discussion we focus on the charged fields. The singlets can be reconstructed by looking at the results of the derivation with the supersymmetric index (see Appendix \ref{qsd2}).

\begin{figure}
\centering
\includegraphics[width = 19.5 cm,angle=90,origin=c]{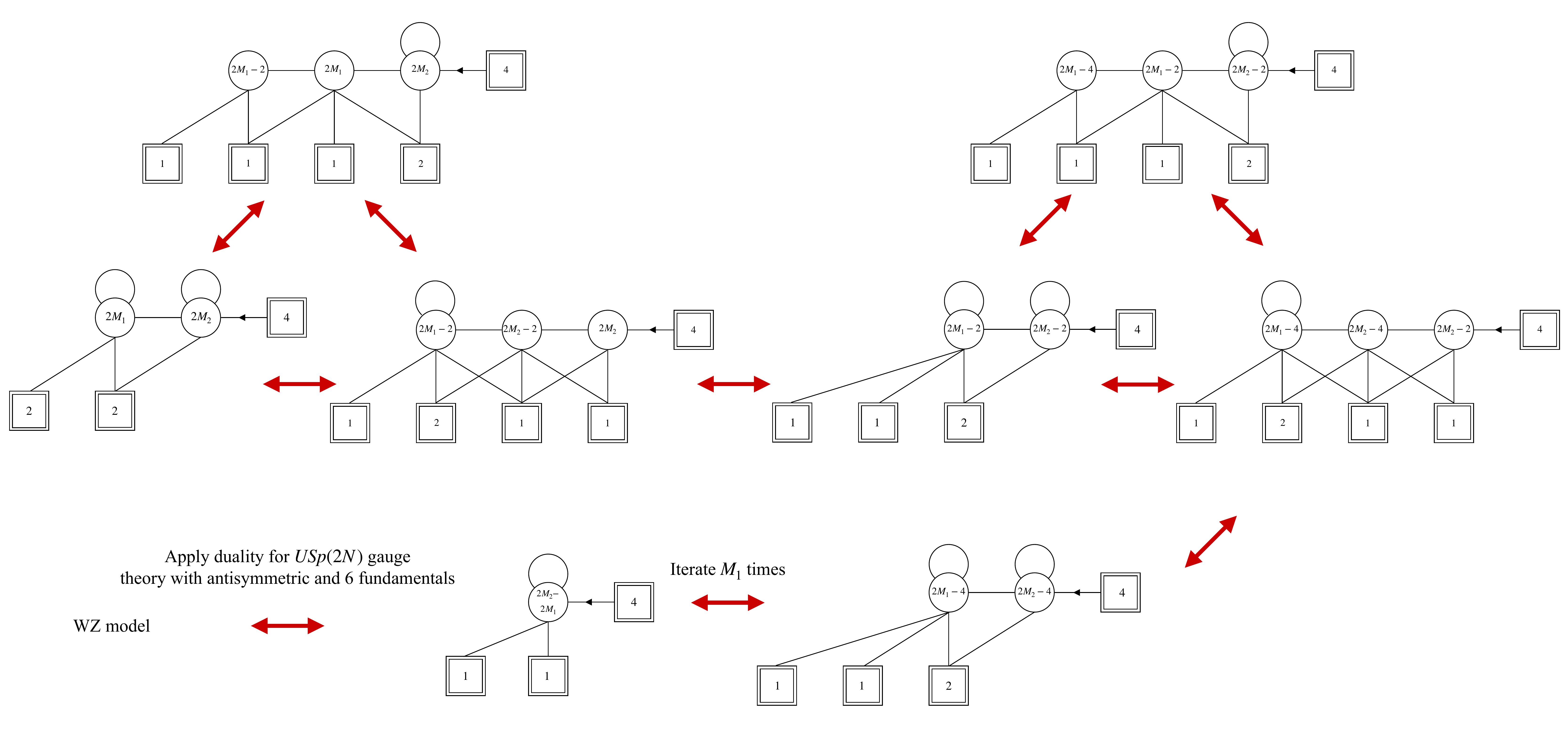}
\caption{Schematic representation of the main steps of the derivation of the duality for $N=2$. We avoid drawing singlets not to clutter the figure.}
\label{qsd}
\end{figure}

Let's consider first the two node case $N=2$ sketched  in Figure \ref{qsd}.
Starting from the auxiliary quiver in the top left corner we can apply the IP duality to the  $USp(2M_1 -2)$ node and move to the original theory, the first quiver in the second line. If we instead apply IP to the  $USp(2M_1)$ node we move to the second quiver in the second line. The antisymmetric chiral of the $USp(2M_2)$ node has been removed so we can IP dualize also this node which confines and we reach the third quiver in the second line. This quiver theory has the same structure of the original one but all ranks are decreased by one unit and some extra singlets have been created
\footnote{Notice that, because of their interactions, one $SU(2)$ flavor symmetry  is broken as in the single node case in Figure \ref{fig6}.}.
We can  say that the original quiver theory is {\it stable} under this sequence of dualizations and we can iterate it to systematically lower the ranks. Indeed we can now consider the auxiliary quiver in the top right corner. If we IP dualise the $USp(2M_1-4)$ node we go back to the third quiver in the second line. If we dualise the $USp(2M_1-2)$ node we move to the last quiver in the second line. Here the $USp(2M_2-2)$ node has no antisymmetric and we can use the IP duality to confine it  and  reach the second quiver in the last line. This quiver has again the same structure of the original quiver theory but all ranks are now decreased by two units and some extra singlets have been created. 
Iterating this sequence of dualizations $M_1$ times we reach the first quiver in the third line.
This is a $USp(2M_2-2M_1)$ theory with one antisymmetric and six fundamental chiral multiplets and using the duality discussed in the previous section we reach the WZ dual frame.

For longer quivers we similarly consider an auxiliary quiver where we trade the  antisymmetric on the first $USp(2M_1)$ gauge node with a new $USp(2M_1 -2)$ gauge node. Dualising this node it confines, with part of the singlets reconstructing the antisymmetric on the $USp(2M_1)$ node, and hence we go back to our original theory. 
If instead we IP dualize the second $USp(2M_1)$ gauge node, we turn it into a $USp(2M_2-2)$ gauge node 
removing the antisymmetric of the  $USp(2M_3)$ node, so that now it can be dualised. We can now continue duailsing all the gauge nodes proceeding to the right of the tail with the result that the rank of each gauge node to which we apply the duality shifts from $USp(2M_i)$ to $USp(2M_{i+1}-2)$. When we finally apply the duality to the last gauge node, this confines and we obtain a quiver with the same structure of the original  theory but with the rank of each gauge node lowered by one unit. 
So the original quiver is {\it stable} under this sequence of dualizations and we can iterate this procedure $k$ times
until we obtain a new quiver where all the ranks are lowered by $k$ units.
For $k=M_1$ we completely confine the first gauge node and obtain a shorter quiver whose first node has group $USp(2M_2-2M_1)$.  We can now perform other  $2M_2-2M_1$ iterations to confine also the second gauge node and obtain a shorter quiver. Continuing with this strategy  we reach a frame with only one remaining $USp(2M_N-2M_{N-1})$ gauge node, an antisymmetric and  six fundamental chirals which 
we know is dual to a WZ model. Using this result, we can completely confine also the last gauge node, which concludes the derivation of the duality between the gauge theory and the WZ model.

At this point it is clear why we required that the ranks are ordered as $M_1\leq M_2\leq \cdots \leq M_N$. This is indeed needed in order for the quiver to start confining from the left. If instead the ranks weren't ordered, one could in principle still apply the deconfinement procedure, but the derivation we just described would need to be modified. In particular, at some point of our derivation we would have a quiver with the lowest rank node being in the middle instead on the very left. This will cause this node to confine before the leftmost one, so that the quiver actually breaks into two subquivers. It would be interesting to investigate further what would be the dual theory obtained from deconfinement in case of non-ordered ranks.

All the manipulations we described can be repeated at the level of the supersymmetric index, keeping track of the singlets produced at each step of the derivation. This allows us to provide a test of the duality by analytically matching the index of the original theory and the one of the final WZ model. In Appendix \ref{qsd2} we give all the details of the $N=2$ case, while here we will give the index identity for generic $N$.

 The index of the gauge theory depends on fugacities for the global symmetries $U(1)_t$ and $U(1)_c$ symmetries, which we denote accordingly by $t$ and $c$, fugacities for the $SU(2)_{y_i}$ symmetries, which we call $y_i$, and finally fugacities for the $SU(4)_x$ symmetry, which we denote $x_a$.  The index identity corresponding to the duality between the quiver gauge theory and the WZ model is then given by
 \begin{align}\label{quiverwzid}
 	&\mathcal{I}^{(N)}_{\text{gauge}} (\vec{M};\vec{y},\vec{x};t,c) =  \prod_{n=1}^N \prod_{i=1}^{M_n-M_{n-1}} \Gamma_e ((pq)^{2-i-M_N-M_{n-1}+M_{N-1}+M_n} c^{-2} t^{-M_{N-1}-M_n+M_N+M_{n-1}+N-n+i-1)} \nonumber \\
  &\times \prod_{n=1}^N \prod_{i=1}^{M_n - M_{n-1}} \Gamma_e ( (pq)^{1-i} t^{i})\oint \prod_{n=1}^N d \vec{z}^{\,(M_n)}\prod_{n=1}^N \Gamma_e(pqt^{-1})^{M_n} \prod_{i<j}^{M_n} \Gamma_e (pqt^{-1} z_i^{(M_n) \pm 1} z_j^{(M_n) \pm 1})  \nonumber \\
 	&\times \prod_{n=1}^{N-1} \prod_{i=1}^{M_n} \prod_{j=1}^{M_{n+1}} \Gamma_e ( t^\frac{1}{2} z_i^{(M_n) \pm 1} z_j^{(M_{n+1}) \pm 1}) \prod_{i=1}^{M_N} \prod_{a=1}^4 \Gamma_e ((pq)^\frac{3+M_{N-1}-2M_{N}}{4} c^{-\frac{1}{2}} t^{\frac{2M_N-M_{N-1}-2}{4}} x_a z_i^{(M_N) \pm 1} ) \nonumber \\
 	&\times \prod_{n=1}^N \prod_{i=1}^{M_n} \Gamma_e ((pq)^\frac{M_N+M_{n-1}-M_{N-1}-M_n}{2} c \, t^\frac{M_{N-1}+M_{n}-M_{N}-M_{n-1}-N+n}{2} y_n^{\pm 1} z_i^{(M_n) \pm 1}) \nonumber \\
 	&\times \prod_{n=1}^{N-1} \prod_{i=1}^{M_n} \Gamma_e ( (pq)^\frac{2+M_{N-1}+M_{n+1}-M_N-M_n}{2}c^{-1}t^\frac{M_N+M_n-M_{N-1}-M_{n+1}+N-n-2}{2} y_{n+1}^{\pm 1}  z_i^{(M_n) \pm 1}) = \nonumber \\
 	&=  \prod_{i=1}^{M_N} \prod_{a<b}^4 \Gamma_e ( (pq)^\frac{1+2i+M_{N-1}-2M_N}{2} c^{-1} t^\frac{2M_N-M_{N-1}-2i}{2} x_a x_b)  \nonumber \\
 	&\times \prod_{n=1}^{N-1} \prod_{i=1}^{M_n - M_{n-1}} \prod_{m=n+1}^{N} \Gamma_e ( (pq)^\frac{2i+M_{n-1}-M_n+M_m -M_{m-1}}{2} t^\frac{M_{m-1} +M_n -M_{n-1} -M_m -2i}{2} y_n^{\pm 1} y_m^{\pm 1}) \nonumber \\
 	&\times  \prod_{n=1}^N \prod_{i=1}^{M_n-M_{n-1}} \prod_{a =1}^4 \Gamma_e ( (pq)^\frac{2M_{n-1}-M_{N-1}-2M_n+4i-1}{4} c^\frac{1}{2} t^\frac{2-4i+2M_n+M_{N-1}-2M_{n-1}}{4} y_n^{\pm 1} x_a)=\mathcal{I}^{(N)}_{\text{WZ}} (\vec{M};\vec{y},\vec{x};t,c) \,.
\end{align} 
 where we put $M_n=0$ if $n \leq 0$ or $n > N$. With $\vec{M} = (M_1, ..., M_N)$ we denote an $N$-dimensional vector containing the ranks of the gauge nodes.
 
\section{Cross-leg duality and $E_6$ symmetry enhancement}
\label{sec:crossleg}

\begin{figure}[t]
\centering
\includegraphics[width = 0.45\textwidth]{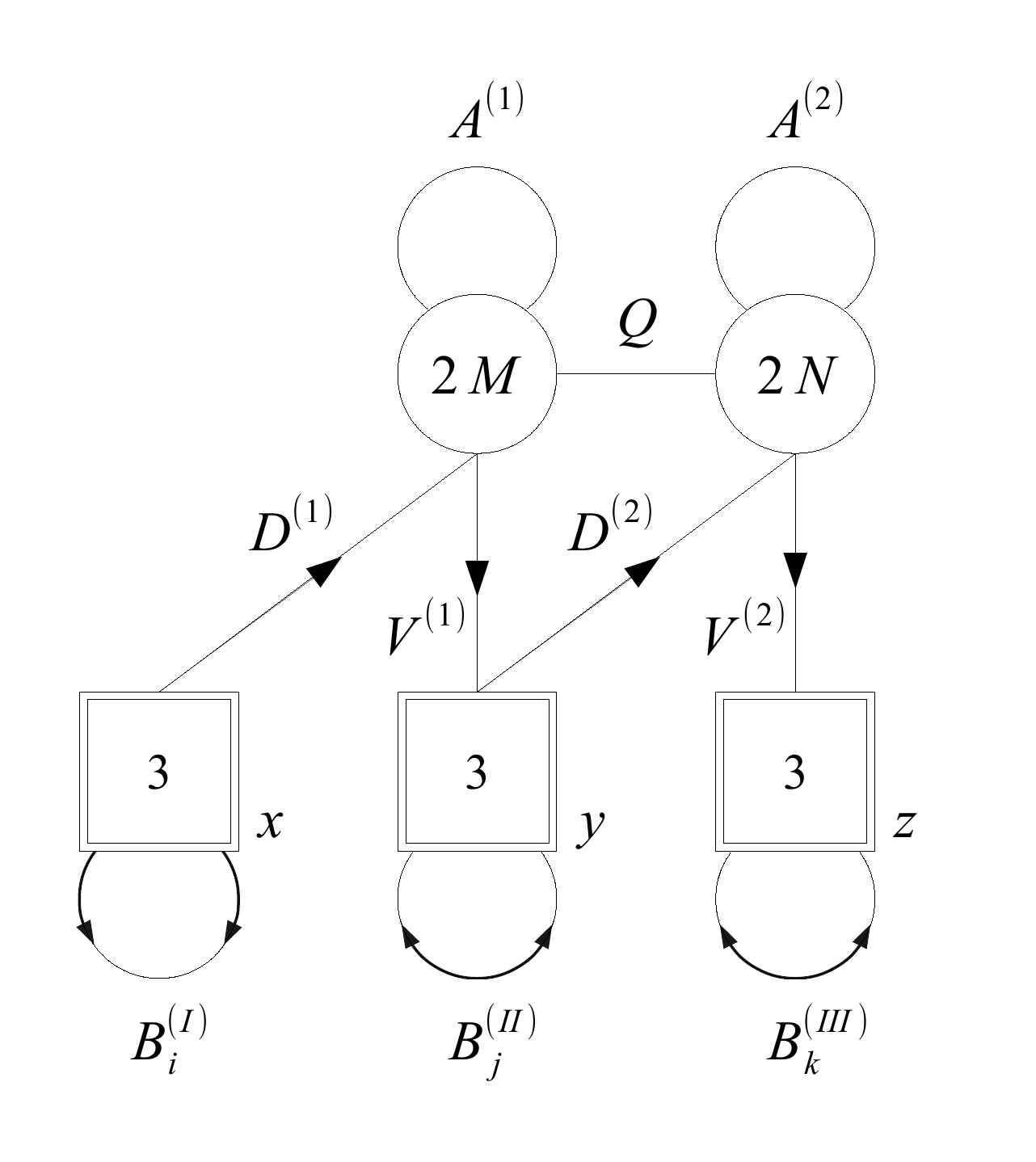}
\caption{\label{fig:TheoryA} The quiver diagram summarizing the field content of the $USp(2 M) \times USp(2 N)$ gauge theory. Later we will also refer to it as Theory A.}
\end{figure}

\subsection{The theory}

In this section, we study a quiver theory with $USp(2 M) \times USp(2 N)$ gauge group 
which enjoys a novel duality, which we call the \emph{cross-leg duality}, and  provide evidence 
of an IR enhancement of the global symmetry to  $E_6$.
Indeed, this theory can be regarded as a multi-nodes extension of the $E_6$ model in \cite{Razamat:2018gbu,Hwang:2020ddr}. Also this new cross-leg duality 
can be proven by the method of sequential deconfinement.
The matter content of the theory, which we shall also call Theory A, consists of chiral fields in the fundamental, bifundamental and antisymmetric representations of the gauge group $USp(2 M) \times USp(2 N)$:
\begin{itemize}
	\item a chiral field $Q$ in the bifundamental representation of $USp(2M) \times USp(2N)$,
	\item six chiral fields $D_\alpha^{(1)}$ and $V_\beta^{(1)}$ in the fundamental representation of $USp(2M)$ with $\alpha, \, \beta=1,2,3$,
	\item six chiral fields $D_\beta^{(2)}$ and $V_\gamma^{(2)}$ in the fundamental representation of $USp(2N)$ with $\beta, \, \gamma=1,2,3$,
	\item two chiral fields $A^{(1)}$ and $A^{(2)}$ in the antisymmetric representation of $USp(2M)$ and $USp(2N)$ respectively,
	\item three towers of $3\times 3$ antisymmetric matrices of gauge singlet fields $B_i^{(I)}, B_j^{(II)}$ and $B_k^{(III)}$ with $i = 1,\dots, M, \, j = 1,\dots,M-N$ and $k=1,\dots,N$ assuming $M-N \geq 0$ without loss of generality,
	\item two towers of gauge singlet fields $\beta^{(1)}_i$ and $\beta^{(2)}_k$ with $i = 1,\dots, M$ and $k = 1,\dots, N$.
\end{itemize}
The content of the theory is schematically represented by the quiver diagram shown in figure \ref{fig:TheoryA}.
\begin{table}[t]
\centering
\begin{tabular}{|c|c|c|c|c|}
\hline
 & $SU(3)_x\times SU(3)_y\times SU(3)_z$ & $U(1)_{R_0}$ & $U(1)_t$ & $U(1)_c$ \\
\hline
$Q$ & $\mathbf{(1,1,1)}$ & 0 & 1/2 & 0 \\
$A^{(1/2)}$ & $\mathbf{(1,1,1)}$ & 2 & $-1$ & 0 \\
$D^{(1)}$ & $\mathbf{(3,1,1)}$ & $-\frac12 M+1$ & $\frac14 M-\frac12$ & 1 \\
$V^{(1)}$ & $\mathbf{(1,\overline{3},1)}$ & $-\frac56 M+\frac23 N+1$ & $\frac{5}{12} M-\frac13 N-\frac{1}{6}$ & $-1$ \\
$D^{(2)}$ & $\mathbf{(1,3,1)}$ & $\frac56 M-\frac23 N+1$ & $-\frac{5}{12} M+\frac13 N-\frac{1}{3}$ & 1 \\
$V^{(2)}$ & $\mathbf{(1,1,\overline{3})}$ & $-\frac16 M-\frac23 N+1$ & $\frac{1}{12} M+\frac13 N-\frac{1}{3}$ & $-1$ \\
$B_i^{(I)}$ & $\mathbf{(3,1,1)}$ & $M-2 i+2$ & $-\frac12 M+i$ & $-2$ \\
$B_{j}^{(II)}$ & $\mathbf{(1,\overline{3},1)}$ & $\frac53 M-\frac43 N-2 j+2$ & $-\frac56 M+\frac23 N+j-\frac23$ & 2 \\
$B_{k}^{(III)}$ & $\mathbf{(1,1,\overline{3})}$ & $\frac13 M+\frac43 N-2 k+2$ & $-\frac16 M-\frac23 N+k-\frac13$ & 2 \\
$\beta^{(1)}_i$ & $\mathbf{(1,1,1)}$ & $2-2 i$ & $i$ & 0 \\
$\beta^{(2)}_k$ & $\mathbf{(1,1,1)}$ & $2-2 k$ & $k$ & 0 \\
\hline
\end{tabular}
\caption{\label{tab:charge A} Charges and representations of the matter fields under the global symmetry, including the trial $R$-charge $R_0$.}
\end{table}
The matter fields interact via four types of interactions preserving the global symmetry
\begin{align}
SU(3)_x\times SU(3)_y\times SU(3)_z \times U(1)_t \times U(1)_c \,,
\end{align}
under which they are charged as shown in Table \ref{tab:charge A}.
Firstly, we have two types of cubic couplings: one between the bifundamentals and the antisymmetrics and the other among the chirals in the triangle of the quiver. Also there is an interaction between $B_k^{(\cdot)}$ and some of the mesons dressed by the antisymmetrics. Lastly, the trace of powers of the antisymmetrics are flipped by $\beta^{(1)}_i$ and $\beta^{(2)}_k$. The total superpotential is explicitly given by
\begin{align}
	\mathcal{W}_A &= \text{Tr}_M \left[ A^{(1)} \text{Tr}_N (Q Q) \right] + \text{Tr}_N \left[ A^{(2)} \text{Tr}_M (Q Q) \right] + \text{Tr}_{y} \text{Tr}_M \text{Tr}_N \left[V^{(1)} Q D^{(2)} \right] \nonumber \\
	&+ \sum_{i = 1}^M \text{Tr}_{x} \left[B_{i}^{(I)} \text{Tr}_M \left( A^{(1) \, i-1} D^{(1)}  D^{(1)} \right)\right] + \sum_{j = 1}^{M-N} \text{Tr}_{y} \left[B_{j}^{(II)} \text{Tr}_M \left( A^{(1) \, j-1} V^{(1)}  V^{(1)} \right)\right] \nonumber \\
	&+ \sum_{k = 1}^N \text{Tr}_{z} \left[B_{k}^{(III)} \text{Tr}_N \left( A^{(2) \, k-1} V^{(2)}  V^{(2)} \right)\right]+ \sum_{i = 1}^M \beta^{(1)}_i \text{Tr}_M A^{(1) i}+ \sum_{k = 1}^N \beta^{(2)}_k \text{Tr}_N A^{(2) k}
\end{align}
where $M \geq N$ is assumed without loss of generality.
Here $\text{Tr}_M$ and $\text{Tr}_N$ denote the traces over the color indices of the $USp(2M)$ and $USp(2N)$ gauge nodes respectively, while $\text{Tr}_{x,y,z}$ denote the traces over the three $SU(3)_{x,y,z}$ flavor symmetries. 

Some examples of the gauge invariant operators are listed in Table \ref{tab:operators}.
\begin{table}[b]
\centering
\begin{tabular}{|c|c|c|c|c|}
\hline
 & $SU(3)_x\times SU(3)_y\times SU(3)_z$ & $U(1)_{R_0}$ & $U(1)_t$ & $U(1)_c$ \\
\hline
$D^{(1)} A^{(1) i-1} V^{(1)}$ & $\mathbf{(3,\overline{3},1)}$ & $-\frac43 M+\frac23 N+2 i$ & $\frac23 M-\frac13 N-i+\frac13$ & 0 \\
$D^{(1)} A^{(1) i-1} Q V^{(2)}$ & $\mathbf{(3,1,\overline{3})}$ & $-\frac23 M-\frac23 N+2 i$ & $\frac13 M+\frac13 N-i+\frac23$ & 0 \\
$D^{(2)} A^{(2) k-1} V^{(2)}$ & $\mathbf{(1,3,\overline{3})}$ & $\frac23 M-\frac43 N+2 k$ & $-\frac13 M+\frac23 N-k+\frac13$ & 0 \\
\hline
\end{tabular}
\caption{\label{tab:operators} Examples of the gauge invariant operators. Contractions of all color indices are understood.}
\end{table}
\begin{figure}[t]
\centering
\includegraphics[width = 0.45\textwidth]{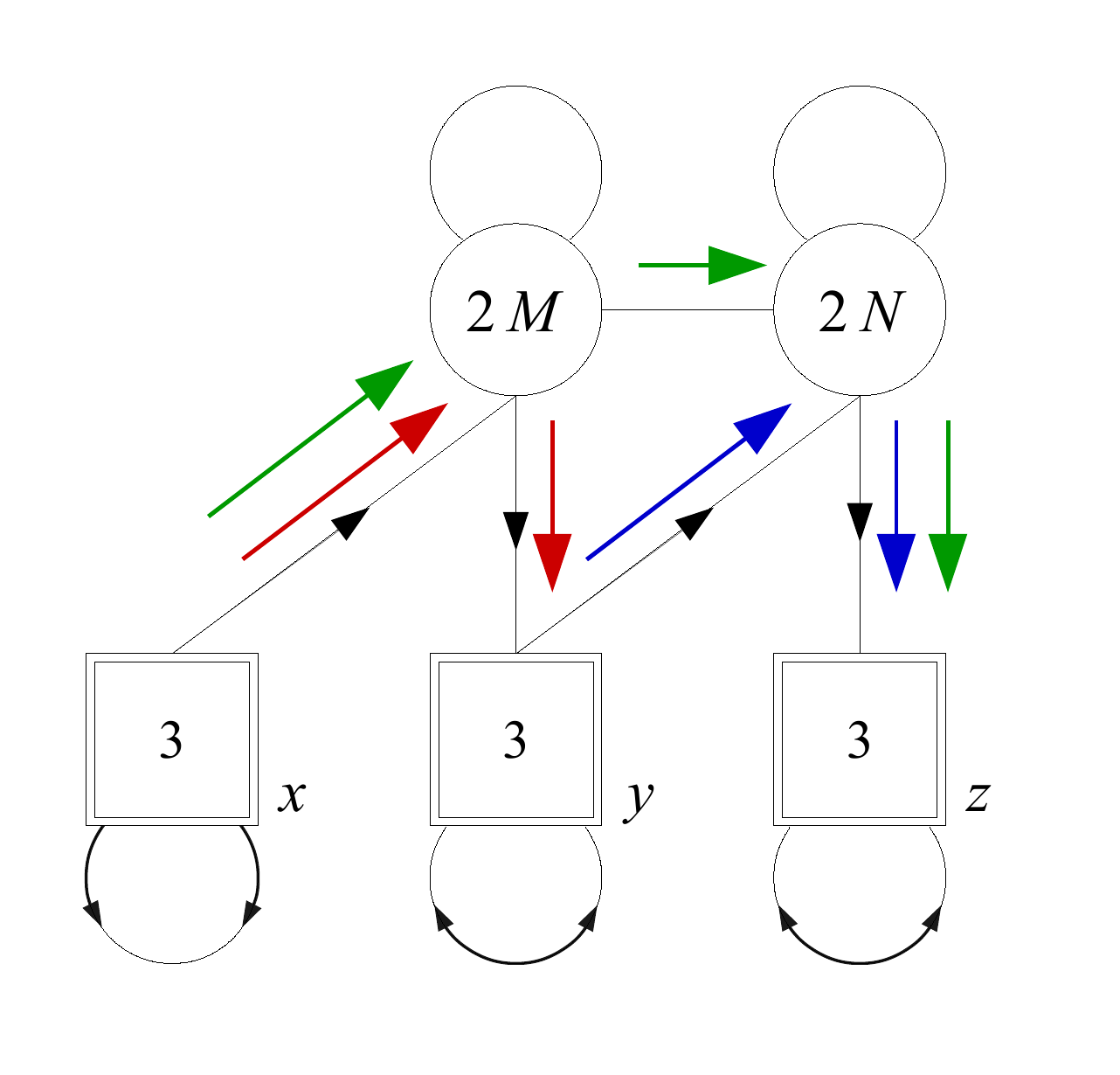}
\caption{\label{fig:operators} Schematic representation of how the gauge invariant operators listed in Table \ref{tab:operators} are constructed.
For example operators $D^{(1)}V^{(1)}$ are indicated in red, $D^{(2)}  V^{(2)}$ in blue and $D^{(1)} Q V^{(2)}$ in green.}
\end{figure}
As depicted in Figure \ref{fig:operators}, they are constructed starting from a diagonal link and ending on a vertical link, which results in the bifundamental representations of all possible pairs of $SU(3)$s. Moreover, also here we can dress all these operators with powers of the antisymmetric of the corresponding gauge node. Note the $D^{(1)}Q V^{(2)}$ meson can be equivalently dressed with $A^{(1)}$ or $A^{(2)}$, as we discussed in the previous section.

\subsection{The duality}

\begin{figure}[tbp]
\centering
\includegraphics[width=0.45\textwidth]{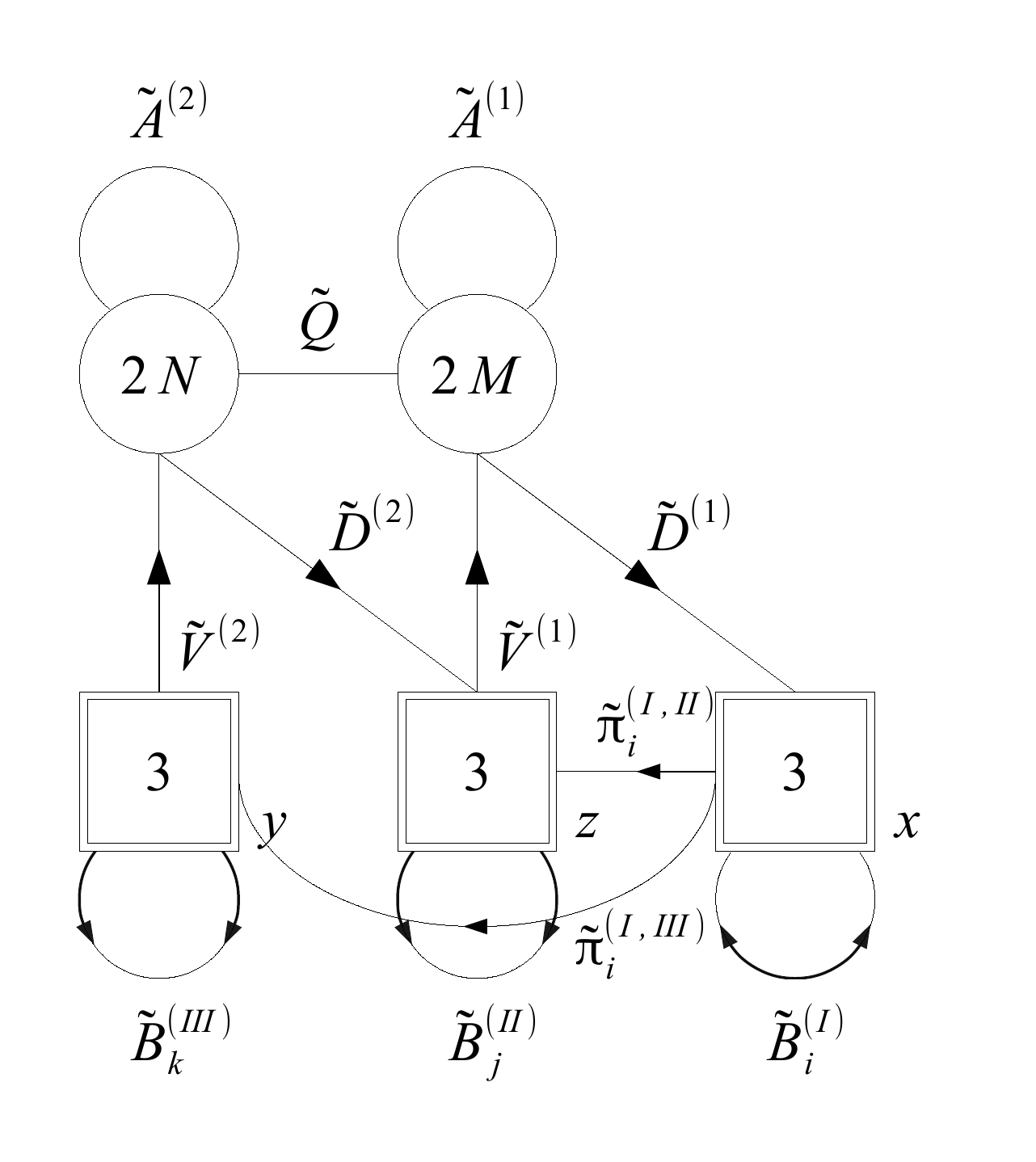}
\caption{The quiver diagram for Theory B.}
\label{fig:TheoryB}
\end{figure}

We claim that Theory A shown in Figure \ref{fig:TheoryA} admits a cross-leg dual description that has the same quiver structure of Theory A with additional gauge singlets $\tilde \pi_i^{(\cdot,\cdot)}$. We call this dual theory Theory B. The corresponding quiver diagram is shown in Figure \ref{fig:TheoryB}. 

The superpotential of Theory B consists of five types of interactions. While the first four are exactly the same as those of Theory A, we have extra terms coupling the singlets $\tilde \pi_i^{(\cdot,\cdot)}$ to some of the dual mesons. The total superpotential of Theory B is then given by
\begin{align}
	\mathcal{W}_B &= \text{Tr}_M \left[ \tilde A^{(1)} \text{Tr}_N (\tilde Q \tilde Q) \right] + \text{Tr}_N \left[ \tilde A^{(2)} \text{Tr}_M (\tilde Q \tilde Q) \right] + \text{Tr}_{z} \text{Tr}_M \text{Tr}_N \left[\tilde{V}^{(1)} \tilde Q \tilde{D}^{(2)} \right] \nonumber \\
	&+ \sum_{i = 1}^M \text{Tr}_{x} \left[\tilde B_{i}^{(I)} \text{Tr}_M \left( \tilde A^{(1) \, k-1} \tilde{D}^{(1)} \tilde{D}^{(1)} \right)\right] +\sum_{j = 1}^{M-N} \text{Tr}_{z} \left[\tilde B_{j}^{(II)} \text{Tr}_M \left( \tilde A^{(2) \, j-1} \tilde{V}^{(2)} \tilde{V}^{(2)} \right)\right] \nonumber \\
	& + \sum_{k = 1}^N \text{Tr}_{y} \left[\tilde B_{k}^{(III)} \text{Tr}_N \left( \tilde A^{(2) \, i-1} \tilde{V}^{(2)} \tilde{V}^{(2)} \right)\right]+ \sum_{i = 1}^M \tilde \beta^{(1)}_i \text{Tr}_M \tilde A^{(1) i}+ \sum_{k = 1}^N \tilde \beta^{(2)}_k \text{Tr}_N \tilde A^{(2) k} \nonumber \\
	& + \sum_{i = 1}^M \text{Tr}_{x} \text{Tr}_{z} \left[\tilde \pi_i^{(I,II)} \text{Tr}_M \left(\tilde{D}^{(1)} \tilde A^{(1) i-1}\tilde{V}^{(1)}\right)\right] + \sum_{i = 1}^M \text{Tr}_{x} \text{Tr}_{y} \left[\tilde \pi_i^{(I,III)} \text{Tr}_N \text{Tr}_M \left(\tilde{D}^{(1)} \tilde A^{(1) i-1} \tilde Q \tilde{V}^{(2)}\right)\right] ,
\end{align}
where the last line shows the extra interactions terms including the $\tilde \pi_i^{(\cdot,\cdot)}$ singlets. The charge assignments for Theory B are summarized in Table \ref{tab:charge B}.

\begin{table}[tbp]
\centering
\begin{tabular}{|c|c|c|c|c|}
\hline
 & $SU(3)_x\times SU(3)_y\times SU(3)_z$ & $R_0$ & $U(1)_t$ & $U(1)_c$ \\
\hline
$\tilde Q$ & $\mathbf{(1,1,1)}$ & 0 & 1/2 & 0 \\
$\tilde A^{(1/2)}$ & $\mathbf{(1,1,1)}$ & 2 & $-1$ & 0 \\
$\tilde{D}^{(1)}$ & $\mathbf{(\overline{3},1,1)}$ & $-\frac12 M+1$ & $\frac14 M-\frac12$ & $-1$ \\
$\tilde{V}^{(1)}$ & $\mathbf{(1,1,3)}$ & $-\frac56 M+\frac23 N+1$ & $\frac{5}{12} M-\frac13 N-\frac{1}{6}$ & 1 \\
$\tilde{D}^{(2)}$ & $\mathbf{(1,1,\overline{3})}$ & $\frac56 M-\frac23 N+1$ & $-\frac{5}{12} M+\frac13 N-\frac{1}{3}$ & $-1$ \\
$\tilde{V}^{(2)}$ & $\mathbf{(1,3,1)}$ & $-\frac16 M-\frac23 N+1$ & $\frac{1}{12} M+\frac13 N-\frac13$ & 1 \\
$\tilde B_{i}^{(I)}$ & $\mathbf{(\overline{3},1,1)}$ & $M-2 i+2$ & $-\frac12 M+i$ & 2 \\
$\tilde B_{j}^{(II)}$ & $\mathbf{(1,1,3)}$ & $\frac{5}{3} M-\frac{4}{3} N-2 j+2$ & $-\frac56 M+\frac23 N+j-\frac23$ & $-2$ \\
$\tilde B_k^{(III)}$ & $\mathbf{(1,3,1)}$ & $\frac13 M+\frac43 N-2 k+2$ & $-\frac16 M-\frac23 N+k-\frac13$ & $-2$ \\
$\tilde \beta^{(1)}_i$ & $\mathbf{(1,1,1)}$ & $2-2 i$ & $i$ & 0 \\
$\tilde \beta^{(2)}_k$ & $\mathbf{(1,1,1)}$ & $2-2 k$ & $k$ & 0 \\
$\tilde \pi_{i}^{(I,II)}$ & $\mathbf{(3,1,\overline{3})}$ & $\frac43 M-\frac23 N-2 i+2$ & $-\frac23 M+\frac13 N+i-\frac13$ & 0 \\
$\tilde \pi_{i}^{(I,III)}$ & $\mathbf{(3,\overline{3},1)}$ & $\frac23 M+\frac23 N-2 i+2$ & $-\frac13 M-\frac13 N+i-\frac23$ & 0 \\
\hline
\end{tabular}
\caption{\label{tab:charge B} Charges and representations of the matter fields of Theory B under the global symmetry.}
\end{table}

We can provide the first non-trivial evidence of the duality by checking that all the anomalies match across the duality. For example, the anomalies for the $U(1)$ global symmetries are given by
\begin{align}
\mathrm{Tr} U(1)_{R_0}^3 &= M^2-M-4 M^2 N^2-2 M N-N+N^2+\frac{2}{3} \left(-5 M^4+4 M^3 N+4 M N^3-5 N^4\right) , \\
\mathrm{Tr} U(1)_{R_0}^2 U(1)_t &= 2 M^2 N^2+M N+\frac{1}{2} \left(-3 M^2+M+N-3 N^2\right) \nonumber \\
&\quad +\frac{1}{3} \left(5 M^4-2 M^3-4 M^3 N+2 M^2 N+2 M N^2-4 M N^3-2 N^3+5 N^4\right) , 
\end{align}
\begin{align}
\mathrm{Tr} U(1)_{R_0} U(1)_t^2 &= -M^2 N^2+\frac{5}{6} \left(-M^4-N^4\right)+\frac{1}{2} \left(-M-N\right) \nonumber \\
&\quad +\frac{2}{3} \left(M^3+M^3 N+M^2-M^2 N-M N^2+N^2+M N^3+N^3\right) , \\
\mathrm{Tr} U(1)_{R_0} U(1)_t^2 &= \frac{5}{12} \left(M^4+N^4\right)+\frac{1}{4} \left(-M^2-N^2\right)+\frac{1}{3} \left(-M^3 N+M+N-M N^3\right) \nonumber \\
&\quad +\frac{1}{2} \left(-M^3+M^2 N+M^2 N^2-M N+M N^2-N^3\right)\,.
\end{align}
All anomalies involving   $U(1)_c$ vanish. 

Stronger evidence is provided by the match of the supersymmetric indices of Theory A and Theory B, which we obtain
by implementing at the level of  the index the sequential deconfinement procedure, allowing us to 
derive  this duality.

\begin{figure}[tbp]
\centering
\includegraphics[width = 19.5 cm,angle=90,origin=c]{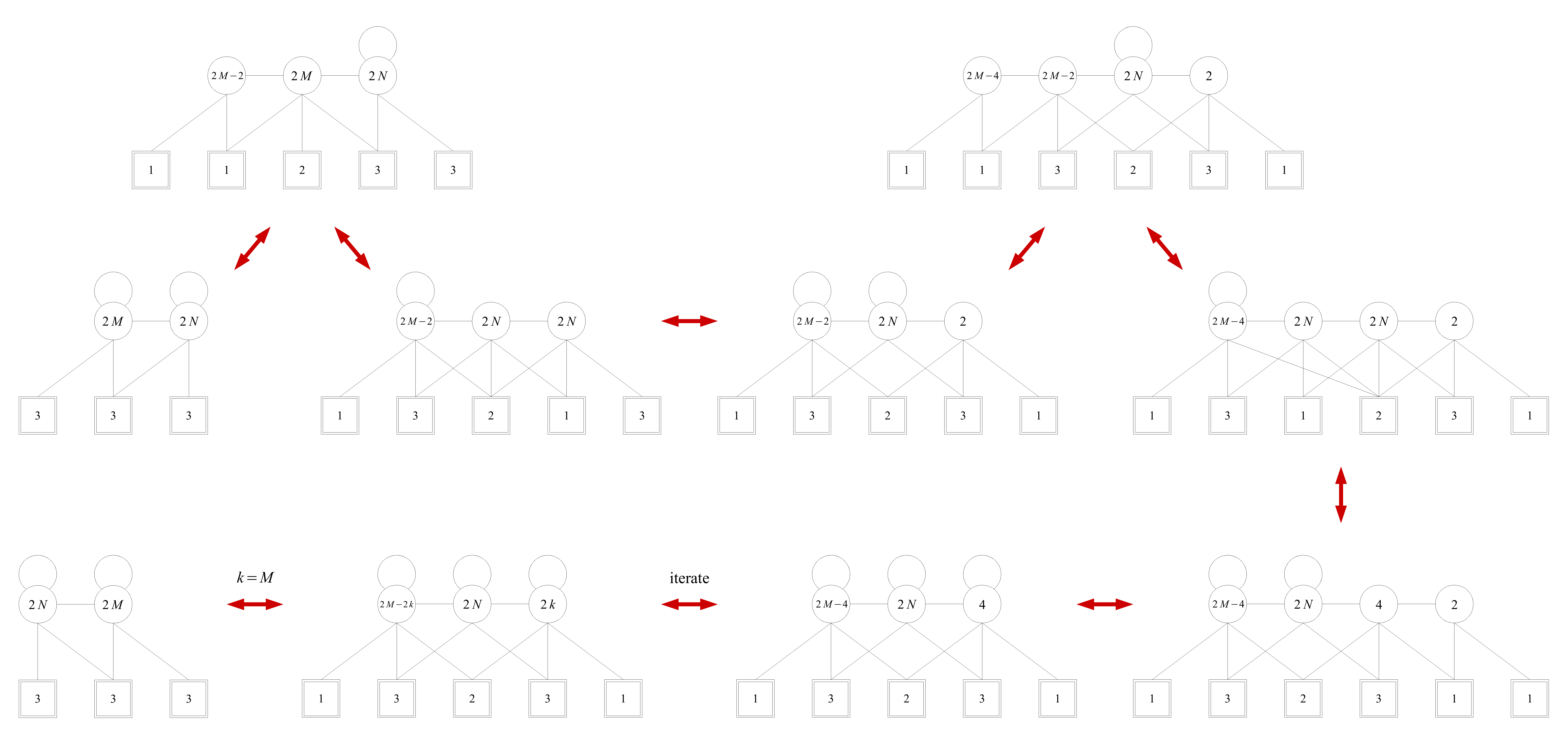}
\caption{Sketch of the manipulations we perform on the original theory to obtain the cross-leg duality. We avoid drawing singlets not to clutter the drawing.}
\label{fig11}
\end{figure}

The sequential deconfinement procedure in this case is sketched in Figure \ref{fig11}.
We start from the auxiliary quiver theory with $USp(2M-2) \times USp(2M) \times USp(2N)$ gauge group in the top left corner. If we IP dualize  the $USp(2M-2)$ gauge node, this confines, the antisymmetric at the $USp(2M)$ node is restored and  we go back to Theory A. If we instead IP dualize the $USp(2M)$ node, this becomes a $USp(2N)$ gauge node and we obtain the second quiver in the second line. In the process, the antisymmetric of the original $USp(2N)$ node becomes massive, so we can IP dualize the $USp(2 N)$ node converting it into a $USp(2)$ gauge node and we obtain the third quiver in the second line. 
Comparing this quiver theory with the original one, we can notice that the effect of these dualizations was to decrease
 by one unit the rank of the first node from left and to add a new $USp(2)$  gauge node on the right. 
 This quiver form is stable, in the sense that if we repeat the whole sequence of dualizations we move to the third quiver in the third line which is still a three-nodes quiver but now  the rank of the first gauge node is lowered by two units, the central node is still $USp(2N)$ and the last node is increased by one unit to $USp(4)$. 
If we iterate the sequence of dualizations  $k$ times, we obtain the second quiver in the third line, 
 which has gauge nodes $USp(2M-2k) \times USp(2N) \times USp(2k)$. This is one of the $k$ \textit{cross-leg dual frames} of our theory.
For $k=M$, we obtain again a  two-node quiver identical to the original one, with restored manifest $SU(3)^3$ symmetry
but with the  $USp(2M)$ and $USp(2N)$ gauge nodes swapped. 
Keeping track of all the singlets generated at each dualization, we find that  this quiver  coincides with our Theory B.

We can repeat all these steps at the level of the index (keeping track of singlets) to obtain the following index identity
for the duality  between the initial and the final frame\footnote{Here we are using the notation  $\vec x \, {}^{-1}=(x_1^{-1},x_2^{-1}, x_2^{-1})$ and similarly for the $\vec{y}$ and $\vec{z}$ fugacities.}:
\begin{align}
& \mathcal I_A(\vec x, \vec y, \vec z;c,t) = \prod_{i = 1}^M \prod_{\alpha = 1}^3 \prod_{\gamma = 1}^3 \Gpq{(pq)^{\frac23 M-\frac13 N-i+1} t^{-\frac23 M+\frac13 N+i-\frac13} x_\alpha z_\gamma^{-1}} \nonumber \\
& \times \prod_{i = 1}^M \prod_{\alpha = 1}^3 \prod_{\beta = 1}^3 \Gpq{(pq)^{\frac13 M+\frac13 N-i+1} t^{-\frac13 M-\frac13 N+i-\frac23} x_\alpha y_\beta^{-1}} \mathcal I_A(\vec x \, {}^{-1}, \vec z\, {}^{-1}, \vec y \, {}^{-1};c^{-1},t) = \mathcal I_B(\vec x, \vec y, \vec z;c,t)
\end{align}
where
\begin{align}
&\mathcal I_A(\vec x, \vec y, \vec z;c,t) \nonumber \\
&= \prod_{i = 1}^M \prod_{\alpha = 1}^3 \Gpq{(pq)^{\frac12 M-i+1} t^{-\frac12 M+i} c^{-2} x_\alpha} \prod_{j = 1}^{M-N} \prod_{\beta = 1}^3 \Gpq{(pq)^{\frac56 M-\frac23 N-j+1} t^{-\frac56 M+\frac23 N+j-\frac23} c^2 y_\beta^{-1}} \nonumber \\
& \times \prod_{k = 1}^N \prod_{\gamma = 1}^3 \Gpq{(pq)^{\frac16 M+\frac23 N-k+1} t^{-\frac16 M-\frac23 N+k-\frac13} c^2 z_\gamma^{-1}} \prod_{i = 1}^M \Gpq{(pq)^{1-i} t^i} \prod_{k = 1}^N \Gpq{(pq)^{1-k} t^k} \nonumber \\
& \times \oint \ud{\vec u_M} \ud{\vec v_N} \Gpq{pq t^{-1}}^{M+N} \prod_{i < j}^M \Gpq{pq t^{-1} u_i^\pm u_j^\pm} \prod_{k < l}^N \Gpq{p q t^{-1} v_k^\pm v_l^\pm} \prod_{i=1}^M \prod_{k = 1}^N \Gpq{t^\frac12 u_i^\pm v_k^\pm} \nonumber \\
& \times \prod_{i = 1}^M \prod_{\alpha = 1}^3 \Gpq{(pq)^{-\frac14 M+\frac12} t^{\frac14 M-\frac12} c u_i^\pm x_\alpha} \prod_{i = 1}^M \prod_{\beta = 1}^3 \Gpq{(pq)^{-\frac{5}{12} M+\frac13 N+\frac12} t^{\frac{5}{12} M-\frac13 N-\frac16} c^{-1} u_i^\pm y_\beta^{-1}} \nonumber \\
& \times \prod_{k = 1}^N \prod_{\beta = 1}^3 \Gpq{(pq)^{\frac{5}{12} M-\frac13 N+\frac12} t^{-\frac{5}{12} M+\frac13 N-\frac13} c v_k^\pm y_\beta} \prod_{k = 1}^N \prod_{\gamma = 1}^3 \Gpq{(pq)^{-\frac{1}{12} M-\frac13 N+\frac12} t^{\frac{1}{12} M+\frac13 N-\frac13} c^{-1} v_k^\pm z_\gamma^{-1}} \nonumber \\
\end{align}
with  $\prod_{\alpha = 1}^3 x_\alpha = \prod_{\beta = 1}^3 y_\beta = \prod_{\gamma = 1}^3 z_\gamma = 1$ (and 
we assumed $M \geq N$ without loss of generality). Here, $x_\alpha, \, y_\alpha, \, z_\alpha, \, c$ and $t$ are the fugacities of $SU(3)_x \times SU(3)_y \times SU(3)_z \times U(1)_c \times U(1)_t$, respectively. 

\subsection{A self-dual frame with $E_6$ global symmetry}

We have seen that Theory A admits several dual frames as shown in Figure \ref{fig11}. Especially, a particular dual frame that we called Theory B has the same quiver structure and the manifest global symmetry as Theory A. In this regard, Theory A is \emph{almost} self-dual because Theory A and Theory B only differ by extra singlets.

Recently it was argued that sometimes such almost self-dual theories can be modified by reorganizing the extra singlets\footnote{By this we mean adding new singlet fields on both sides of the duality, such that on one frame they flip some operators while on the other they flip singlet fields that are mapped to such operators across the cross-leg duality. In the latter frame, both the new and the old singlets are massive and can be integrated out. Overall, the effect was of moving some singlets from one side of the duality to the other.} so that they become \emph{exactly} self-dual,
meaning that the dual frames have exactly the same matter content, including singlets and superpotential, but there is still    a non-trivial  map of the operators across the duality. In particular it has been observed that such exactly self-dual theories often  enjoy non-trivial enhancements of the global symmetry \cite{Razamat:2017hda,Razamat:2018gbu,Hwang:2020ddr}. We will shortly see that our model is also such a case: the gauge singlets of a cross-leg dual pair can be reorganized such that the duality becomes an exact self-duality. The modified theory will then exhibit an interesting symmetry enhancement to the $E_6$ group in the IR.

According to \cite{Razamat:2018gbu,Hwang:2020ddr}, such redistribution of the singlets can be easily determined by looking at the marginal operators and their relations. Let us first consider the simplest case: $M = N = 1$. To examine the marginal operators and their relations, it is useful to compute the superconformal index. Then we first need to perform the $a$-maximization, which determines the mixing coefficients of various $U(1)$ symmetries with the $R$-symmetry, which are input parameters for the index. Sometimes there can be an operator $\mathcal O$ violating the unitarity bound, i.e.~having the conformal dimension less than 1, in which case this operator decouples and  should be flipped by introducing an additional singlet field $\mathcal F$ and with interaction $\Delta \mathcal{W} = \mathcal O \mathcal F$ \cite{Benvenuti:2017lle}. For $M = N = 1$, the $a$-maximization gives the following mixing coefficient of $U(1)_t$ with the $R$-symmetry:
\begin{align}
\mathfrak t = 3-\sqrt{\frac{11}{3}} \approx 1.1 \approx \frac{12}{11} \,,
\end{align}
where $\frac{12}{11}$ is an approximate rational value we use for the index computation. Note that $U(1)_c$ doesn't contribute to the $a$-function because all the anomalies involving $U(1)_c$ identically vanish as we have seen in the previous subsection. 
 With this value of the mixing coefficient, there is no unitarity violating operator. Thus, one can obtain the index of the interacting theory without any further flip of operators. The resulting index is then given by
\begin{align}
\label{eq:(1,1)}
\mathcal I_A^{(1,1)} &= 1+\left(\chi_\mathbf{3}(x) \chi_\mathbf{\overline{3}}(y)+\chi_\mathbf{3}(y) \chi_\mathbf{\overline{3}}(z)\right) t^{-\frac13} (pq)^\frac{16}{33}+\chi_\mathbf{3}(x) \chi_\mathbf{\overline{3}}(z) \, t^{\frac13} (pq)^\frac{17}{33}+t (pq)^\frac{18}{33}+\dots \nonumber \\
&\quad +\left(-1-\chi_\mathbf{8}(x)-\chi_\mathbf{8}(y)-\chi_\mathbf{8}(z)+\chi_\mathbf{6}(x) \chi_\mathbf{\overline{3}}(y) \chi_\mathbf{\overline{3}}(z)+\chi_\mathbf{3}(x) \chi_\mathbf{3}(y) \chi_\mathbf{\overline{6}}(z)\right) pq+\dots \,.
\end{align}
Note that the index is written in terms of $\chi_\mathbf{n}$, which is the character of an $\mathbf n$-dimensional representation of the $SU(3)$ group. From \eqref{eq:(1,1)}, one can detect the contribution of one of the components of the conserved current multiplet \cite{Beem:2012yn}
\begin{align}
\label{eq:currents}
-\left(1+\chi_\mathbf{8}(x)+\chi_\mathbf{8}(y)+\chi_\mathbf{8}(z)\right) pq \,,
\end{align}
which shows that the global symmetry of the theory is
\begin{align}
SU(3)_x \times SU(3)_y \times SU(3)_z \times U(1)_t \,,
\end{align}
which agrees with the manifest symmetry of the theory up to the $U(1)_c$ factor. The fact that the expanded index is independent of $c$ and that all the anomalies involving $U(1)_c$ vanish signals that this is not a faithful symmetry in the IR.
From the expansion \eqref{eq:(1,1)} we see that there are 108 marginal operators, whose contribution to the index is given by
\begin{align}
\label{eq:marginal operators}
\left(\chi_\mathbf{6}(x) \chi_\mathbf{\overline{3}}(y) \chi_\mathbf{\overline{3}}(z)+\chi_\mathbf{3}(x) \chi_\mathbf{3}(y) \chi_\mathbf{\overline{6}}(z)\right) pq \,.
\end{align}

In order to read the relations among the multiplets counted by the index, it is convenient to evaluate its Plethystic Logarithm (PL) \cite{Benvenuti:2006qr}\footnote{The PL is defined as
\begin{align}
\mathrm{PL}[f(x)] = \sum_{n = 1}^\infty \frac{\mu(n)}{n} \log(f(x^n)) \,,
\end{align}
where $\mu(n)$ is the Möbius function
\begin{align}
\mu(n) = \left\{\begin{array}{ll}
0 \,, & \qquad \text{$n$ has repeated prime factors,} \\
1 \,, & \qquad n = 1, \\
(-1)^k \,, & \qquad \text{$n$ is a product of $k$ distinct primes.}
\end{array}\right.
\end{align}
The PL is the inverse function of the Plethystic Exponential (PE)
\begin{align}
\mathrm{PE}[g(x)] = \exp\left[\sum_{n = 1}^\infty \frac1n g(x^n)\right] \,.
\end{align}}, which captures the single trace operators and their relations. For $M = N = 1$, the PL of the index is given by
\begin{align}
\label{eq:PL}
\mathrm{PL}\left[\mathcal I_A^{(1,1)}\right] &= \left(\chi_\mathbf{3}(x) \chi_\mathbf{\overline{3}}(y)+\chi_\mathbf{3}(y) \chi_\mathbf{\overline{3}}(z)\right) t^{-\frac13} (pq)^\frac{16}{33}+\chi_\mathbf{3}(x) \chi_\mathbf{\overline{3}}(z) \, t^{\frac13} (pq)^\frac{17}{33}+t (pq)^\frac{18}{33} \nonumber \\
&\quad -\left(\chi_\mathbf{\overline{3}}(x) \chi_\mathbf{3}(y)+\chi_\mathbf{\overline{3}}(y) \chi_\mathbf{3}(z)+\chi_\mathbf{3}(x) \chi_\mathbf{\overline{3}}(z)\right) t^{-\frac23} (pq)^{\frac{32}{33}} \nonumber \\
&\quad -\left(1+\chi_\mathbf{8}(x)+\chi_\mathbf{8}(y)+\chi_\mathbf{8}(z)+\chi_\mathbf{3}(x) \chi_\mathbf{3}(y) \chi_\mathbf{3}(z)+\chi_\mathbf{\overline{3}}(x) \chi_\mathbf{\overline{3}}(y) \chi_\mathbf{\overline{3}}(z)\right) pq+\dots \,.
\end{align}
In particular, the last line of \eqref{eq:PL} is the contribution of the fermionic single trace operators belonging to the conserved current multiplet as well as the relations of the marginal operators. More precisely, we have seen that the first four terms come from the current multiplet, see \eqref{eq:currents}, and so the other two correspond to the relations of the marginal operators.

Indeed, if we look at the first line of \eqref{eq:(1,1)}, we find  the contributions of the  following operators
\begin{gather}
\text{Tr}_{M=1} \left[D^{(1)} V^{(1)}\right] , \qquad \text{Tr}_{N=1} \left[D^{(2)} V^{(2)}\right] , \label{eq:op1} \\
\text{Tr}_{M=1} \text{Tr}_{N=1} \left[D^{(1)} Q V^{(2)}\right] , \label{eq:op2} \\
\mathrm{Tr}_{M=1} \mathrm{Tr}_{N=1} \left[QQ\right] \label{eq:op3}
\end{gather}
in the $SU(3)_x\times SU(3)_y\times SU(3)_z \times U(1)_t$ representations
\begin{gather}
\left(\mathbf{3},\mathbf{\overline{3}},\mathbf{1}\right)_{-\frac13} \oplus \left(\mathbf{1},\mathbf{3},\mathbf{\overline{3}}\right)_{-\frac13} \,, \\
\left(\mathbf{3},\mathbf{1},\mathbf{\overline{3}}\right)_{\frac13} \,, \\
\left(\mathbf{1},\mathbf{1},\mathbf{1}\right)_{1} \,,
\end{gather}
respectively. 
We can then take the  product of the operators in the first two lines to obtain 
 marginal operators uncharged under the abelian symmetry in  the representation
\begin{align}\label{wouldbemarginal}
\left(\mathbf{6},\mathbf{\overline{3}},\mathbf{\overline{3}}\right)_0\oplus\left(\mathbf{3},\mathbf{3},\mathbf{\overline{6}}\right)_0\oplus\left(\mathbf{3},\mathbf{3},\mathbf{3}\right)_0\oplus\left(\mathbf{\overline{3}},\mathbf{\overline{3}},\mathbf{\overline{3}}\right)_0 \,.
\end{align}
However, from \eqref{eq:marginal operators}, we see that only the first two representations appear in the index, which implies the latter two are truncated since the marginal operators satisfy the relations corresponding to those representations. These relations should be encoded in the following negative contributions in the index
\begin{align}
-\left(\chi_\mathbf{3}(x) \chi_\mathbf{3}(y) \chi_\mathbf{3}(z)+\chi_\mathbf{\overline{3}}(x) \chi_\mathbf{\overline{3}}(y) \chi_\mathbf{\overline{3}}(z)\right) pq\,,
\end{align}
which indeed cancel those of the last two operators in \eqref{wouldbemarginal}. These precisely coincide with the last two terms of the PL of the index \eqref{eq:PL}.

As argued in \cite{Razamat:2018gbu,Hwang:2020ddr}, one can remove the marginal operators of the theory by flipping some of their ingredients, which are \eqref{eq:op1} and \eqref{eq:op2} in our case. Once the marginal operators are removed in this way, the flip fields also provide extra conserved currents having the same representation as the relations of the original marginal operators and therefore lead to the enhancement of the global symmetry. 
Here we introduce $\pi^{(I,II)}$ and $\pi^{(II,III)}$ flipping $\text{Tr}_{M=1} \left[D^{(1)} V^{(1)}\right]$ and $\text{Tr}_{N=1} \left[D^{(2)} V^{(2)}\right]$ in \eqref{eq:op1} respectively to remove the marginal operators. Then we also find that $\text{Tr}_{M=1} \text{Tr}_{N=1} \left[Q Q\right]$ in \eqref{eq:op3} hits the unitarity bound and becomes free. Once we flip all those operators in \eqref{eq:op1} and \eqref{eq:op3}, we obtained the {\it flipped} theory 
with index:
\begin{align}
\mathcal I_{flipped}^{(1,1)} &= 1+\chi^{E_6}_\mathbf{27} \, t^{\frac13} (pq)^\frac{4}{9} (1+p+q)+t^{-1} (pq)^\frac23+\chi^{E_6}_\mathbf{351} \, t^\frac23 (pq)^\frac89-\left(1+\chi^{E_6}_\mathbf{78}\right) pq+\dots \,,
\end{align}
where we have used a new mixing coefficient which is exactly (that is with no approximation) the one coming from $a$-maximization
\begin{align}
\mathfrak t = \frac23 \,.
\end{align}
As before, $U(1)_c$ doesn't contribute to the $a$-function because it has the vanishing anomalies.
Now all the terms are neatly organized into the characters $\chi^{E_6}_\mathbf{n}$ of the $E_6$ group. For example, the second term is contributed by the operators
\begin{align}
\label{eq:operators}
\pi^{(I,II)} \,, \qquad \pi^{(II,III)} \,, \qquad \text{Tr}_{M=1} \text{Tr}_{N=1} \left[D^{(1)} Q V^{(2)}\right] ,
\end{align}
which are in the representation
\begin{align}
\left(\mathbf{\overline 3},\mathbf{3},\mathbf{1}\right)_{\frac13} \oplus \left(\mathbf{1},\mathbf{\overline{3}},\mathbf{3}\right)_{\frac13} \oplus \left(\mathbf{3},\mathbf{1},\mathbf{\overline{3}}\right)_{\frac13} \quad \longrightarrow \quad \mathbf{27}_\frac13 \,,
\end{align}
of $SU(3)_x \times SU(3)_y \times SU(3)_z \times U(1)_t \subset E_6 \times U(1)_t$. Moreover, we find the contribution of 79 conserved currents in the adjoint representation of $E_6 \times U(1)_t$. This proves that the flipped theory exhibits the global symmetry enhancement
\begin{align}
SU(3)_x \times SU(3)_y \times SU(3)_z \times U(1)_t \qquad \longrightarrow \qquad E_6 \times U(1)_t \,.
\end{align}

Notice that, as expected, the flipped theory is exactly self-dual under the cross-leg duality
and we  have a non-trivial duality map of the operators; e.g.~in \eqref{eq:operators} the operators $\pi^{(I,II)}$ and $\text{Tr}_{M=1} \text{Tr}_{N=1} \left[D^{(1)} Q V^{(2)}\right]$ are exchanged under the duality
\begin{align}
\pi^{(I,II)} \quad \longleftrightarrow \quad \text{Tr}_{M=1} \text{Tr}_{N=1} \left[D^{(1)} Q V^{(2)}\right] ,
\end{align}
while $\pi^{(II,III)}$ is mapped to itself.

The next example we consider is $(M,N) = (2,1)$. In this case, the $a$-maximization gives the mixing coefficient
\begin{align}
\mathfrak t = \frac19 \left(23-\sqrt{73}\right) \approx 1.6 \approx \frac85 \,.
\end{align}
The index of the original theory is given by
\begin{align}
\label{eq:(2,1)}
\mathcal I_A^{(2,1)} &= 1+\chi_\mathbf{3}(x) \chi_\mathbf{\overline{3}}(y) t^{\frac13} (pq)^\frac{4}{15}+\left(\chi_\mathbf{3}(x) \chi_\mathbf{\overline{3}}(y)+\chi_\mathbf{3}(y) \chi_\mathbf{\overline{3}}(z)\right) t^{-\frac23} (pq)^\frac{7}{15} \nonumber \\
&\quad+54 \, t^\frac23 (pq)^\frac{8}{15}+t^2 (pq)^\frac35+\dots \,,
\end{align}
where the terms higher than $(pq)^\frac12$ are evaluated without the refinement for the $SU(3)^3$ fugacities for computational simplicity.
The second and third terms are the contributions of the following mesonic operators:
\begin{align}
\text{Tr}_2 \left[D^{(1)} V^{(1)}\right] \,, \qquad \text{Tr}_2 \left[D^{(1)} A^{(1)} V^{(1)}\right] \,, \qquad \text{Tr}_1 \left[D^{(2)} V^{(2)}\right] ,
\end{align}
which should be flipped to obtain the self-dual  theory where the marginal operators are removed leading to the potential enhancement of the global symmetry. In addition, we find that once those operators are flipped, $\beta^{(2)}_2$ corresponding to $t^2 (pq)^\frac35$ hits the unitarity bound. Thus, we also flip this operator and obtain the index of the flipped theory as follows:
\begin{align}
\mathcal I_{flipped}^{(2,1)} &= 1+\chi^{E_6}_\mathbf{27} \, t^{\frac23} (pq)^\frac{4}{9} (1+p+q)+t^{-2} (pq)^\frac23+351 \, t^\frac43 (pq)^\frac89-\left(1+78\right) pq+\dots \,,
\end{align}
with the exact mixing coefficient
\begin{align}
\mathfrak t = \frac43 \,.
\end{align}
Again the terms higher than $(pq)^\frac12$ are evaluated without the $SU(3)^3$ fugacities for computational simplicity. The last term of the index is consistent with the enhanced IR global symmetry
\begin{align}
E_6 \times U(1)_t \,.
\end{align}

Lastly, let us consider the example $(M,N) = (2,2)$. The $a$-maximization gives the mixing coefficient
\begin{align}
\mathfrak t = \frac94-\frac{\sqrt{73}}{12} \approx 1.5 \approx \frac{17}{11} \,.
\end{align}
The index of the original theory is given by
\begin{align}
\label{eq:(2,1)}
\mathcal I_A^{(2,2)} &= 1+18 \, (pq)^\frac13+9 \, t (pq)^\frac{29}{66}+2 \, t^2 (pq)^\frac{6}{11}+18 \, t^{-1} (pq)^\frac{37}{66}+180 \, (pq)^\frac23+\dots \,,
\end{align}
where all the $SU(3)^3$ global symmetry fugacities are turned off for computational simplicity. Note that the product of the second and sixth terms and that of the third and fifth terms give rise to the marginal operators. In particular, the second and fifth terms are the contributions of the following mesonic operators:
\begin{align}
\label{eq:flipped operators}
\text{Tr}_2 \left[D^{(1)} V^{(1)}\right] \,, \qquad \text{Tr}_2 \left[D^{(1)} A^{(1)} V^{(1)}\right] \,, \qquad \text{Tr}_1 \left[D^{(2)} V^{(2)}\right] \,, \qquad \text{Tr}_1 \left[D^{(2)} A^{(2)} V^{(2)}\right] .
\end{align}
The self-dual theory can be obtained by flipping those operators. No other operators are necessarily flipped because, unlike the previous examples, there is no decoupled operator when the operators given in \eqref{eq:flipped operators} are flipped. Nevertheless, we have found that it is more convenient to flip $\beta_2^{(i)}$ for $i=1,2$ to expand the index because in that case the mixing coefficient of $U(1)_t$ has a rational value, and we can expand the index with the exact powers without any approximation. Once the operators in \eqref{eq:flipped operators} and $\beta_2^{(i)}$ are flipped, the index of the self-dual theory is given by
\begin{align}
\mathcal I_{flipped}^{(2,2)} &= 1+27 \, t (pq)^\frac25 (1+p+q)+(1+1) \, t^{-2} (pq)^\frac{8}{15}+ 27 \, (pq)^\frac23+t r^\frac{11}{15}+(351+27) \, t^2 (pq)^\frac45 \nonumber \\
&\quad +27 \, t^{-1} (pq)^\frac{14}{15}-(78+1+1+1) pq+\dots
\end{align}
with the exact mixing coefficient
\begin{align}
\mathfrak t = \frac{22}{15} \,.
\end{align}
As before all the $SU(3)^3$ global symmetry fugacities are omitted for computational simplicity. We observe that the coefficients of the expanded index naturally fit the dimensions of representations of $E_6$. Naively, there are 81 conserved currents, which are expected to be in the adjoint representation of
\begin{align}
E_6 \times U(1)_t \times G \,,
\end{align}
with an emergent symmetry group $G$ of dimension 2 or larger. Let us first assume $G$ doesn't include any abelian factor. The simplest candidate of $G$ is then $SU(2)_a$. In that case, we expect the term $-81 \, pq$ will be refined as follows
\begin{align}
\left(1-\chi^{E_6}_\mathbf{78}-1-\chi^{SU(2)_a}_\mathbf{3}\right) pq \,.
\end{align}
The first positive term indicates that there is one marginal operator, while the rest negative terms are contributed by the conserved current of the enhanced symmetry $E_6 \times U(1)_t \times SU(2)_a$. Note that the non-abelian emergent symmetry $G = SU(2)_a$ doesn't affect the IR superconformal R-symmetry.

On the other hand, if $G$ includes an abelian factor $U(1)_a$, in principle, we cannot completely trust the term $-81 \, pq$ because some of the contributions could be charged under $U(1)_a$, and their powers are shifted if $U(1)_a$ mixes with the R-symmetry. In that case, we have at least one emergent $U(1)$ symmetry, but the entire symmetry is not clear.

\begin{figure}[tbp]
\centering
\includegraphics[width=.45\textwidth]{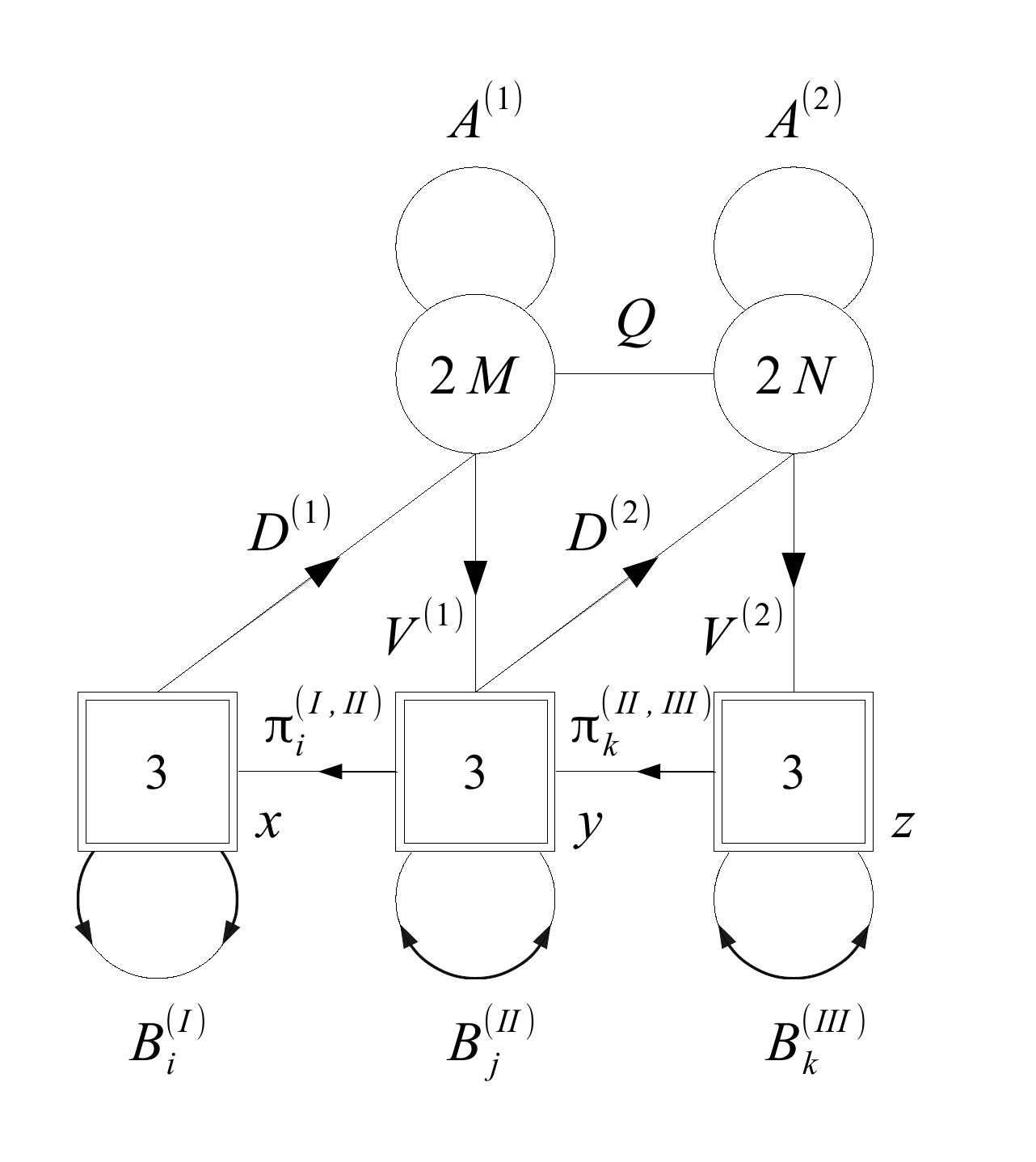}
\caption{\label{fig:self-dual} The quiver representation of the self-dual theory, which we propose to have $SU(3)^3\to E_6$ enhancement. Two sets of singlets $\pi^{(I,II)}_i$ and $\pi^{(I,III)}_k$ for $i = 1,\dots,M$ and $k = 1,\dots,N$ have been added to Theory A.}
\end{figure}

We have also other evidence that the $E_6$  enhancement can appear for generic $(M,N)$.
Indeed  for arbitrary $(M,N)$  we can obtain a self-dual theory by flipping the following operators:
\begin{align}
\text{Tr}_M \left[D^{(1)} (A^{(1)})^{i-1} V^{(1)}\right] , &\qquad i = 1, \dots, M \,, \\
\text{Tr}_N \left[D^{(2)} (A^{(2)})^{k-1} V^{(2)}\right] , &\qquad k = 1, \dots, N \,.
\end{align}
The resulting theory is represented by a quiver diagram in Figure \ref{fig:self-dual}. We expect this to have an $E_6$ enhanced symmetry for any $(M,N)$.

We can provide an additional piece of evidence for our claim that the theory in Figure \ref{fig:self-dual} enjoys an $E_6$ global symmetry in the IR. The following discussion holds for generic ranks $(M,N)$ and avoids any tedious index computation. The idea is that there is a necessary condition on the anomalies of the $SU(3)_x\times SU(3)_y\times SU(3)_z$ global symmetries that these should satisfy in order for the $E_6$ enhancement to be possible.

Suppose that we have a $4d$ theory with global symmetry $G\times U(1)$ and consider the subgroup $H\times\tilde{H}\subset G$. The anomalies of $H$ and $\tilde{H}$ can be uniquely determined in terms of those of $G$ as follows. Let us assume for simplicity that $G$, $H$ and $\tilde{H}$ are non-abelian, as in the case of interest for us, so that the only non-trivial anomalies are those with the $U(1)$. Then we have
\be\label{anomemb}
\Tr\,H^2\,U(1)=I(H\hookrightarrow G)\,\Tr\,G^2\,U(1), \quad \Tr\,\tilde{H}^2\,U(1)=I(\tilde{H}\hookrightarrow G)\,\Tr\,G^2\,U(1)\,,
\ee
where $I(H\hookrightarrow G)$ and $I(\tilde{H}\hookrightarrow G)$ are the \emph{embedding indices} of $H$ and $\tilde{H}$ in $G$. If a representation $\bf r$ of $G$ decomposes into $\oplus_i {\bf r}_i$ under $H$, then the embedding index of $H$ in $G$ is defined as
\be
I_{H \hookrightarrow G} = \frac{\sum_i T({\bf r}_i)}{T({\bf r})}\,,
\ee
where $T({\bf r})$ denotes the Dynkin index.

Conversely, if we have a $4d$ theory with manifest global symmetry $H\times\tilde{H}\times U(1)$, a necessary condition for the $H\times\tilde{H}$ part to enhance in the IR to $G$ is that the relations \eqref{anomemb} hold true. Since it is not possible to compute the anomalies for the IR symmetry $G$ from the UV description, the condition that should be checked is
\be\label{condemb}
\frac{\Tr\,H^2\,U(1)}{I(H\hookrightarrow G)}=\frac{\Tr\,\tilde{H}^2\,U(1)}{I(\tilde{H}\hookrightarrow G)}\,.
\ee

In our case we have the decomposition $SU(3)_x\times SU(3)_y\times SU(3)_z\subset E_6$ with the embedding 
\begin{align}
\mathbf{27}\to\left(\mathbf{\overline 3},\mathbf{3},\mathbf{1}\right) \oplus \left(\mathbf{1},\mathbf{\overline{3}},\mathbf{3}\right)\oplus \left(\mathbf{3},\mathbf{1},\mathbf{\overline{3}}\right) \,.
\end{align}
We then find the embedding indices\footnote{We use that the Dynkin index of the (anti-)fundamental representation of $SU(3)$ is $\frac{1}{2}$ and that of the $\bf 27$ of $E_6$ is $3$.}
\be
I(SU(3)_x\hookrightarrow E_6)=I(SU(3)_y\hookrightarrow E_6)=I(SU(3)_z\hookrightarrow E_6)=\frac{3\times\frac{1}{2}+3\times\frac{1}{2}}{3}=1\,.
\ee
In particular what is relevant for us is that the embedding indices for the three $SU(3)_{x,y,z}$ subgroups are equal. Hence, a necessary condition for the $E_6$ enhancement is that the mixed anomalies with all the abelian symmetries are equal for the different $SU(3)_{x,y,z}$ groups, which we can easily check
\be
&&\Tr\,SU(3)_x^2\,U(1)_t=\Tr\,SU(3)_y^2\,U(1)_t=\Tr\,SU(3)_z^2\,U(1)_t=\frac{MN}{2},\nn\\
&&\Tr\,SU(3)_x^2\,U(1)_{R_0}=\Tr\,SU(3)_y^2\,U(1)_{R_0}=\Tr\,SU(3)_z^2\,U(1)_{R_0}=-MN\,.
\ee
From this we can also deduce the anomalies for the enhanced $E_6$ global symmetry
\be
\Tr\,\left(E_6\right)^2\,U(1)_t=\frac{MN}{2},\qquad\Tr\,\left(E_6\right)^2\,U(1)_{R_0}=-MN\,.
\ee

Finally we can try to look at how the action of the  Weyl of the conjectured $E_6$ IR  global symmetry group is realised. For this it is useful to look at 
 index identity corresponding to the  self-duality:
\begin{align}
\mathcal I_\text{self-dual}(\vec x, \vec y, \vec z;c,t) = \mathcal I_\text{self-dual}(\vec x \, {}^{-1}, \vec z\, {}^{-1}, \vec y \, {}^{-1};c^{-1},t)
\end{align}
where
\begin{align}
&\mathcal I_\text{self-dual}(\vec x, \vec y, \vec z;c,t) = \prod_{i = 1}^M \prod_{\alpha = 1}^3 \prod_{\beta = 1}^3 \Gpq{(pq)^{\frac23 M-\frac13 N-i+1} t^{-\frac23 M+\frac13 N+i-\frac13} x_\alpha^{-1} y_\beta} \nonumber \\
& \times \prod_{k = 1}^N \prod_{\beta = 1}^3 \prod_{\gamma = 1}^3 \Gpq{(pq)^{-\frac13 M+\frac23 N-k+1} t^{\frac13 M-\frac23 N+k-\frac13} y_\beta^{-1} z_\gamma} \times \mathcal I_A(\vec x, \vec y, \vec z;c,t) \,.
\end{align}
The identity shows that the self-duality implies an emergent $\mathbb Z_2$ symmetry in the IR, acting on the the global symmetry fugacities as follows:
\begin{align}
\label{eq:Z2}
x_\alpha \quad &\longrightarrow \quad x^{-1}_\alpha \,, \nn\\
y_\alpha \quad &\longrightarrow \quad z^{-1}_\alpha \,, \nn\\
z_\alpha \quad &\longrightarrow \quad y^{-1}_\alpha \,,\nn \\
c \quad &\longrightarrow \quad c^{-1} \,, \nn\\
t \quad &\longrightarrow \quad t \,,
\end{align}
Namely, it consists of the permutation of $SU(3)_y$ and $SU(3)_z$ and the charge conjugation on $SU(3)_x \times SU(3)_y \times SU(3)_z \times U(1)_c$\footnote{The charge conjugation on $U(1)_c$ is trivial because it is not a faithful symmetry.}.

The $\mathbb Z_2$ symmetry \eqref{eq:Z2} turns out to be part of the Weyl group of the enhanced $E_6$ symmetry of the theory in the IR. To show this, we first recall that the root system of $E_6$ can be represented as vectors in a six-dimensional hyperplane in the Euclidean space $\mathbb R^9$ orthogonal to the following three vectors:
\begin{align}
\label{eq:vectors}
e_1+e_2+e_3 \,, \quad e_4+e_5+e_6 \,, \quad e_7+e_8+e_9 \,, 
\end{align}
where a set of nine $e_i$ is an orthonormal basis of $\mathbb R^9$. The $E_6$ roots are then given by
\begin{align}
(1,-1,0;0,0,0;0,0,0) \,,  &\quad \text{and permutations of $(1,-1,0)$} \,, \\
(0,0,0;1,-1,0;0,0,0) \,,  &\quad \text{and permutations of $(1,-1,0)$} \,, \\
(0,0,0;0,0,0;1,-1,0) \,,  &\quad \text{and permutations of $(1,-1,0)$} \,, \\
\frac 13 \left(-2,1,1;-2,1,1;-2,1,1\right) , &\quad \text{and permutations among each $\left(-2,1,1\right)$} \,, \\
\frac13 \left(2,-1,-1;2,-1,-1;2,-1,-1\right) , &\quad \text{and permutations among each $\left(2,-1,-1\right)$} \,,
\end{align}
where the first three lines give the roots of $SU(3)^3 \subset E_6$.
The Weyl group of $E_6$ is generated by the reflections $s_\alpha$ with respect to the root vectors $\alpha$ as follows:
\begin{align}
s_\alpha(v) = v-2 \frac{(v,\alpha)}{(\alpha,\alpha)} \alpha \,, \qquad v \in \mathbb R^9
\end{align}
where $(\cdot,\cdot)$ is the inner product on $\mathbb R^9$. For example, the reflection for $\alpha = (1,-1,0;0,0,0;0,0,0)$ gives the following linear transformation
\begin{align}
\left(\begin{array}{ccccccccc}
0 & 1 & 0 & 0 & 0 & 0 & 0 & 0 & 0 \\
1 & 0 & 0 & 0 & 0 & 0 & 0 & 0 & 0 \\
0 & 0 & 1 & 0 & 0 & 0 & 0 & 0 & 0 \\
0 & 0 & 0 & 1 & 0 & 0 & 0 & 0 & 0 \\
0 & 0 & 0 & 0 & 1 & 0 & 0 & 0 & 0 \\
0 & 0 & 0 & 0 & 0 & 1 & 0 & 0 & 0 \\
0 & 0 & 0 & 0 & 0 & 0 & 1 & 0 & 0 \\
0 & 0 & 0 & 0 & 0 & 0 & 0 & 1 & 0 \\
0 & 0 & 0 & 0 & 0 & 0 & 0 & 0 & 1
\end{array}\right)
\end{align}
acting on $\mathbb R^9$.

We have found that the cross-leg duality corresponds to the linear transformation
\begin{align}
\label{eq:cross-leg transformation}
\frac13 \left(\begin{array}{ccccccccc}
-1 & 2 & 2 & 0 & 0 & 0 & 0 & 0 & 0 \\
2 & -1 & 2 & 0 & 0 & 0 & 0 & 0 & 0 \\
2 & 2 & -1 & 0 & 0 & 0 & 0 & 0 & 0 \\
0 & 0 & 0 & 1 & 1 & 1 & -2 & 1 & 1 \\
0 & 0 & 0 & 1 & 1 & 1 & 1 & -2 & 1 \\
0 & 0 & 0 & 1 & 1 & 1 & 1 & 1 & -2 \\
0 & 0 & 0 & -2 & 1 & 1 & 1 & 1 & 1 \\
0 & 0 & 0 & 1 & -2 & 1 & 1 & 1 & 1 \\
0 & 0 & 0 & 1 & 1 & -2 & 1 & 1 & 1
\end{array}\right) ,
\end{align}
which is the combination of the reflections with respect to the following four root vectors:
\begin{gather}
(1,-1,0;0,0,0;0,0,0) \,, \\
\frac13 (1,1,-2;1,1,-2,1,1,-2) \,, \\
\frac13 (1,1,-2;1,-2,1,1,-2,1) \,, \\
\frac13 (1,1,-2;-2,1,1,-2,1,1) \,.
\end{gather}
Notice that the transformation \eqref{eq:cross-leg transformation} maps a generic point in the six-dimensional hyperplane in $\mathbb R^9$ orthogonal to the vectors in \eqref{eq:vectors}
\begin{align}
(x_1,x_2,-x_1-x_2;y_1,y_2,-y_1-y_2;z_1,z_2,-z_1-z_2)
\end{align}
to
\begin{align}
(-x_1,-x_2,x_1+x_2;-z_1,-z_2,z_1+z_2;-y_1,-y_2,y_1+y_2) \,,
\end{align}
which is the permutation of $SU(3)_y$ and $SU(3)_z$ with the charge conjugation of all three $SU(3)$. This is exactly how the cross-leg duality acts on the $SU(3)_x \times SU(3)_y \times SU(3)_z$ symmetry, as shown in \eqref{eq:Z2}. Therefore, the emergent $\mathbb Z_2$ symmetry due to the cross-leg duality corresponds to the Weyl element \eqref{eq:cross-leg transformation} of the enhanced $E_6$ symmetry of the self-dual theory. Note that this is true for generic ranks $M$ and $N$, since the cross-leg duality holds for any rank.

Indeed, one can find four more cross-leg dual frames permuting three $SU(3)$ in different ways, which are also part of the $E_6$ Weyl group. First recall that we applied the dualization procedure starting from the $SU(3)_x$ flavor node of the original quiver in Figure \ref{fig:TheoryA}. This gives the dual frame in Figure \ref{fig:TheoryB}, where $SU(3)_y$ and $SU(3)_z$ are swapped and all three $SU(3)$ charges are conjugated, which can be represented by the fugacity map: $(x,y,z) \rightarrow (x^{-1},z^{-1},y^{-1})$. One can repeat the procedure, now starting from $SU(3)_y$, the left most node of the dual quiver in Figure \ref{fig:TheoryB}, which leads to another dual frame with the map: $(x,y,z) \rightarrow (z,x,y)$. One can keep repeating this procedure until going back to the original frame. The sequence of the duality frames we obtain in this way are associated with the following fugacity maps:
\begin{align}
(x,y,z) &\rightarrow (x,y,z) \, \\
(x,y,z) &\rightarrow (x^{-1},z^{-1},y^{-1}) \, \\
(x,y,z) &\rightarrow (z,x,y) \, \\
(x,y,z) &\rightarrow (z^{-1},y^{-1},x^{-1}) \, \\
(x,y,z) &\rightarrow (y,z,x) \, \\
(x,y,z) &\rightarrow (y^{-1},x^{-1},z^{-1}) \,,
\end{align}
respectively. Thus, we can identify $1296 = 216 \times 6$ elements of the $E_6$ Weyl group, 216 from the Weyl group of the manifest $SU(3)_x \times SU(3)_y \times SU(3)_z$ symmetry and 6 from the cross-leg dual frames above. On the other hand, the other elements of the $E_6$ Weyl group do not preserve $SU(3)_x \times SU(3)_y \times SU(3)_z$, which may signal the existence of  new dual frames breaking the original $SU(3)_x \times SU(3)_y \times SU(3)_z$ symmetry.  We leave the investigation of these extra dual frames for the future.

\section{Rank stabilization duality}
\label{sec:rankstab}

\subsection{The duality}

\begin{figure}[t]
	\centering
  	\includegraphics[width=.9\textwidth]{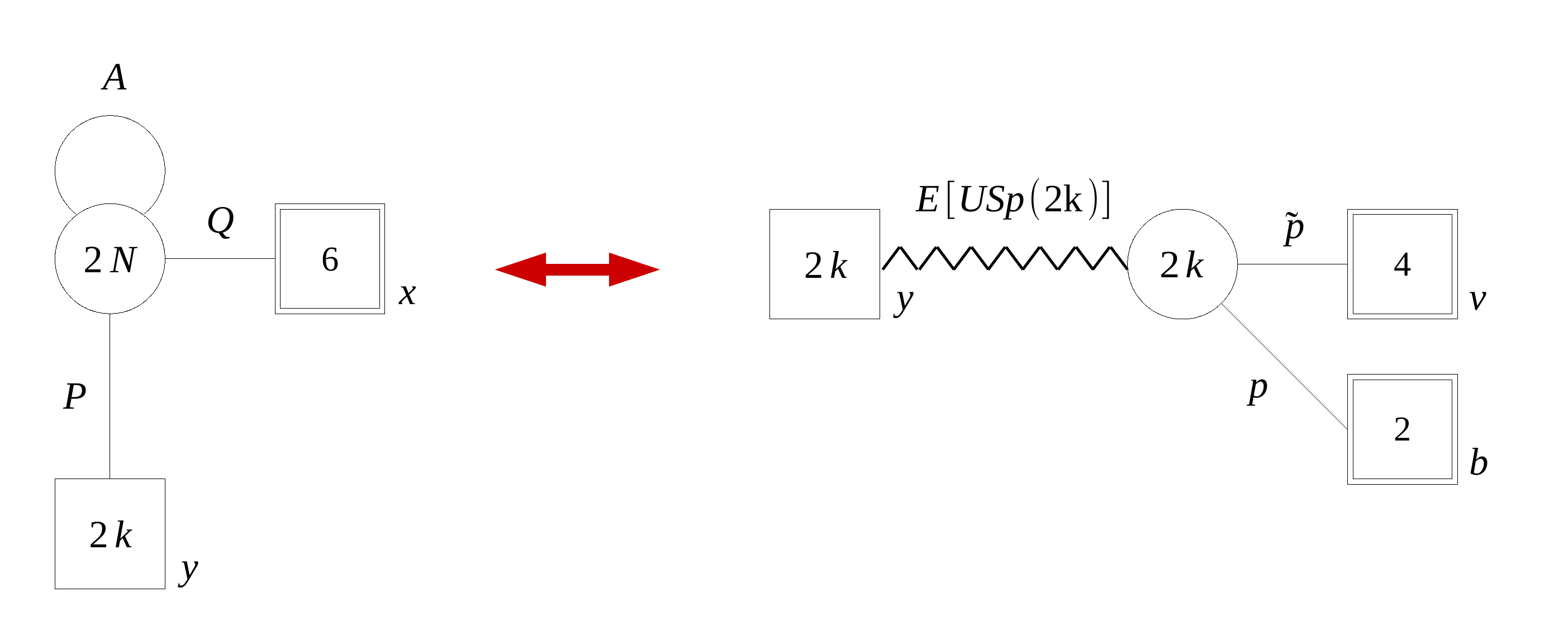} 
	 \caption{Schematic representation of the $4d$ rank stabilization duality. We don't draw singlets, which are specified in the main text.}
        \label{fig:4drankstab}
\end{figure}

In this section, we present a generalization of the confining duality for the $USp(2N)$ gauge theory in Section \ref{sec:confining} to a higher number of flavors. More specifically, we introduce additional fundamental chirals interacting with the antisymmetric via a cubic superpotential. Therefore, we now have a $USp(2N)$ gauge theory with one antisymmetric and $2k+6$ fundamental chirals of which $2k$ form a cubic superpotential with the antisymmetric, plus some gauge singlet flipping fields. We will show that this theory is dual to the $E[USp(2k)]$ theory \cite{Pasquetti:2019hxf} whose manifest $USp(2k)$ symmetry is gauged with 6 fundamental chirals, plus some gauge singlet fields. This duality is the $4d$ version of the rank stabilization duality that was proposed in $3d$ in \cite{Pasquetti:2019tix}. The reason for the name ``rank stabilization" is that the rank $N$ of the original theory doesn't appear anymore as the rank in the dual, but it only appears in the number of the singlet fields and in the charges of the various fields under the abelian symmetries. The duality is schematically depicted in Figure \ref{fig:4drankstab}.

The first theory, which we will refer to as Theory A, is a $USp(2N)$ gauge theory with one antisymmetric chiral $A$, $2k+6$ fundamental chirals $P_n$ for $n=1,\cdots,2k$ and $Q_a$ for $a=1,\dots, 6$, and $N-k$ chiral singlets $\gb_i$. The superpotential of the theory is
\be
\mathcal{W}_A=\sum_{n=1}^{2k}\Tr_N AP_n^2+\sum_{i=1}^{N-k}\gb_j\Tr_N A^j\,.
\label{eq:superpotrankstabA}
\ee
Because of the first superpotential term, the flavor symmetry rotating the chirals $P_n$ is $USp(2k)$. 
The chirals $Q_a$, instead, don't enter in the superpotential and are rotated by an $SU(6)$ global symmetry, as in the $k=0$ case. Finally, there is one non-anomalous abelian symmetry preserved by the superpotential. The full manifest global symmetry of the model is thus
\be
\label{eq:F_A}
F_A=USp(2k)_y\times SU(6)_x\times U(1)_t\,.
\ee
The transformation rules of the matter fields under these symmetries and a possible choice of non-anomalous $R$-symmetry are summarized in Table \ref{tab:rankstabchargesTA}.

\begin{table}[h]
\centering
\scalebox{1}{\setlength{\extrarowheight}{2pt}
\begin{tabular}{|c|cccc|}\hline
{} & $USp(2k)_y$ & $SU(6)_x$ & $U(1)_t$  & $U(1)_{R_0}$\\ \hline
$A$ & $\bf 1$ & $\bf 1$ & $1$ & 0 \\
$P$ & $\bf 2k$ & $\bf 1$ & $-\frac{1}{2}$ & 1 \\
$Q$ & $\bf 1$ & $\bf 6$ & $\frac{k+2-2N}{6}$ & $\frac{1}{3}$ \\\hline
$\gb_i$ & $\bf 1$ & $\bf 1$ & $-i$ & 2\\\hline
\end{tabular}}
\caption{Transformation properties under the global symmetry $F_A$ of the matter fields of Theory A. The horizontal line separates fields charged under the gauge symmetry from the singlets.}
\label{tab:rankstabchargesTA}
\end{table}

The dual theory, which we will refer to as Theory B, is constructed by gauging the manifest $USp(2k)$ symmetry\footnote{Recall that the $E[USp(2 k)]$ theory has two $USp(2 k)$ global symmetries in the IR, only one of which is manifest in the UV. On the other hand, the other is an emergent one enhanced from the $SU(2)^k$ symmetry in the UV. See Appendix \ref{app:eusp}.} of $E[USp(2k)]$ with the addition of 6 chirals in the fundamental representation of such gauge group $p_\ga$ for $\ga=1,2$ and $\tilde{p}_a$ for $a=1,\cdots,4$. In the theory there are also $15N-5k+1$ chiral singlets that we are going to denote as follows:
\be
&&\mu_{ab;i},\qquad a<b=1,\cdots,4,\, i=1,\cdots,N\nn\\
&&\mu_{56;i}\qquad i=1,\cdots,N-k\nn\\
&&\mu_{a\ga;i},\qquad a=1,\cdots,4,\,\ga=1,2,\,i=1,\cdots,N-k\nn\\
&&\nu_{\ga n},\qquad \ga=1,2,\,n=1,\cdots,2k\nn\\
&&b_k\,.
\ee
The reason for these names is because in order to write the superpotential in a compact form it is useful to collect some of them together with some gauge invariant operators that we can construct from the gauge charged matter fields. First, we define the set of operators $\hat{\mu}_{ab;i}=-\hat{\mu}_{ba;i}$ for $a<b=1,\cdots,6$ and $i=1,\cdots,N$ as follows:
\be
&&\hat{\mu}_{ab;i}=\mu_{ab;i},\qquad a<b=1,\cdots,4,\, i=1,\cdots,N\nn\\
&&\hat{\mu}_{56;i}=\begin{cases}\mu_{56;i},\quad i=1,\cdots,N-k \\ b_{N-i+1},\quad i=N-k+1,\cdots,N\end{cases}\nn\\
&&\hat{\mu}_{a,\ga+4;i}=\begin{cases}\mu_{a\ga;i},\quad i=1,\cdots,N-k \\ \Tr_k\left[\mathsf{H}^{k+i-N-1}\tilde{p}_ap_\ga\right] ,\quad i=N-k+1,\cdots,N\end{cases}\quad a=1,\cdots,4,\,\ga=1,2\,,
\label{eq:hatmuop}\nn\\
\ee
where $b_n$ and $\mathsf{H}$ are some of the gauge invariant operators of the $E[USp(2k)]$ theory, with the former ones being singlets under the non-abelian symmetries while the latter being in the antisymmetric representation of the gauged $USp(2k)$ symmetry (see Appendix \ref{app:eusp}). The superpotential of Theory B can then be written in the following compact form\footnote{It is interesting to notice that the object $\hat{\mu}_{ab;i}$ is analogous to $\mu_{ab;i}$ that we had in the $k=0$ case in Section \ref{sec:confining}. The difference is that now the dual is not a WZ model but a gauge theory and accordingly some of the singlets in $\mu_{ab;i}$ are replaced by gauge invariant operators in $\hat{\mu}_{ab;i}$.}
\be
\mathcal{W}_B&=&\mathcal{W}_{E[USp(2k)]}+b_k\sum_{n=1}^{2k}\Tr_k\left[\Pi^n\Pi_n\right]+\sum_{n=1}^{2k}\sum_{\ga=1,2}\nu_{\ga n}\Tr_k\left[p^\ga\Pi^n\right]+\nn\\
&+&\sum_{i,j,l=1}^N\sum_{a,b,c,d,e,f=1}^6\epsilon_{abcdef}\hat{\mu}_{ab;i}\hat{\mu}_{cd;j}\hat{\mu}_{ef;l}\gd_{i+j+l,2N-k+1}\,,
\label{eq:superpotrankstabB}
\ee
where $\mathcal{W}_{E[USp(2k)]}$ is the superpopential of $E[USp(2k)]$ and again $\Pi$ is an operator of $E[USp(2k)]$ that transforms in the bifundamental representation of its two $USp(2k)$ symmetries (see Appendix \ref{app:eusp}). 

Notice that even if the second line of the superpotential is the same that we had for $k=0$ in \eqref{sspp} and so it naively seems to preserve a $U(6)$ structure, this symmetry rotating the 6 chirals in the fundamental representation of the $USp(2k)$ gauge node is explicitly broken to $SU(4)\times SU(2)\times U(1)^2$ by the last term in the first line, which treats differently 2 out of the 6 chirals with respect to the remaining 4. In addition, from the $E[USp(2 k)]$ block, we also have two abelian symmetries as well as the manifest $SU(2)^k$ symmetry, which is enhanced to $USp(2 k)$ in the IR. Among the four abelian symmetries in total, the $USp(2k)$ gauging makes one combination gauge anomalous\footnote{In order to work out which combination of the four abelian symmetries is anomalous one needs to know the contribution of the $E[USp(2k)]$ block to the mixed anomaly between the $USp(2k)$ gauge symmetry and the most general trial $R$-symmetry. This can be computed using the quiver description of $E[USp(2k)]$ (see eq.~(A.14) of \cite{Garozzo:2020pmz}).} and the superpotential explicitly breaks another one. Thus, the full manifest non-anomalous global symmetry is
\be
F_B = \prod_{i = 1}^k SU(2)_{y_i} \times SU(4)_v\times SU(2)_b\times U(1)_s\times U(1)_t\,.
\ee
As mentioned, $\prod_{i = 1}^k SU(2)_{y_i}$ of the $E[USp(2 N)]$ block is enhanced to $USp(2k)_y$ in the IR. Moreover, the duality tells us that there should be another enhancement of symmetry:
\be
SU(4)_v\times SU(2)_b\times U(1)_s \quad \rightarrow \quad SU(6)_x
\ee
so that the IR global symmetry of Theory B becomes
\begin{align}
USp(2k)_y\times SU(6)_x\times U(1)_t\,,
\end{align}
which is the same as the manifest symmetry \eqref{eq:F_A} of Theory A.
We will give other evidence for this momentarily. The transformation rules of the matter fields under these symmetries and a possible choice of non-anomalous $R$-symmetry are summarized in Table \ref{tab:rankstabchargesTB}.

\begin{table}[t]
\centering
\scalebox{1}{\setlength{\extrarowheight}{2pt}
\begin{tabular}{|c|cccccc|}\hline
{} & $USp(2k)_y$ & $SU(4)_v$ & $SU(2)_b$ & $U(1)_s$ & $U(1)_t$  & $U(1)_{R_0}$\\ \hline
$p$ & $\bf 1$ & $\bf 1$ & $\bf 2$ & 0 & $\frac{N-k+1}{2}$ & 0\\
$\tilde{p}$ & $\bf 1$ & $\bf 4$ & $\bf 1$ & $-1$ & $\frac{1-N-k}{6}$ & $\frac{2}{3}$ \\
$\mathsf{H}$ & $\bf 1$ & $\bf 1$ & $\bf 1$ & 0 & 1 & 0 \\
$\Pi$ & $\bf 2k$ & $\bf 1$ & $\bf 1$ & 2 & $\frac{2k-N-2}{6}$ & $\frac{2}{3}$ \\ \hline
$\mu_{ab;i}$ & $\bf 1$ & $\bf 6$ & $\bf 1$ & 2 & $\frac{3i+k-2N-1}{3}$ & $\frac{2}{3}$\\
$\mu_{56;i}$ & $\bf 1$ & $\bf 1$ & $\bf 1$ & $-4$ & $\frac{3i+k-2N-1}{3}$ & $\frac{2}{3}$ \\
$\mu_{a\ga;i}$ & $\bf 1$ & $\bf 4$ & $\bf 2$ & $-1$ & $\frac{3i+k-2N-1}{3}$ & $\frac{2}{3}$ \\
$\nu_{\ga n}$ & $\bf 2k$ & $\bf 1$ & $\bf 2$ & $-2$ & $\frac{k-2N-1}{6}$ & $\frac{4}{3}$ \\
$\mathsf{C}$ & $\bf 1$ & $\bf 1$ & $\bf 1$ & 0 & $-1$ & 2 \\ \hline
\end{tabular}}
\caption{Transformation properties under the global symmetry of the matter fields of Theory B. We are also including some important operators of the $E[USp(2k)]$ block. The horizontal line separates operators charged under the $USp(2 k)$ gauge symmetry from the singlets.}
\label{tab:rankstabchargesTB}
\end{table}

Before presenting some tests for the duality, let us briefly comment on some special cases which are related to known dualities. For low $k$, indeed, Theory B significantly simplifies and we can reasonably expect to recover dualities that already appeared in the literature. 
We previously pointed out that for $k=0$ the dual Theory B is just a WZ model and we recover the confining duality of \cite{Csaki:1996zb} that we discussed in Section \ref{sec:confining}.
For $k=1$ the $E[USp(2k)]$ theory consists simply of an $SU(2)\times SU(2)$ bifundamental plus a flip of the associated quadratic operator, without any gauge group. One of the two $SU(2)$ is gauged to construct Theory B, which is thus a simple single gauge node theory. Hence, modulo singlet fields the rank stabilization duality for $k=1$ relates a $USp(2N)$ gauge theory with one antisymmetric and 8 fundamental flavors to an $SU(2)$ gauge theory with 8 fundamental flavors. This duality corresponds to the ``rank changing" duality of \cite{Razamat:2017hda}. If we further set $N=1$ both Theory A and Theory B are $SU(2)$ gauge theories with 8 fundamental chirals and differ only for singlet fields and superpotential terms involving them. We thus recover a pair of the known 72 self-dual frames of $SU(2)$ with 8 chirals \cite{Csaki:1997cu,Spiridonov:2008zr,Dimofte:2012pd,Razamat:2017hda,Hwang:2020ddr}.

\subsection{Tests}

We now move on to the discussion of some tests for the $4d$ rank stabilization duality. The strongest one will be the analytic matching of the supersymmetric indices of the dual theories. Even if they are implied by the equality of the indices, in order to better understand how the duality works we will first discuss the mapping of some gauge invariant operators and the matching of some of the anomalies.

\subsubsection*{Mapping some operators}

It is interesting to look at how the map of some of the gauge invariant operators works, since they are very simple on the side of Theory A, while they map to some more complicated operators in Theory B. Moreover, on the side of Theory B we only manifestly see the $SU(4)_v\times SU(2)_b\times U(1)_s$ subgroup of the $SU(6)_x$ global symmetry, so we expect to be able to collect its gauge invariant operators into representations of the larger symmetry. 

On the side of Theory A we have two types of operators. We can have the mesons constructed with the fundamental chirals $Q$ and $P$ possibly dressed with the antisymmetric $A$ and the gauge invariant combinations that we can construct with powers of $A$. Let us start considering the former ones. We can have three different kinds of mesons, depending on which combinations of $Q$ and $P$ we use. For example, we can have the following dressed mesons:
\be
Q_aA^{i-1}Q_b,\qquad i=1,\cdots,N\,,
\ee
where as usual the contraction of gauge indices is understood.
These operators transform in the antisymmetric $\bf 15$ representation of $SU(6)_x$, while they are singlets of $USp(2k)_y$. On the side of Theory B there is a combination of singlets and gauge invariant operators which is a natural candidate for mapping to these mesons. This is the set of operators $\hat{\mu}_{ab;i}$ that we already introduced in \eqref{eq:hatmuop}. Since the last term of the superpotential \eqref{eq:superpotrankstabB} containing $\hat \mu_{ab;i}$ is written in an $SU(6)$ invariant way, we already expect that these operators indeed form an antisymmetric representation of the enhanced $SU(6)_x$, but we can also check it explicitly. Indeed, from Table \ref{tab:rankstabchargesTB} we can see that the operators in \eqref{eq:hatmuop} all have the same charges under $U(1)_t$ and $U(1)_{R_0}$, while their transformation properties under $SU(4)_v\times SU(2)_b\times U(1)_s\subset SU(6)_x$ are compatible with the branching rule
\be
{\bf 15}\to ({\bf 6},{\bf 1})^2\oplus({\bf 4},{\bf 2})^{-2}\oplus({\bf 1},{\bf 1})^{-4}\,.
\ee
Moreover, the charges of $\hat{\mu}_{ab;i}$ under the abelian symmetry $U(1)_t$ and the reference $R$-symmetry $U(1)_{R_0}$ are precisely equal to those of the operators $Q_aA^{i-1}Q_b$, so we expect the two to be mapped to each other under the duality (see Table \ref{tab:rankstabopmap}).

We can also have mesons constructed from $P$
\be
P_nP_m\,.
\ee
In this case there is no dressing, since the F-term equations associated to $P_n$ set the dressed mesons to zero in the chiral ring. These operators form the antisymmetric $\bf k(2k-1)-1$ representation of $USp(2k)_y$, while they are singlets of $SU(6)_x$. The natural candidate on the side of Theory B for the operators that should map to this meson is the gauge invariant operator $\mathsf{C}$ of the $E[USp(2k)]$ block. We can indeed check that the two operators transform in the same way under all the global symmetries and the reference $R$-symmetry (see Table \ref{tab:rankstabopmap}).

\begin{table}[t]
\centering
\scalebox{1}{\setlength{\extrarowheight}{2pt}
\begin{tabular}{|c|cccc|c|}\hline
Theory A & $USp(2k)_y$ & $SU(6)_x$ & $U(1)_t$  & $U(1)_{R_0}$ & Theory B \\ \hline
$Q_aA^{i-1} Q_b$ & ${\bf 1}$ & ${\bf 15}$ & $\frac{k+3i-2N-1}{3}$ & $\frac{2}{3}$ & $\hat{\mu}_{ab;i}$ \\
$P_n P_m$ & $\bf k(2k-1)-1$ & $\bf 1$ & $-1$ & 2 & $\mathsf{C}$ \\
$Q_a P_n$ & $\bf 2k$ & $\bf 6$ & $\frac{k-2N-1}{6}$ & $\frac{4}{3}$ & $\hat{\nu}_{an}$ \\
$A^{N-k+i}$ & $\bf 1$ & $\bf 1$ & $N-k+i$ & 0 & $p_\ga\mathsf{H}^{i-1}p_\gb$
\\ \hline
\end{tabular}}
\caption{Transformation properties under the global symmetry $F_A$ of some of the gauge invariant operators of Theory A and B, from which we can understand how these map under the duality.}
\label{tab:rankstabopmap}
\end{table}

The last type of meson that we can construct in Theory B is
\be
Q_aP_n\,,
\ee
where the dressed mesons are again set to zero in the chiral ring for the same reason as before. This operator transforms in the fundamental $\bf 2k$ of $USp(2k)_y$ and also in the fundamental $\bf 6$ of $SU(6)_x$. The corresponding operator in Theory B should then be constructed collecting gauge invariant operators that have the same $U(1)_t$ and $U(1)_{R_0}$ charges, that are all in the fundamental of $USp(2k)_y$ and that transform under the $SU(4)_v\times SU(2)_b\times U(1)_s$ subgroup of $SU(6)_x$ according to the branching rule
\be\label{BR6SU6}
{\bf 6}\to({\bf 4},{\bf 1})^1\oplus({\bf 1},{\bf 2})^{-2}\,.
\ee
The desired operator, which we will denote by $\hat{\nu}_{na}$ for $n=1,\cdots,2k$ and for $a=1,\cdots,6$, is defined as follows:
\be
\hat{\nu}_{an}=\begin{cases}\tilde{p}_a\Pi_n,\quad a=1,\cdots,4 \\ \nu_{a-4, n},\quad a=5,6\end{cases}\quad n=1,\cdots,2k\,.
\ee
One can check that this operator $\hat{\nu}_{na}$ has the same transformation properties under the global symmetry and the $R$-symmetry as the meson $Q_a P_n$ of Theory A (see Table \ref{tab:rankstabopmap}).

Finally, on the side of Theory A we can construct gauge invariant combinations of the antisymmetric chiral $A$. Those that are independent and that are not set to zero in the chiral ring by the F-term equations of the singlets $\gb_i$ are
\be
A^{N-k+i},\qquad i=1,\cdots,k\,.
\ee
The operators of Theory B that have the correct charges to be mapped to it are (see Table \ref{tab:rankstabopmap})
\be
p_\ga\mathsf{H}^{i-1}p_\gb,\qquad i=1,\cdots,k\,.
\ee
This concludes the mapping of the main gauge invariant operators\footnote{Similarly to the $k=0$ case, the singlets $\gb_j$ are expected not to get a VEV at the quantum level.}

\subsubsection*{Anomalies}

Another possible test that we can perform is matching the anomalies of the theories. In particular, we will focus on the $a$ and $c$ trial central charges.
We stress again that this is expected to follow from the matching of the supersymmetric indices, but it is useful to compute the $a$ and $c$ central charges to have a better understanding of the dynamics of the theories.

On the side of Theory A we only have one abelian symmetry $U(1)_t$. The most general non-anomalous $R$-symmetry $U(1)_R$ is thus parametrized by the mixing coefficient of $U(1)_t$ with the reference $R$-symmetry $U(1)_{R_0}$
\be
R(\mathfrak{t})=R_0+q_t\mathfrak{t}\,,
\ee
where $R$ is the $R$-charge under the generic $R$-symmetry, $R_0$ is the $R$-charge under the reference $R$-symmetry $U(1)_{R_0}$, $q_t$ is the charge under $U(1)_t$ and $\mathfrak{t}$ is the mixing coefficient. The trial central charges are then
\be
a_A(\mathfrak{t})=\frac{3}{32}\left(3\Tr R(\mathfrak{t})^3-\Tr R(\mathfrak{t})\right),\qquad c_A(\mathfrak{t})=\frac{1}{32}\left(9\Tr R(\mathfrak{t})^3-5\Tr R(\mathfrak{t})\right)\,.
\ee
Computing these for Theory A we find
\be
&&a_A(\mathfrak{t})=\frac{1}{128} \left(-9 k^4 \mathfrak{t}^3+2 k^3 \mathfrak{t}^2 ((19 N+9) \mathfrak{t}-18)-3 k^2 \mathfrak{t} \left(\mathfrak{t} \left(22 N^2 \mathfrak{t}+14 N (\mathfrak{t}-2)+3 (\mathfrak{t}-6)\right)+16\right)+\right.\nn\\
&&\qquad\quad\left.+6 k (\mathfrak{t} (N (\mathfrak{t} (N (10 N \mathfrak{t}+\mathfrak{t}-2)+4 \mathfrak{t}-34)+32)-3 \mathfrak{t}+8)-4)+\right.\nn\\
&&\qquad\quad\left.+5 N \left(8-(N-1) \mathfrak{t}^2 (N ((5 N-1) \mathfrak{t}+12)-4 \mathfrak{t}+6)\right)\right)\nn\\
&&c_A(\mathfrak{t})=\frac{1}{128} \left(-9 k^4 \mathfrak{t}^3+2 k^3 \mathfrak{t}^2 ((19 N+9) \mathfrak{t}-18)-k^2 \mathfrak{t} \left((6 N (11 N+7)+9) \mathfrak{t}^2-6 (14 N+9) \mathfrak{t}+44\right)+\right.\nn\\
&&\qquad\quad\left.+ k (\mathfrak{t} (N (3 \mathfrak{t} (N (10 N \mathfrak{t}+\mathfrak{t}-2)+4 \mathfrak{t}-34)+92)-9 \mathfrak{t}+22)-8)+\right.\nn\\
&&\qquad\quad\left.-5 N ((N-1) \mathfrak{t}+2) ((N-1) \mathfrak{t} ((5 N+4) \mathfrak{t}+2)-8)\right)
\label{acrankstabA}
\ee

On the side of Theory B there are two abelian symmetries that can in principle mix with the $R$-symmetry, $U(1)_t$ and $U(1)_s$. Hence, the most general non-anomalous $R$-symmetry depends on two parameters, the two mixing coefficients $\mathfrak{t}$ and $\mathfrak{s}$
\be
R(\mathfrak{t},\mathfrak{s})=R_0+q_t\mathfrak{t}+q_s\mathfrak{s}\,.
\ee
The trial central charges of Theory B will then be functions of both $\mathfrak{t}$ and $\mathfrak{s}$. Specifically, we find
\be
&&a_B(\mathfrak{t},\mathfrak{s})=\frac{1}{128} \left(-9 k^4 \mathfrak{t}^3+2 k^3 \mathfrak{t}^2 ((19 N+9) \mathfrak{t}-18)-3 k^2 \mathfrak{t} \left(\mathfrak{t} \left(22 N^2 \mathfrak{t}+14 N (\mathfrak{t}-2)+3 (\mathfrak{t}-6)\right)+16\right)+\right.\nn\\
&&\qquad\quad\left.+6 k \left(\mathfrak{t} \left(N \left(\mathfrak{t} (N (10 N \mathfrak{t}+\mathfrak{t}-2)+4 \mathfrak{t}-34)+72 \mathfrak{s}^2+32\right)-3 \mathfrak{t}+8\right)-4\right)+\right.\nn\\
&&\qquad\quad\left.+(N-1) N \mathfrak{t} \left(-5 \mathfrak{t} (N ((5 N-1) \mathfrak{t}+12)-4 \mathfrak{t}+6)-864 \mathfrak{s}^2\right)-8 N \left(108 \mathfrak{s}^2 (\mathfrak{s}+2)-5\right)\right)\nn\\
&&c_B(\mathfrak{t},\mathfrak{s})=\frac{1}{128} \left(-9 k^4 \mathfrak{t}^3+2 k^3 \mathfrak{t}^2 ((19 N+9) \mathfrak{t}-18)-k^2 \mathfrak{t} \left((6 N (11 N+7)+9) \mathfrak{t}^2-6 (14 N+9) \mathfrak{t}+44\right)+\right.\nn\\
&&\qquad\quad\left.+ k \left(\mathfrak{t} \left(N \left(3 \mathfrak{t} (N (10 N \mathfrak{t}+\mathfrak{t}-2)+4 \mathfrak{t}-34)+216 \mathfrak{s}^2+92\right)-9 \mathfrak{t}+22\right)-8\right)+\right.\nn\\
&&\qquad \quad\left.+N \left(-864 \mathfrak{s}^2 ((N-1) \mathfrak{t}+2)-5 ((N-1) \mathfrak{t}+2) ((N-1) \mathfrak{t} (5 N \mathfrak{t}+4 \mathfrak{t}+2)-8)-864 \mathfrak{s}^3\right)\right)\,.
\label{acrankstabB}
\ee
Despite the explicit dependence on $\mathfrak{s}$, one can check that the local maximum of $a_B(\mathfrak{t},\mathfrak{s})$ corresponds to $\mathfrak{s}=0$ and to a value of $\mathfrak{t}$ which is the same that we can find maximizing the trial $a_A(\mathfrak{t})$ central charge of Theory A. This means that $U(1)_s$ doesn't mix with the $R$-symmetry in the IR, which is compatible with the fact that it participates to the enhancement to the non-abelian $SU(6)_x$ symmetry. Moreover, setting $\mathfrak{s}=0$ one finds that the $a$ and $c$ central charges of Theory A and B coincide as functions of $\mathfrak{t}$
\be
a_A(\mathfrak{t})=a_B(\mathfrak{t},\mathfrak{s}=0),\qquad c_A(\mathfrak{t})=c_B(\mathfrak{t},\mathfrak{s}=0)\,,
\ee
which is another test of the duality.

\subsubsection*{Supersymmetric index}

The strongest test that we can provide for the duality is the matching of the supersymmetric indices. This amounts to proving the following non-trivial integral identity:
\be
&&\oint\udl{\vec{z}_N}\Gamma_e(t)^N \prod_{i<j}^N \Gamma_e (t z_i^{\pm 1} z_j^{\pm 1})\times\nn\\
&&\times\prod_{i=1}^N\prod_{a=1}^6\Gpq{(pq)^{\frac{1}{6}}t^{-\frac{2N-k-2}{6}}x_az_i^{\pm1}}\prod_{j=1}^{k}\Gpq{\left(pq\right)^{1/2}t^{-1/2}z_i^{\pm1}y_j^{\pm1}}=\nn\\
&&\qquad=\Gpq{(pq)^{\frac{1}{3}}t^{\frac{N-2k+2}{3}}s^{-4}}\prod_{i=1}^N\prod_{\ga<\gb=1}^4\Gpq{(pq)^{\frac{1}{3}}t^{\frac{3i+k-2N-1}{3}}s^2v_\ga v_\gb}\times\nn\\
&&\qquad\times\prod_{j=1}^{N-k}\Gpq{t^j}\prod_{\ga=1}^4\Gpq{(pq)^{1/3}t^{\frac{3j+k-2N-1}{3}}s^{-1}b^{\pm1}v_\ga}\times\nn\\
&&\qquad\times\Gpq{(pq)^{1/3}t^{\frac{3j+k-2N-1}{3}}s^{-4}}\prod_{j=1}^k\Gpq{(pq)^{2/3}t^{\frac{-1+k-2N}{6}}s^{-2}b^{\pm1}y_j^{\pm1}}\times\nn\\
&&\qquad\times\oint\udl{\vec{w}_k}\mathcal{I}_{E[USp(2k)]}\left(\vec{w};\vec{y};t;(pq)^{1/3}t^{\frac{2k-N-2}{6}}s^2\right)\times\nn\\
&&\qquad\times\Gpq{t^{\frac{N-k+1}{2}}b^{\pm1}w_i^{\pm1}}\prod_{i=1}^{k}\prod_{\ga=1}^4\Gpq{(pq)^{1/3}t^{\frac{1-N-k}{6}}s^{-1}v_\ga w_i^{\pm1}}\, .
\label{4drankstabfinal}
\ee
In this identity $x_a$ are fugacities for $SU(6)_x$ and satisfy the constraint $\prod_{a=1}^6x_a=1$, $v_\ga$ are fugacities of $SU(4)_v$ and satisfy the constraint $\prod_{\ga=1}^4v_\ga=1$, $b$ is the fugacity for $SU(2)_b$, $t$ is the fugacity for $U(1)_t$ and $s$ is the fugacity for $U(1)_s$\footnote{Notice that in order to recover \eqref{csaki} from \eqref{4drankstabfinal} for $k=0$ we have to redefine $x_a\to(pq)^{-\frac{1}{6}}t^{\frac{N-1}{3}}x_a$.}. $\mathcal{I}_{E[USp(2k)]}$ denotes the index of the $E[USp(2k)]$ theory, which we explicitly define in Appendix \ref{app:eusp}. The identity \eqref{4drankstabfinal} holds provided that the parameters on the two sides are identified as follows:
\be
x_a=\begin{cases}s v_a & a=1,\cdots, 4 \\ s^{-2}b^{\pm1} & a=5,6\end{cases}\,.
\ee
This is compatible with the embedding \eqref{BR6SU6} of $SU(4)_v\times SU(2)_b\times U(1)_s$ inside $SU(6)_x$ that we used in the operator map. We prove this identity for arbitrary $N$ and $k$ in Appendix \ref{app:newid} by using and extending the results of \cite{2014arXiv1408.0305R}.

\acknowledgments
We are grateful to Stephane Bajeot and Sergio Benvenuti for coordinating the release of their work \cite{stephane-sergio} with us. The research of C.H. is partially supported by the STFC consolidated grant ST/T000694/1. S.P. is partially supported by the  INFN. MS is partially supported by the ERC Consolidator Grant \#864828 “Algebraic Foundations of Supersymmetric Quantum Field Theory (SCFTAlg)” and by the Simons Collaboration for the Nonperturbative Bootstrap under grant \#494786 from the Simons Foundation. 

\appendix

\section{Quiver sequential deconfinement }\label{qsd2}

In this appendix we explicitly show how to derive the index identity \eqref{quiverwzid} for the duality of Section \ref{sec:confquiv} using the strategy summarized in Figure being a straightforward generalization. For simplicity we also consider the quiver gauge theory without any additional singlets. We then expect to recover the $\beta_i^{(n)}$ and the $\gamma_i^{(n)}$  singlets all on  the r.h.s of the duality with flipped charges together with the singlets entering the WZ superpotential. 

The index of the electric theory with no singlets is given by (to distinguish it from the one with the singlets appearing in \eqref{quiverwzid} we denote it by $\tilde{\mathcal{I}}^{(N)}_{\text{gauge}}$)
\begin{align}
  &\tilde{\mathcal{I}}^{(2)}_{\text{gauge}}(\vec{M};\vec{y},\vec{x};c,t) = \Gamma_e(pqt^{-1})^{M_1+M_2} \oint d \vec{z}^{\,(M_1)} d \vec{z}^{\,(M_2)} \prod_{i<j}^{M_1} \Gamma_e (pqt^{-1}z_i^{(M_1)\pm1} z_j^{(M_1)\pm1})  \nonumber \\
  &\times  \prod_{i<j}^{M_2} \Gamma_e (pqt^{-1}z_i^{(M_2)\pm1} z_j^{(M_2)\pm1}) \prod_{i=1}^{M_1} ((pq)^{\frac{M_2-2M_1}{2}} c t^\frac{2M_1-M_2-1}{2} y_1^{\pm 1} z_i^{(M_1)\pm1} ) 
  \nonumber \\
  &\times  \prod_{i=1}^{M_1} (pq c^{-1} t^{-\frac{1}{2}} y_2^{\pm 1} z_i^{(M_1)\pm1} ) \prod_{i=1}^{M_1} \prod_{j=1}^{M_2} \Gamma_e (t^\frac{1}{2} z_i^{(M_1)\pm1} z_j^{(M_2)\pm1} ) \prod_{i=1}^{M_2} \Gamma_e (c y_2^{\pm 1} z_i^{(M_2)\pm1} ) \nonumber \\
  &\times \prod_{i=1}^{M_2} \prod_{a=1}^4 \Gamma_e ( (pq)^\frac{3+M_1-2M_2}{4} c^{-\frac{1}{2}} t^\frac{2M_2-M_1-2}{4} x_a z_i^{(M_2)\pm1}) \,,
\end{align}
The index of the auxiliary gauge theory is obtained deconfining the antisymmetric on the first $USp(2M_1)$ gauge node and trading its contribution with an additional integral over $USp(2M_1-2)$ gauge fugacities. Specifically, the index of the auxiliary theory reads
\begin{align} \label{index_aux}
&\mathcal{I}^{(2)}_{\text{aux}} = \Gamma_e (pqt^{-1})^{M_2} \oint d \vec{z}^{(M_1-1)} d\vec{z}^{(M_1)} d\vec{z}^{(M_2)} \prod_{i=1}^{M_1-1} \Gamma_e ( (pq)^\frac{M_2-2M_1-1}{2} c t^\frac{2M_1-M_2}{2} y_1^{-1} z_i^{(M_1-1) \, \pm 1} ) \nonumber \\
& \prod_{i=1}^{M_1-1} \Gamma_e ( (pq)^\frac{3-M_2}{2} c^{-1} t^\frac{M_2}{2} y_1 z_i^{(M_1-1) \, \pm 1} ) \prod_{i=1}^{M_1-1} \prod_{j=1}^{M_1} \Gamma_e ( (pq)^\frac{1}{2} t^{-\frac{1}{2}} z_i^{(M_1-1) \, \pm 1} z_j^{(M_1) \, \pm 1} ) \nonumber \\
&\times \prod_{i=1}^{M_1} \Gamma_e ( (pq)^\frac{M_2-2}{2} c t^{\frac{1-M_2}{2}} y_1^{-1} z_i^{(M_1) \, \pm 1} ) \prod_{i=1}^{M_1} ((pq)^{\frac{M_2-2M_1}{2}} c t^\frac{2M_1-M_2-1}{2} y_1 z_i^{(M_1)\pm1} ) \nonumber \\
&\times  \prod_{i=1}^{M_1} (pq c^{-1} t^{-\frac{1}{2}} y_2^{\pm 1} z_i^{(M_1)\pm1} ) \prod_{i=1}^{M_1} \prod_{j=1}^{M_2} \Gamma_e (t^\frac{1}{2} z_i^{(M_1)\pm1} z_j^{(M_2)\pm1} ) \prod_{i=1}^{M_2} \Gamma_e (c y_2^{\pm 1} z_i^{(M_2)\pm1} ) \nonumber \\
  &\times \prod_{i=1}^{M_2} \prod_{a=1}^4 \Gamma_e ( (pq)^\frac{3+M_1-2M_2}{4} c^{-\frac{1}{2}} t^\frac{2M_2-2-M_1}{4} x_a z_i^{(M_2)\pm1}) \prod_{i<j}^{M_2} \Gamma_e (pqt^{-1}z_i^{(M_2)\pm1} z_j^{(M_2)\pm1}) \,.
\end{align} 
Applying the identity \eqref{IP} for the IP duality to the $USp(2M_1-2)$ node corresponds to the evaluation formula
\begin{align}
 &\oint d \vec{z}^{(M_1-1)} \prod_{i=1}^{M_1-1} \Gamma_e ( (pq)^\frac{M_2-2M_1-1}{2} c t^\frac{2M_1-M_2}{2} y_1^{-1} z_i^{(M_1-1) \, \pm 1} ) \prod_{i=1}^{M_1-1} \Gamma_e ( (pq)^\frac{3-M_2}{2} c^{-1} t^\frac{M_2}{2} y_1 z_i^{(M_1-1) \, \pm 1} )  \nonumber \\
 &\times \prod_{i=1}^{M_1-1} \prod_{j=1}^{M_1} \Gamma_e ( (pq)^\frac{1}{2} t^{-\frac{1}{2}} z_i^{(M_1-1) \, \pm 1} z_j^{(M_1) \, \pm 1} ) = \prod_{i=1}^{M_1} ((pq)^{\frac{M_2-2M_1}{2}} c t^\frac{2M_1-M_2-1}{2} y_1^{-1} z_i^{(M_1)\pm1} )  \nonumber \\
 &\times \prod_{i=1}^{M_1} ((pq)^{\frac{4-M_2}{2}} c^{-1} t^\frac{M_2-1}{2} y_1 z_i^{(M_1)\pm1} ) \Gamma_e (pqt^{-1})^{M_1} \prod_{i<j}^{M_1} \Gamma_e (pqt^{-1}z_i^{(M_1)\pm1} z_j^{(M_1)\pm1})  \Gamma_e ((pq)^{1-M_1} t^{M_1}) \,,
\end{align}
and it is easy to check that plugging this into \eqref{index_aux} we obtain the index of the original theory plus the the contribution of a singlet $\Gamma_e ((pq)^{1-M_1} t^{M_1})$. 
 
Now we apply the IP duality to the original $USp(2M_1)$ gauge node of the auxiliary quiver, which becomes a $USp(2M_2-2)$ node. This corresponds to the integral identity
\begin{align}
 &\oint d \vec{z}^{(M_1)} \prod_{i=1}^{M_1-1} \prod_{j=1}^{M_1} \Gamma_e ( (pq)^\frac{1}{2} t^{-\frac{1}{2}} z_i^{(M_1-1)\, \pm 1} z_j^{(M_1) \,\pm 1} ) \prod_{i=1}^{M_1} \Gamma_e ( (pq)^\frac{M_2-2}{2} c t^{\frac{1-M_2}{2}} y_1^{-1} z_i^{(M_1) \, \pm 1} ) \nonumber \\
&\times  \prod_{i=1}^{M_1} ((pq)^{\frac{M_2-2M_1}{2}} c t^\frac{2M_1-M_2-1}{2} y_1 z_i^{(M_1)\pm1} )  \prod_{i=1}^{M_1} (pq c^{-1} t^{-\frac{1}{2}} y_2^{\pm 1} z_i^{(M_1)\pm1} ) \prod_{i=1}^{M_1} \prod_{j=1}^{M_2} \Gamma_e (t^\frac{1}{2} z_i^{(M_1)\pm1} z_j^{(M_2)\pm1} ) = \nonumber \\
&= \Gamma_e (pqt^{-1})^{M_1-1} \prod_{i<j}^{M_1-1}\Gamma_e (pqt^{-1}z_i^{(M_1-1)\pm1} z_j^{(M_1-1)\pm1}) \Gamma_e ((pq)^2 c^{-2} t^{-1} ) \Gamma_e ((pq)^{M_2-M_1-1} c^{2} t^{M_1-M_2} ) \nonumber \\
&\times \Gamma_e ((pq)^\frac{M_2}{2} t^{-\frac{M_2}{2}} y_1^{-1} y_2^{\pm 1} ) \Gamma_e ((pq)^\frac{2+M_2-2M_1}{2} t^{\frac{2M_1-M_2-2}{2}} y_1 y_2^{\pm 1} ) \prod_{i=1}^{M_1-1} ((pq)^\frac{M_2-1}{2} c t^{-\frac{M_2}{2}} y_1^{- 1} z_i^{(M_1-1)\pm1} ) \nonumber \\
&\times \prod_{i=1}^{M_1-1} ((pq)^\frac{M_2-2M_1+1}{2} c t^{\frac{2M_1-M_2-2}{2}} y_1 z_i^{(M_1-1)\pm1} ) \prod_{i=1}^{M_1-1} ((pq)^\frac{3}{2} c^{-1} t^{-1} y_2^{\pm 1} z_i^{(M_1-1)\pm1} ) \Gamma_e (t)^{M_2}  \nonumber \\
&\times  \prod_{i<j}^{M_2}\Gamma_e (tz_i^{(M_2)\pm1} z_j^{(M_2)\pm1}) \prod_{i=1}^{M_2}\Gamma_e ((pq)^\frac{M_2-2}{2}c t^\frac{2-M_2}{2} y_1^{-1} z_i^{(M_2)\pm1}) \prod_{i=1}^{M_2}\Gamma_e ((pq)^\frac{M_2-2M_1}{2}c t^\frac{2M_1-M_2}{2} y_1 z_i^{(M_2)\pm1}) \nonumber \\
&\times  \prod_{i=1}^{M_2}\Gamma_e (pqc^{-1}  y_2^{\pm 1} z_i^{(M_2)\pm1}) \oint d \vec{z}^{(M_2-1)} \prod_{i=1}^{M_2-1} \Gamma_e ( (pq)^\frac{3-M_2}{2} c^{-1} t^\frac{M_2-1}{2} y_1 z_i^{(M_2-1)\, \pm 1}) \nonumber \\
&\times \prod_{i=1}^{M_2-1} \Gamma_e ( (pq)^\frac{2M_1-M_2+1}{2} c^{-1} t^\frac{1+M_2-2M_1}{2} y_1^{-1} z_i^{(M_2-1)\, \pm 1})  \prod_{i=1}^{M_2-1} \Gamma_e ( (pq)^{-\frac{1}{2}} c t^\frac{1}{2} y_2^{\pm 1} z_i^{(M_2-1)\, \pm 1}) \nonumber \\
&\times \prod_{i=1}^{M_2-1} \prod_{j=1}^{M_2}  \Gamma_e ( (pq)^\frac{1}{2}  t^{-\frac{1}{2}}  z_i^{(M_2-1)\, \pm 1} z_j^{(M_2)\, \pm 1})  \prod_{i=1}^{M_2-1} \prod_{j=1}^{M_1-1}  \Gamma_e ( t^{\frac{1}{2}}  z_i^{(M_2-1)\, \pm 1} z_j^{(M_1-1)\, \pm 1}) \,.
\end{align}   
Plugging the above identity into \eqref{index_aux}, we obtain the index of a new quiver theory with three gauge nodes $USp(2M_1-2)$, $USp(2M_2-2)$ and $USp(2M_2)$. Notice that the contribution to the index of the antisymmetric on the $USp(2M_2)$ gauge node cancels out in this last step, so that now we can apply the IP duality to it. We then collect all the $z_i^{(M_2)}$ terms and apply the duality to the last node, which confines; this corresponds to the evaluation formula
\begin{align}
&\oint d \vec{z}^{(M_2)} \prod_{i=1}^{M_2-1} \prod_{j=1}^{M_2}  \Gamma_e ( (pq)^\frac{1}{2}  t^{-\frac{1}{2}}  z_i^{(M_2-1)\, \pm 1} z_j^{(M_2)\, \pm 1}) \prod_{i=1}^{M_2} \Gamma_e ( (pq)^\frac{M_2-2}{2} c t^\frac{2-M_2}{2} y_1^{-1} z_i^{(M_2) \pm 1}) \nonumber \\
&\times \prod_{i=1}^{M_2} \Gamma_e ( (pq)^\frac{M_2-2M_1}{2} c t^\frac{2M_1-M_2}{2} y_1 z_i^{(M_2) \pm 1}) 
 \prod_{i=1}^{M_2} \prod_{a=1}^4 \Gamma_e ( (pq)^\frac{3+M_1-2M_2}{4} c^{-\frac{1}{2}} t^\frac{2M_2-2-M_1}{4} x_a z_i^{(M_2) \pm 1}) = \nonumber \\
&= \Gamma_e (pqt^{-1})^{M_2-1} \prod_{i<j}^{M_2-1}\Gamma_e (pqt^{-1}z_i^{(M_2-1)\pm1} z_j^{(M_2-1)\pm1}) \prod_{a<b}^4\Gamma_e ((pq)^\frac{3+M_1-2M_2}{2} c^{-1} t^\frac{2M_2-2-M_1}{2} x_a x_b) \nonumber \\
&\times \Gamma_e ((pq)^{M_2-M_1-1} c^{2} t^{1+M_1-M_2} ) \prod_{a=1}^4\Gamma_e ((pq)^\frac{M_1-1}{4} c^\frac{1}{2} t^\frac{2-M_1}{4} y_1^{-1} x_a ) \prod_{a=1}^4\Gamma_e ((pq)^\frac{3-3M_1}{4} c^\frac{1}{2} t^\frac{3M_1-2}{4} y_1 x_a ) \nonumber \\
&\times \prod_{i=1}^{M_2-1}\Gamma_e ((pq)^\frac{M_2-1}{2} c t^\frac{1-M_2}{2} y_1^{-1} z_i^{(M_2-1) \pm 1}  ) \prod_{i=1}^{M_2-1}\Gamma_e ((pq)^\frac{M_2-2M_1+1}{2} c t^\frac{2M_1-M_2-1}{2} y_1 z_i^{(M_2-1) \pm 1} ) \nonumber \\
&\times \prod_{i=1}^{M_2-1} \prod_{a=1}^4\Gamma_e (pq)^\frac{5+M_1-2M_2}{4} c^{-\frac{1}{2}} t^\frac{2M_2-4-M_1}{4}  x_a z_i^{(M_2-1) \pm 1}) \,.
\end{align}

Collecting all the remaining pieces we finally obtain the following expression for the index
\begin{align}
  &\tilde{\mathcal{I}}^{(2)}_{\text{gauge}} = \Gamma_e ((pq)^{M_1} t^{-M_1}) \Gamma_e ((pq)^2 c^{-2} t^{-1} ) \Gamma_e ((pq)^{M_2-M_1-1} c^{2} t^{M_1-M_2} ) \Gamma_e ((pq)^\frac{M_2}{2} t^{-\frac{M_2}{2}} y_1^{-1} y_2^{\pm 1} ) \nonumber \\
&\times \Gamma_e ((pq)^\frac{2+M_2-2M_1}{2} t^{\frac{2M_1-M_2-2}{2}} y_1 y_2^{\pm 1} )  \prod_{a<b}^4\Gamma_e ((pq)^\frac{3+M_1-2M_2}{2} c^{-1} t^\frac{2M_2-2-M_1}{2} x_a x_b) \nonumber \\
&\times \Gamma_e ((pq)^{M_2-M_1-1} c^{2} t^{1+M_2-M_2} ) \prod_{a=1}^4\Gamma_e ((pq)^\frac{M_1-1}{4} c^\frac{1}{2} t^\frac{2-M_1}{4} y_1^{-1} x_a ) \prod_{a=1}^4\Gamma_e ((pq)^\frac{3-3M_1}{4} c^\frac{1}{2} t^\frac{3M_1-2}{4} y_1 x_a ) \nonumber \\
&\times \Gamma_e (pq t^{-1})^{M_1+M_2-2} \oint d \vec{z}^{(M_2-1)} d \vec{z}^{(M_2-1)} \prod_{i<j}^{M_1-1}\Gamma_e (pqt^{-1}z_i^{(M_1-1)\pm1} z_j^{(M_1-1)\pm1})  \nonumber \\
&\times \prod_{i<j}^{M_2-1}\Gamma_e (pqt^{-1}z_i^{(M_2-1)\pm1} z_j^{(M_2-1)\pm1}) \prod_{i=1}^{M_1-1} (pq^\frac{M_2-2M_1}{2} c t^{\frac{2M_1-M_2-1}{2}} ( (pq)^\frac{1}{2} t^{-\frac{1}{2} }y_1)^{\pm 1} z_i^{(M_1-1)\pm1} ) \nonumber \\
&\times \prod_{i=1}^{M_1-1} (pq^\frac{3}{2} c^{-1} t^{-1} y_2^{\pm 1} z_i^{(M_1-1)\pm1} ) \prod_{i=1}^{M_2-1} \prod_{j=1}^{M_1-1}  \Gamma_e ( t^{\frac{1}{2}}  z_i^{(M_2-1)\, \pm 1} z_j^{(M_1-1)\, \pm 1}) \nonumber \\
&\times \prod_{i=1}^{M_2-1} \Gamma_e ( (pq)^{-\frac{1}{2}} c t^\frac{1}{2} y_2^{\pm 1} z_i^{(M_2-1)\, \pm 1}) \prod_{i=1}^{M_2-1} \prod_{a=1}^4\Gamma_e (pq)^\frac{5+M_1-2M_2}{4} c^{-\frac{1}{2}} t^\frac{2M_2-4-M_1}{4}  x_a z_i^{(M_2-1) \pm 1}) \,.
\end{align}
Notice that if we neglect the contribution of the singlet fields, the expression above corresponds to the index of the same quiver theory we started with, but with the rank of each gauge node lowered by one unit.

At this point we can repeat the whole procedure that we just described, \textit{i.e.} construct the auxiliary quiver and apply sequentially the Intriligator--Pouliot duality to each gauge node in it. At this stage we will reach a quiver theory which is the same we started with, but with the rank of each gauge node lowered by two units, besides producing an additional bunch of singlets. 

Then it is easy to understand what happens if we iterate the whole procedure $M_1$ times: we will completely confine the first gauge node, produce additional singlets and obtain the expression
  \begin{align} 
&\tilde{\mathcal{I}}^{(2)}_{\text{gauge}}= \prod_{i=1}^{M_1} \Gamma_e((pq)^{M_1+1-i} t^{i-M_1-1}) \prod_{i=1}^{M_1}  \Gamma_e ( (pq)^{1+i} c^{-2} t^{-i}) \prod_{i=1}^{M_1}  \Gamma_e((pq)^{M_2-M_1-i} c^2 t^{M_1-M_2+i-1}) \nonumber \\
&\times \prod_{i=1}^{M_1} \Gamma_e((pq)^{\frac{M_2-2M_1+2i}{2}} t^\frac{2M_1-M_2-2i}{2} y_1^{\pm 1} y_2^{\pm 1}) \prod_{i=1}^{M_1} \prod_{a<b}^4 \Gamma_e ((pq)^\frac{1+2i+M_1-2M_2}{2} c^{-1} t^\frac{2M_2-M_1-2i}{2} x_a x_b) \nonumber \\
&\times \prod_{i=1}^{M_1} \Gamma_e((pq)^{M_2-M_1-i} c^2 t^{M_1-M_2+i}) \prod_{i=1}^{M_1} \prod_{a=1}^4 \Gamma_e ((pq)^\frac{4i-3M_1-1}{4}c^\frac{1}{2} t^\frac{3M_1+2-4i}{4} y_1^{\pm 1} x_a )   \nonumber \\
&\times \Gamma_e (pqt^{-1})^{2M_2-2M_1} \oint d \vec{z}^{M_2-M_1} \prod_{i=1}^{M_2-M_1} \Gamma_e ( (pq)^{-\frac{M_1}{2}} c t^\frac{M_1}{2} y_2^{\pm 1} z_i^{(M_2-M_1) \pm 1}) \nonumber \\&\times \prod_{i=1}^{M_2-M_1}  \prod_{a=1}^4 \Gamma_e ( (pq)^\frac{3+3M_1-2M_2}{4} c^{-\frac{1}{2}} t^\frac{2M_2-3M_1-2}{4} x_a z_i^{(M_2-M_1) \pm 1}) \prod_{i<j}^{M_2-M_1} \Gamma_e ( pq t^{-1} z_i^{(M_2-M_1) \pm 1} z_j^{(M_2-M_1) \pm 1}) \,.
\end{align}
Notice that the integral over $\vec{z}^{(M_2-M_1)}$ in the expression above corresponds to the index of a gauge theory with $USp(2M_2-2M_1)$ gauge group, one antisymmetric chiral and six fundamental chirals. Then we can simply plug in the known expression for the WZ model dual to this gauge theory \eqref{csaki} and finally obtain (again $\tilde{\mathcal{I}}_{\text{WZ}}$ differs from $\mathcal{I}_{\text{WZ}}$ used in \eqref{quiverwzid} for singlet fields)
 \begin{align}
&\tilde{\mathcal{I}}_{\text{WZ}}(\vec{x};\vec{y};c,t) =\prod_{i=1}^{M_1} \Gamma_e((pq)^{M_1+1-i} t^{i-M_1-1}) \prod_{i=1}^{M_1}  \Gamma_e ( (pq)^{1+i} c^{-2} t^{-i}) \prod_{i=1}^{M_1} \Gamma_e((pq)^{M_2-M_1-i} c^2 t^{M_1-M_2+i-1}) \nonumber \\
&\times \prod_{i=1}^{M_1}\Gamma_e((pq)^{\frac{M_2-2M_1+2i}{2}} t^\frac{2M_1-M_2-2i}{2} y_1^{\pm 1} y_2^{\pm 1}) \prod_{i=1}^{M_1} \prod_{a<b}^4 \Gamma_e ((pq)^\frac{1+2i+M_1-2M_2}{2} c^{-1} t^\frac{2M_2-M_1-2i}{2} x_a x_b) \nonumber \\
&\times \prod_{i=1}^{M_1} \Gamma_e((pq)^{M_2-M_1-i} c^2 t^{M_1-M_2+i}) \prod_{i=1}^{M_1} \prod_{a=1}^4 \Gamma_e ((pq)^\frac{4i-3M_1-1}{4}c^\frac{1}{2} t^\frac{3M_1+2-4i}{4} y_1^{\pm 1} x_a )  \nonumber \\
&\times \prod_{j=1}^{M_2-M_1} \Gamma_e ((pq)^j t^{-j})  \prod_{j=1}^{M_2-M_1} \prod_{a<b}^4 \Gamma_e ( (pq)^\frac{1+3M_1-2M_2+2j}{2} c^{-1} t^\frac{2M_2-3M_1-2j}{2} x_a x_b)  \nonumber \\
&\times  \prod_{j=1}^{M_2-M_1} \Gamma_e ( (pq)^{j-1-M_1} c^2 t^{M_1+1-j}) \prod_{j=1}^{M_2-M_1}  \prod_{a=1}^4 \Gamma_e ( (pq)^\frac{M_1-2M_2+4j-1}{4} c^\frac{1}{2} t^\frac{2M_2-M_1+2-4j}{4} y_2^{\pm 1} x_a) \,.
\end{align}
Simplifying the contribution of the massive fields, this reads
\begin{align}  \label{WZ_manynodes}
&\tilde{\mathcal{I}}_{\text{WZ}}(\vec{x};\vec{y};c,t) = \prod_{n=1}^2\prod_{i=1}^{M_{n}-M_{n-1}} \Gamma_e((pq)^{i} t^{-i})\nonumber \\
&\times \prod_{n=1}^2\prod_{i=1}^{M_{n}-M_{n-1}} \Gamma_e((pq)^{M_2+M_{n-1}-M_1-M_n+i-1} c^2 t^{M_1+M_n-M_2-M_{n-1}-1+n-i}) \nonumber \\
&\times \prod_{i=1}^{M_1}\Gamma_e((pq)^{\frac{M_2-2M_1+2i}{2}} t^\frac{2M_1-M_2-2i}{2} y_1^{\pm 1} y_2^{\pm 1}) \prod_{i=1}^{M_2} \prod_{a<b}^4 \Gamma_e ((pq)^\frac{1+2i+M_1-2M_2}{2} c^{-1} t^\frac{2M_2-M_1-2i}{2} x_a x_b) \nonumber \\
&\times \prod_{n=1}^2 \prod_{a=1}^4 \Gamma_e ((pq)^\frac{4i-1+2M_{n-1}-M_1-2M_n}{4} c^\frac{1}{2} t^\frac{2-4i+2M_n+M_1-2M_{n-1}}{4} y_n^{\pm 1} x_a) \,.
\end{align}
Notice that the singlets in the first line reconstruct exactly the $\beta_i^{(n)}$ singlets of the electric theory when brought on the l.h.s. Similarly, the singlets in the second line reconstruct exactly the filipping fields $\gamma_i^{(n)}$ for the dressed diagonal mesons when brought to the l.h.s side. All the other singlets appearing in \eqref{WZ_manynodes} are precisely the contribution to the index of the chiral fields appearing in the WZ that we described above. Then we have proven $\mathcal{I}^{(2)}_{\text{gauge}} (\vec{M};\vec{y},\vec{x};t,c) = \mathcal{I}_{\text{WZ}}^{(2)} (\vec{M};\vec{x};\vec{y};c,t)$ as given in \eqref{quiverwzid}.

\section{A lightening review of $E[USp(2N)]$}
\label{app:eusp}

In this appendix we review a few aspects of the $E[USp(2N)]$ theory that are needed in the main text. This theory was first introduced in \cite{Pasquetti:2019hxf}, see also \cite{Hwang:2020wpd,Garozzo:2020pmz,Hwang:2021xyw,Bottini:2021vms} for other reviews. In this paper we will use a slightly different definition for it than that originally used in \cite{Pasquetti:2019hxf}, which is instead the one used in \cite{Hwang:2020wpd,Garozzo:2020pmz,Bottini:2021vms}. 

\begin{figure}[t]
	\centering
  	\includegraphics[scale=0.55]{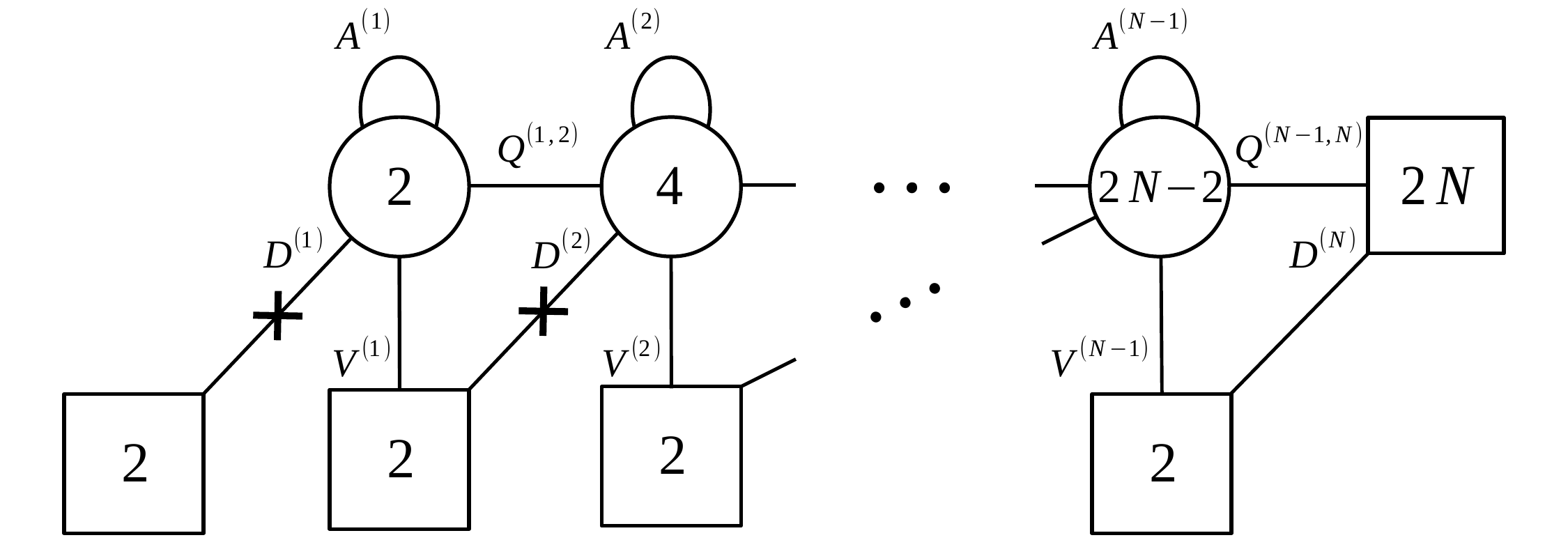} 
   \caption{The quiver diagram for $E[USp(2N)]$. The crosses represent the singlets $b_j$ that flip the diagonal mesons.}
  	\label{euspfields}
\end{figure}

The $E[USp(2N)]$ theory is the $4d$ $\mathcal{N}=1$ quiver theory whose field content is summarized in Figure \ref{euspfields}. The full superpotential is
\begin{align}
\mathcal{W}_{E[USp(2N)]}&=\sum_{j=1}^{N-1}\Tr_{j}\left[A^{(j)}\left(\Tr_{j+1}Q^{(j,j+1)}Q^{(j,j+1)}-\Tr_{j-1}Q^{(j-1,j)}Q^{(j-1,j)}\right)\right]\nn\\
&+\sum_{j=1}^{N-1}\Tr_{y_{j+1}}\Tr_{j}\Tr_{j+1}\left(V^{(j)}Q^{(j,j+1)}D^{(j+1)}\right)+\nn\\
&+\sum_{j=1}^{N-1} b_j\Tr_{y_j}\Tr_{j}\left(D^{(j)}D^{(j)}\right)\, .
\label{superpoteusp}
\end{align}

\begin{figure}[t]
	\centering
  	\includegraphics[scale=0.55]{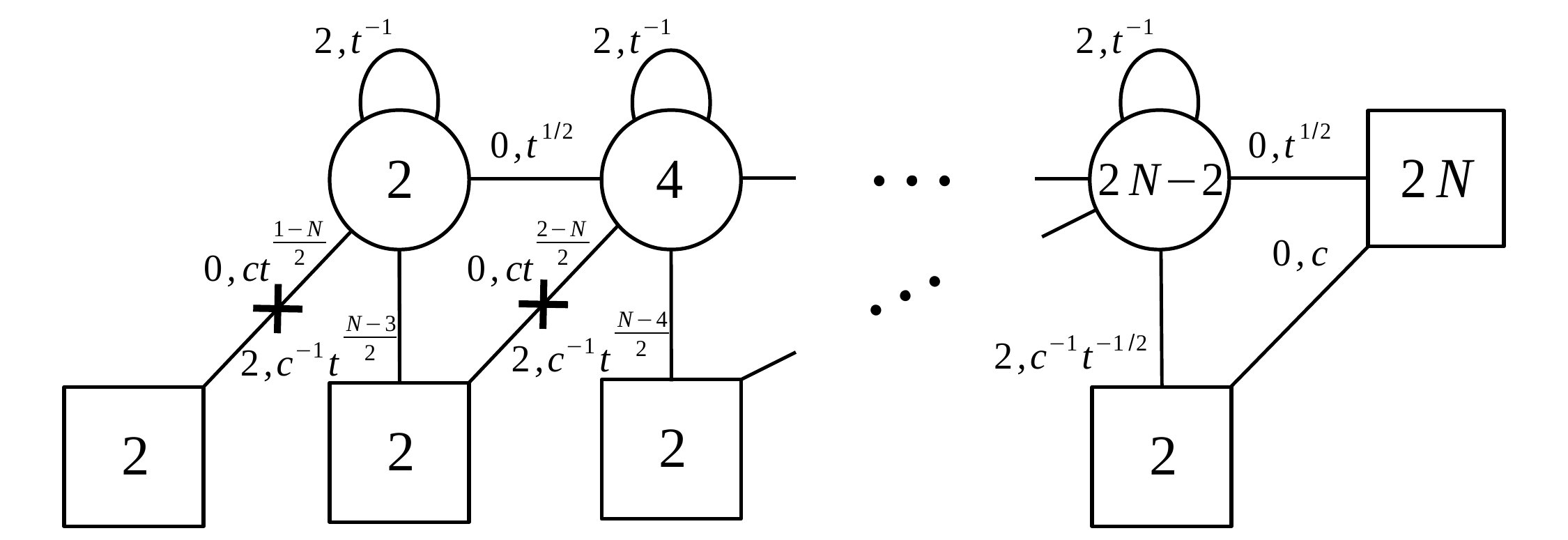} 
   \caption{Charges of the matter fields of $E[USp(2N)]$ under the abelian symmetries and the $R$-symmetry.}
  	\label{euspfugacities}
\end{figure}

It is possible to show \cite{Pasquetti:2019hxf} using dualities or the superconformal index that the manifest global symmetry 
\be
\label{eq:manifest sym}
USp(2N)_x\times\prod_{j=1}^NSU(2)_{y_j}\times U(1)_t\times U(1)_c
\ee
is enhanced in the IR to
\be
USp(2N)_x\times USp(2N)_y\times U(1)_t\times U(1)_c\, .
\ee
The charges of all the chiral fields under the two $U(1)$ symmetries as well as their trial R-charges in our conventions are summarized in Figure \ref{euspfugacities}. 

The gauge invariant operators of $E[USp(2N)]$ that are important for our discussion in the main text are the following:
\begin{itemize}
\item  two operators, which we denote by $\mathsf{H}$ and $\mathsf{C}$, in the traceless antisymmetric representation of $USp(2N)_x$ and $USp(2N)_y$ respectively;
\item an operator $\Pi$ in the bifundamental representation of $USp(2N)_x\times USp(2N)_y$;
\end{itemize}
Those transforming non-trivially under the enhanced $USp(2N)_y$ global symmetry are constructed by collecting several gauge invariant operators transforming properly under the manifest $\prod_{j=1}^NSU(2)_{y_j}$ and with the same charges under all the other symmetries, see \cite{Pasquetti:2019hxf} for more details. The charges and representations of all these operators under the enhanced global symmetry are given in Table \ref{eusptable}.

\begin{table}[t]
\centering
\scalebox{1}{
\begin{tabular}{|c|cccc|c|}\hline
{} & $USp(2N)_x$ & $USp(2N)_y$ & $U(1)_t$ & $U(1)_c$ & $U(1)_{R_0}$ \\ \hline
$\mathsf{H}$ & ${\bf N(2N-1)-1}$ & $\bf 1$ & $1$ & 0 & 0 \\
$\mathsf{C}$ & $\bf1$ & ${\bf N(2N-1)-1}$ & $-1$ & 0 & 2 \\
$\Pi$ & $\bf N$ & $\bf N$ & 0 & $+1$ & 0 \\
\hline
\end{tabular}}
\caption{Trasnformation properties of some of the $E[USp(2N)]$ operators.}
\label{eusptable}
\end{table}

The supersymmetric index of $E[USp(2N)]$ can be defined recursively as follows:
\begin{equation}
\makebox[\linewidth][c]{\scalebox{1}{$
\begin{split}
&\mathcal{I}_{E[USp(2N)]}(\vec x;\vec y;t;c)=  \\
&=\Gpq{pq\,c^{-2}t}\prod_{j=1}^N\Gpq{c\,y_N^{\pm1}x_j^{\pm1}}\oint\udl{\vec{z}_{N-1}^{(N-1)}} \Gamma_e(pq/t)^{N-1} \prod_{a<b}^{N-1} \Gamma_e (pq/t \,z_a^{(N-1)\pm 1} z_b^{(N-1)\pm 1}) \\
&\times\prod_{a=1}^{N-1}\frac{\prod_{j=1}^N\Gpq{t^{1/2}z^{(N-1)}_a{}^{\pm1}x_j^{\pm1}}}{\Gpq{t^{1/2}c\,y_N^{\pm1}z^{(N-1)}_a{}^{\pm1}}}   \mathcal{I}_{E[USp(2(N-1))]}\left(z^{(N-1)}_1,\cdots,z^{(N-1)}_{N-1};y_1,\cdots,y_{N-1};t;t^{-1/2}c\right)\, ,
\label{indexEN}
\end{split}$}}
\end{equation}
with the base of the iteration defined as
\be
\mathcal{I}_{E[USp(2)]}(x;y;c) =\Gpq{c\,x^{\pm1}y^{\pm1}}\,.
\ee

This index coincides up to some prefactor corresponding to singlet fields with the interpolation kernel $\mathcal{K}_c(x,y)$ studied in \cite{2014arXiv1408.0305R}, where many integral identities for this function were proven. Most of them were then interpreted in \cite{Pasquetti:2019hxf} as field theory properties enjoyed by $E[USp(2N)]$. We will now review very quickly those results that are important for us in the present paper.

First of all, $E[USp(2N)]$ enjoys various IR dualities. One, called \emph{mirror duality} in \cite{Hwang:2020wpd}, is actually a self-duality that acts non-trivially on the spectrum of local operators of $E[USp(2N)]$. More precisely, $E[USp(2N)]$ is dual to $E[USp(2N)]^\vee$, which is the same theory but with the $USp(2N)_x$ and $USp(2N)_y$ symmetries exchanged and the $U(1)_t$ fugacity mapped to
\be
t\rightarrow\frac{pq}{t}\,.
\label{redeft}
\ee
Accordingly we have the operator map
\be
\mathsf{H}\quad&\leftrightarrow&\quad \mathsf{C}^\vee\nn\\
\mathsf{C}\quad&\leftrightarrow&\quad \mathsf{H}^\vee\nn\\
\Pi\quad&\leftrightarrow&\quad \Pi^\vee\, .
\label{opmap4dmirror}
\ee
At the level of the index we have the following identity (Theorem 3.1 of \cite{2014arXiv1408.0305R}):
\be
\mathcal{I}_{E[USp(2N)]}(\vec x;\vec y;t,c)=\mathcal{I}_{E[USp(2N)]}(\vec y;\vec x;p q/t,c)\, .
\label{selfduality}
\ee
The reason for the name mirror duality is that in a suitable $3d$ limit it reduces to the well-known self-duality of the $3d$ $\mathcal{N}=4$ $T[SU(N)]$ theory under mirror symmetry \cite{Gaiotto:2008ak}.

There is another duality of $E[USp(2N)]$ called the flip-flip duality. The dual theory, denoted by $FFE[USp(2N)]$, is defined as $E[USp(2N)]$ plus two sets of singlets $\mathsf{O_H}$ and $\mathsf{O_C}$ flipping the two operators $\mathsf{H}^{FF}$ and $\mathsf{C}^{FF}$
\be
\mathcal{W}_{FFE[USp(2N)]}=\mathcal{W}_{E[USp(2N)]}+\Tr_x\left(\mathsf{O_H}\mathsf{H}^{FF}\right)+\Tr_y\left(\mathsf{O_C}\mathsf{C}^{FF}\right)\, .
\ee
In this case the $USp(2N)_x$ and $USp(2N)_y$ symmetries are left unchanged, while only the $U(1)_t$ fugacity transforms as in \eqref{redeft}. The operator map is
\be
\mathsf{H}\quad&\leftrightarrow&\quad \mathsf{O}_{\mathsf{H}}\nn\\
\mathsf{C}\quad&\leftrightarrow&\quad \mathsf{O}_{\mathsf{C}}\nn\\
\Pi\quad&\leftrightarrow&\quad\Pi^{FF}\, .
\ee
At the level of the supersymmetric index we have the following identity (Proposition 3.5 of \cite{2014arXiv1408.0305R}):
\begin{align}
\mathcal{I}_{E[USp(2N)]}(\vec x;\vec y;t;c)&=\prod_{j<l}^N\Gpq{t x_j^{\pm1}x_l^{\pm1}}\Gpq{p q t^{-1} y_j^{\pm1}y_l^{\pm1}}\mathcal{I}_{E[USp(2N)]}(\vec x;\vec y;p q/t;c)\,.
\label{flipflipselfduality}
\end{align}
As shown in \cite{Hwang:2020wpd}, this identity can be derived by iterating \eqref{IP}, meaning that the flip-flip duality can be derived using the IP duality only. Upon reduction to $3d$, we recover the flip-flip duality of $T[SU(N)]$ discussed in \cite{Aprile:2018oau}, which can also be derived by iteratively applying a more fundamental duality, in this case the Aharony duality \cite{Aharony:1997gp}, as shown in \cite{Hwang:2020wpd} (see also Appendix B of \cite{Giacomelli:2020ryy}).

Another property of the index of $E[USp(2N)]$ that will important for us was proven in Theorem 2.16 of \cite{2014arXiv1408.0305R}. This wasn't considered before in the physics literature and we will need it in the next appendix to derive the index identity \eqref{4drankstabfinal} for the rank stabilization duality
\begin{align}
&&\mathcal{I}_{E[USp(2N)]}(\vec{x};\vec{y},t^{k-1}v,\cdots,v;t;c\,d)=\frac{\Gpq{c^2 d^2}}{\Gpq{c^2} \Gpq{d^2}\prod_{i=1}^{k}\frac{\Gpq{t^{1-i}c^2}}{\Gpq{t^{1-i}c^2d^2}}}\prod_{j=1}^N\frac{\Gpq{d\,c\,v\,x_j^{\pm1}}}{\Gpq{(c\,v/d)x_i^{\pm1}}}\times\nn\\
&&\times\prod_{a=1}^{N-k}\frac{\Gpq{c^2v\,y_a^{\pm1}}}{\Gpq{t^kv\,y_a^{\pm1}}}\oint\udl{\vec{z}_{N-k}}\prod_{a=1}^{N-k}\frac{\Gpq{(t^kv/d)z_a^{\pm1}}}{\Gpq{c^2d\,v\,z_a^{\pm1}}}\mathcal{I}_{E[USp(2(N-k))]}(\vec{z};\vec{y};t;d)\times\nn\\
&&\times\mathcal{I}_{E[USp(2N)]}(\vec{x};\vec{z},t^{k-1}v/d,\cdots,v/d;t;c)\, .
\label{generalizedbraid}
\end{align}
We call this identity \emph{generalized braid relation}, since for $k=0$ it coincides with the braid relation appearing in Proposition 2.12 of \cite{2014arXiv1408.0305R}. While the braid relation was interpreted as a field theory duality in \cite{Pasquetti:2019hxf} and derived by iterative application of the IP duality in \cite{Bottini:2021vms}, we lack of both such things for the generalized braid. It would be interesting to investigate more these field theory aspects of the generalized braid relation.

There are two other results of \cite{2014arXiv1408.0305R} that we are going to need (see \cite{Pasquetti:2019hxf,Hwang:2020wpd} for their field theory interpretation). These are basically two evaluation formulas for the supersymmetric index of $E[USp(2N)]$ when some of its fugacities are properly specialized. For example, if we specialize one of the two sets of parameters to a geometric progression, we have (Corollary 2.8 of \cite{2014arXiv1408.0305R})\footnote{This can also be understood as a special case of \eqref{quiverwzid}.}
\begin{align}
\mathcal{I}_{E[USp(2N)]}(\vec{x};t^{N-1}a,t^{N-2}a,\cdots;t;c)=\frac{\Gamma_e(t)^N \prod_{j<l}^{N} \Gamma_e (t\, x_j^{\pm 1} x_l^{\pm 1})}{\prod_{i=1}^{N-1}\Gpq{t^{-i}c^2}}\prod_{j=1}^N\frac{\Gpq{a\,c\,x_j^{\pm1}}\Gpq{\frac{c}{a\,t^{N-1}}x_j^{\pm1}}}{\Gpq{t^j}}\, .\nn\\
\label{conf1}
\end{align}
Another evaluation formula is obtained by properly specifying the fugacity $c$ in terms of the fugacity $t$ (Proposition 2.10 of \cite{2014arXiv1408.0305R})
\begin{align}
\mathcal{I}_{E[USp(2N)]}(\vec{x};\vec{y};t;(pq/t)^{1/2})=\Gamma_e(t)^{N-1} \prod_{j<l}^{N} \Gamma_e (t\, x_j^{\pm 1} x_l^{\pm 1})\prod_{j,l=1}^N\Gpq{(pq/t)^{1/2}x_j^{\pm1}y_l^{\pm1}}\, .
\label{conf2}
\end{align}

\section{Proof of some new index identities}
\label{app:newid}

In this appendix we derive some new elliptic integral identities involving the supersymmetric index of the $E[USp(2N)]$ theory by starting from the results of \cite{2014arXiv1408.0305R} that we have just review. Eventually, we will get the index identity \eqref{4drankstabfinal} for the rank stabilization duality.

The most general identity that we will prove is the following:
\be
&&\oint\udl{\vec{z}_N}\mathcal{I}_{E[USp(2N)]}(\vec{x};\vec{z};t;c)\mathcal{I}_{E[USp(2N)]}(\vec{z};\vec{y},t^{k-1}v,\cdots,v;t;d)\times\nn\\
&&\qquad\qquad\times\prod_{i=1}^N\frac{\Gpq{\frac{v}{d\,u^2}z_i^{\pm1}}}{\Gpq{d\,vz_i^{\pm1}}}\Gpq{s_0z_i^{\pm1}}\Gpq{s_1z_i^{\pm1}}=\nn\\
&&=\frac{\Gpq{c^2}\Gpq{d^2}}{\Gpq{c^2u^{-2}}\Gpq{d^2u^2}}\prod_{i=1}^k\frac{\Gpq{t^{1-i}\frac{1}{u^2}}}{\Gpq{t^{1-i}d^2}}\Gpq{\frac{s_0}{u}\left(\frac{t^{i-1}v}{u\,d}\right)^{\pm1}}\Gpq{\frac{s_1}{u}\left(\frac{t^{i-1}v}{u\,d}\right)^{\pm1}}\times\nn\\
&&\qquad\qquad\times\prod_{i=1}^{N-k}\frac{\Gpq{\frac{v}{u^2}y_i^{\pm1}}}{\Gpq{t^kv\,y_i^{\pm1}}}\prod_{i=1}^N\Gpq{c\,s_0x_i^{\pm1}}\Gpq{c\,s_1x_i^{\pm1}}\times\nn\\
&&\qquad\qquad\times\oint\udl{\vec{w}_{N-k}}\mathcal{I}_{E[USp(2N)]}\left(\vec{x};\vec{w},\frac{t^{k-1}v}{u\,d},\cdots,\frac{v}{u\,d};t;\frac{c}{u}\right)\mathcal{I}_{E[USp(2(N-k))]}(\vec{w};\vec{y};t;d\,u)\times\nn\\
&&\qquad\qquad\times\prod_{i=1}^{N-k}\frac{\Gpq{\frac{t^kv}{u\,d}w_i^{\pm1}}}{\Gpq{\frac{d\,v}{u}w_i^{\pm1}}}\Gpq{\frac{s_0}{u}w_i^{\pm1}}\Gpq{\frac{s_1}{u}w_i^{\pm1}}\, ,\nn\\
\label{newid1}
\ee
which holds when the following balancing condition is satisfied:
\be
s_0s_1=\frac{pq}{c^2}u^2\, .
\label{balancingold}
\ee
Notice that this is a sort of generalization of the generalized braid relation \eqref{generalizedbraid} where we have pairs of integrated kernel functions $\mathcal{I}_{E[USp(2N)]}$, possibly of different lengths, on both sides. 

The proof of \eqref{newid1} is very simple and consists of an iterative application of the generalized braid \eqref{generalizedbraid}. The first step is to rewrite the contribution of the kernel $\mathcal{I}_{E[USp(2N)]}(\vec{z};\vec{y},t^{k-1}v,\cdots,v;t;d)$ on the l.h.s.~of \eqref{newid1} using the generalized braid relation \eqref{generalizedbraid} from right to left
\be
&&\oint\udl{\vec{z}_N}\mathcal{I}_{E[USp(2N)]}(\vec{x};\vec{z};t;c)\mathcal{I}_{E[USp(2N)]}(\vec{z};\vec{y},t^{k-1}v,\cdots,v;t;d)\times\nn\\
&&\times\prod_{i=1}^N\frac{\Gpq{\frac{v}{d\,u^2}z_i^{\pm1}}}{\Gpq{d\,v\,z_i^{\pm1}}}\Gpq{s_0z_i^{\pm1}}\Gpq{s_1z_i^{\pm1}}=\nn\\
&&=\frac{\Gpq{d^2}}{\Gpq{u^{-2}}\Gpq{d^2u^2}}\prod_{i=1}^k\frac{\Gpq{t^{1-i}\frac{1}{u^2}}}{\Gpq{t^{1-i}d^2}}\prod_{i=1}^{N-k}\frac{\Gpq{\frac{v}{u^2}y_i^{\pm1}}}{\Gpq{t^kv\,y_i^{\pm1}}}\oint\udl{\vec{z}_N}\oint\udl{\vec{w}_{N-k}}\mathcal{I}_{E[USp(2N)]}(\vec{x};\vec{z};t;c)\times\nn\\
&&\times\mathcal{I}_{E[USp(2N)]}\left(\vec{z};\vec{w},\frac{t^{k-1}v}{u\,d},\cdots,\frac{v}{u\,d};t;\frac{1}{u}\right)\mathcal{I}_{E[USp(2(N-k))]}(\vec{w};\vec{y};t;d\,u)\times\nn\\
&&\times\prod_{i=1}^N\Gpq{s_0z_i^{\pm1}}\Gpq{s_1z_i^{\pm1}}\prod_{i=1}^{N-k}\frac{\Gpq{\frac{t^kv}{u\,d}w_i^{\pm1}}}{\Gpq{\frac{d\,v}{u}w_i^{\pm1}}}\,.\nn\\
\ee
Now we can remove the original integral by applying the braid relation, that is the generalized braid \eqref{generalizedbraid} in the special case of $k=0$. The result is precisely the claimed identity \eqref{newid1}.

In the identity \eqref{newid1} there are free parameters. We can then try to specialize some of them and use the evaluation formulas \eqref{conf1} and \eqref{conf2} to get some simpler identity with less parameters. The $4d$ rank stabilization identity \eqref{4drankstabfinal} is obtained specializing the $N$ parameters $\vec{x}$ to a geometric progression and fixing the parameter $d$ in terms of $t$
\be
\vec{x}=(t^{N-1}a,t^{N-2}a,\cdots,a),\qquad d=\left(\frac{pq}{t}\right)^{\frac{1}{2}}\, .
\ee
First we combine \eqref{conf1} with the mirror duality \eqref{selfduality} and the flip-flip duality \eqref{flipflipselfduality} to obtain
\begin{align}
&\mathcal{I}_{E[USp(2N)]}(t^{N-1}a,t^{N-2}a,\cdots;\vec{x};t;c) \nonumber \\
&=\frac{\Gpq{t}^N \prod_{j<l}^N\Gpq{t (t^{j-1} a)^{\pm1}(t^{l-1} a)^{\pm1}}}{\prod_{i=1}^{N-1}\Gpq{t^{-i}c^2}}\prod_{j=1}^N\frac{\Gpq{a\,c\,x_j^{\pm1}}\Gpq{\frac{c}{a\,t^{N-1}}x_j^{\pm1}}}{\Gpq{t^j}}\, .\nn\\
\label{eq:flipped_conf1}
\end{align}
Using the identities \eqref{eq:flipped_conf1} and \eqref{conf2}, we can evaluate both of the kernel functions on the l.h.s. and one of those on the r.h.s as follows:
\be
&&\oint\udl{\vec{z}_N}\prod_{i=1}^N\Gpq{a\,c\,z_i^{\pm1}}\Gpq{\frac{c}{a\,t^{N-1}}z_i^{\pm1}}
\frac{\Gpq{\frac{v}{u^2}\left(\frac{t}{pq}\right)^{\frac{1}{2}}z_i^{\pm1}}}{\Gpq{\left(\frac{pq}{t}\right)^{\frac{1}{2}}v\, z_i^{\pm1}}}\Gpq{s_0z_i^{\pm1}}\Gpq{s_1z_i^{\pm1}}\times\nn\\
&&\times\prod_{j=1}^k\Gpq{\left(\frac{pq}{t}\right)^{\frac{1}{2}}\left(t^{j-1}v\right)^{\pm1}z_i^{\pm1}}\prod_{j=1}^{N-k}\Gpq{\left(\frac{pq}{t}\right)^{\frac{1}{2}}z_i^{\pm1}y_j^{\pm1}}\Gpq{t}^N \prod_{i<j}^N \Gpq{t z_i^{\pm 1} z_j^{\pm 1}}=\nn\\
&&=\prod_{i=1}^k\frac{\Gpq{\frac{t^{1-i}}{u^2}}}{\Gpq{\frac{pq}{t^i}}}\Gpq{\frac{s_0}{u}\left(\frac{t^{i-1}v}{u\left(\frac{pq}{t}\right)^{1/2}}\right)^{\pm1}}\Gpq{\frac{s_1}{u}\left(\frac{t^{i-1}v}{u\left(\frac{pq}{t}\right)^{1/2}}\right)^{\pm1}}\Gpq{\frac{a\,c}{u}\left(\frac{t^{i-1}v}{u\left(\frac{pq}{t}\right)^{1/2}}\right)^{\pm1}}\times\nn\\
&&\times\Gpq{\frac{c}{u\,a\,t^{N-1}}\left(\frac{t^{i-1}v}{u\left(\frac{pq}{t}\right)^{1/2}}\right)^{\pm1}}\prod_{i=1}^N\frac{\Gpq{t^{1-i}c^2}}{\Gpq{t^{1-i}\frac{c^2}{u^2}}}\Gpq{c\,s_0\left(t^{i-1}a\right)^{\pm1}}\Gpq{c\,s_1\left(t^{i-1}a\right)^{\pm1}}\times\nn\\
&&\times\prod_{i=1}^{N-k}\frac{\Gpq{\frac{v}{u^2}y_i^{\pm1}}}{\Gpq{t^kv\,y_i^{\pm1}}}\oint\udl{\vec{w}}_{N-k}\mathcal{I}_{E[USp(2(N-k))]}\left(\vec{w};\vec{y};t;\left(\frac{pq}{t}\right)^{\frac{1}{2}}u\right)\times\nn\\
&&\Gpq{t\,u^{-2}}\times\prod_{i=1}^{N-k}\Gpq{\frac{a\,c}{u}w_i^{\pm1}}\Gpq{\frac{c}{a\,t^{N-1}u}w_i^{\pm1}}\frac{\Gpq{\frac{t^kv}{u\left(\frac{pq}{t}\right)^{1/2}}w_i^{\pm1}}}{\Gpq{\frac{\left(\frac{pq}{t}\right)^{1/2}v}{u}w_i^{\pm1}}}\Gpq{\frac{s_0}{u}w_i^{\pm1}}\Gpq{\frac{s_1}{u}w_i^{\pm1}}\, .\nn\\
\label{4drankstabold}
\ee
Now we redifine the parameters $(a,c,v,u,s_0,s_1)$ in the following way:
\be
\begin{cases}
x_1=a\,c\\
x_2=\frac{c}{a\,t^{N-1}}\\
x_3=\frac{v}{u^2}\left(\frac{t}{pq}\right)^{1/2}\\
x_4=(pq)^{1/2}t^{1/2-k}v^{-1}\\
x_5=s_0\\
x_6=s_1
\end{cases}
\label{reparametrization}
\ee
Notice that the balancing condition \eqref{balancingold} becomes
\be
t^{N+k-2}\prod_{a=1}^6x_a=pq\, .
\label{balancingnew}
\ee
The system \eqref{reparametrization} can be inverted to get
\be
\begin{cases}
a=t^{\frac{1-N}{2}}\sqrt{\frac{x_1}{x_2}}\\
c=t^{\frac{N-1}{2}}\sqrt{x_1x_2}\\
v=(pq)^{1/2}t^{1/2-k}x_4^{-1}\\
u=\frac{t^{\frac{1-k}{2}}}{\sqrt{x_3x_4}}\\
s_0=x_5\\
s_1=x_6
\end{cases}
\label{reparametrizationinverse}
\ee
With this redefinition of the parameters and taking into account \eqref{balancingnew}, we can rewrite \eqref{4drankstabold} as (also redefine $k\to N-k$ to compare with Section \ref{sec:rankstab})
\be
&&\oint\udl{\vec{z}}_N\prod_{i=1}^N\prod_{a=1}^6\Gpq{x_az_i^{\pm1}}\prod_{j=1}^{k}\Gpq{\left(pq\right)^{1/2}t^{-1/2}z_i^{\pm1}y_j^{\pm1}} \Gpq{t}^N \prod_{i<j}^N \Gpq{t z_i^{\pm 1} z_j^{\pm 1}}=\nn\\
&&\qquad=\prod_{j=1}^{N-k}\Gpq{t^j}\prod_{a<b}^6\Gpq{t^{j-1}x_ax_b}\prod_{j=N-k+1}^{N}\prod_{a<b=1,2,5,6}\Gpq{t^{j-1}x_ax_b}\times\nn\\
&&\qquad\qquad\times\Gpq{t^{N-k} x_3 x_4}\prod_{j=1}^k\Gpq{(pq)^{1/2}t^{-1/2}x_3y_j^{\pm1}}\Gpq{(pq)^{1/2}t^{-1/2}x_4y_j^{\pm1}}\times\nn\\
&&\qquad\qquad\times\oint\udl{\vec{w}}_{k}\mathcal{I}_{E[USp(2k)]}\left(\vec{w};\vec{y};t;\left(\frac{pq}{t^{N-k}}\right)^{1/2}\frac{1}{\sqrt{x_3x_4}}\right)\prod_{i=1}^{k}\Gpq{t^{\frac{N-k-1}{2}}x_1\sqrt{x_3x_4}w_i^{\pm1}}\times\nn\\
&&\qquad\qquad\times\Gpq{t^{\frac{N-k-1}{2}}x_2\sqrt{x_3x_4}w_i^{\pm1}}\Gpq{t^{\frac{N-k-1}{2}}x_5\sqrt{x_3x_4}w_i^{\pm1}}\Gpq{t^{\frac{N-k-1}{2}}x_6\sqrt{x_3x_4}w_i^{\pm1}}\times\nn\\
&&\qquad\qquad\times\Gpq{t^{\frac{N-k+1}{2}}\left(\frac{x_3}{x_4}\right)^{\pm1/2}w_i^{\pm1}}\, ,\nn\\
\label{4drankstabnew}
\ee
with balancing condition
\be
t^{2N-k-2}\prod_{a=1}^6x_a=pq\, .
\label{balancingnewnew}
\ee

This identity is already the one for the rank stabilization and it is written in a parametrization of the fugacities wuch that for $k=0$ we precisely recover \eqref{csaki}. Nevertheless, since in the dual theory the $SU(6)_x$ symmetry is enhanced in the IR while in the UV only an $SU(4)_{v}\times SU(2)_b\times U(1)_s$ subgroup is visible, it is useful to write the r.h.s.~in terms of different fugacities so to make the UV symmetry explicitly manifest
\be
\begin{cases}
s=(pq)^{1/12}t^{-\frac{2N-k-2}{12}}(x_3 x_4)^{-1/4} \\
b=\left(x_3/x_4\right)^{-1/2} \\
v_a=\left(\frac{t^{2N-k-2}}{pq}\right)^{1/4} (x_3 x_4)^{1/4} \, x_a & a=1,2,5,6
\end{cases}
\ee
Notice that the balancing condition \eqref{balancingnewnew} translates into the $SU(4)_{v}$ condition $\prod_{a=1,2,5,6}v_a=1$. Moreover, by replacing redefinition on the r.h.s.~of \eqref{4drankstabnew} we can see that the fugacity $b$ indeed appears as an $SU(2)_b$ fugacity
\be
&&\oint\udl{\vec{z}_N}\Gamma_e(t)^N \prod_{i<j}^N \Gamma_e (t z_i^{\pm 1} z_j^{\pm 1})\prod_{i=1}^N\prod_{a=1}^6\Gpq{x_az_i^{\pm1}}\prod_{j=1}^{k}\Gpq{\left(pq\right)^{1/2}t^{-1/2}z_i^{\pm1}y_j^{\pm1}}=\nn\\
&&\qquad=\Gpq{(pq)^{\frac{1}{3}}t^{\frac{N-2k+2}{3}}s^{-4}}\prod_{i=1}^N\prod_{a<b=1,2,5,6}\Gpq{(pq)^{1/3}t^{\frac{3i+k-2N-1}{3}}s^2v_av_b}\times\nn\\
&&\qquad\qquad\times\prod_{j=1}^{N-k}\Gpq{t^j}\prod_{a=1,2,5,6}\Gpq{(pq)^{1/3}t^{\frac{3j+k-2N-1}{3}}s^{-1}b^{\pm1}v_a}\times\nn\\
&&\qquad\qquad\times\Gpq{(pq)^{1/3}t^{\frac{3j+k-2N-1}{3}}s^{-4}}\prod_{j=1}^k\Gpq{(pq)^{2/3}t^{\frac{-1+k-2N}{6}}s^{-2}b^{\pm1}y_j^{\pm1}}\times\nn\\
&&\qquad\qquad\times\oint\udl{\vec{w}_k}\mathcal{I}_{E[USp(2k)]}\left(\vec{w};\vec{y};t;(pq)^{1/3}t^{\frac{2k-N-2}{6}}s^2\right)\times\nn\\
&&\qquad\qquad\times\Gpq{t^{\frac{N-k+1}{2}}b^{\pm1}w_i^{\pm1}}\prod_{i=1}^{k}\prod_{a=1,2,5,6}\Gpq{(pq)^{1/3}t^{\frac{1-N-k}{6}}s^{-1}v_aw_i^{\pm1}}\, .\nn\\
\ee
Up to the reparametrization: $v_{1,2,5,6}\to v_{1,2,3,4}$ and $x_a \rightarrow (pq)^{\frac{1}{6}}t^{-\frac{2N-k-2}{6}}x_a$, this is exactly \eqref{4drankstabfinal}.

\bibliographystyle{ytphys}
\bibliography{refs}

\end{document}